\documentclass[conference]{IEEEtran}
\usepackage{graphicx}
\usepackage{amsthm}
\usepackage{epsfig}
\usepackage{latexsym}
\usepackage{amsfonts}
\usepackage{here}
\usepackage{rawfonts}
\usepackage[latin1]{inputenc}
\usepackage[T1]{fontenc}
\usepackage{calc}
\usepackage{capitalgreekitalic}
\usepackage{url}
\usepackage{enumerate}
\usepackage{color}
\usepackage[tbtags]{amsmath}
\usepackage{amssymb}
\usepackage{upref}
\usepackage{epic,eepic}
\usepackage{times}
\usepackage{dsfont}
\usepackage{comment}
\usepackage{cite}
\usepackage{multirow}
\usepackage{rotating}

\usepackage{dsfont}

\newcounter{step}
\newlength{\totlinewidth}
  {\end{list}%
  \rule{\linewidth}{1pt}}
\newcounter{substep}

  {\end{list}}

\newlength{\aligntop}
\newlength{\alignbot}
 

\IEEEoverridecommandlockouts

\usepackage{algorithm}
\usepackage{algorithmic}
\usepackage{subfigure}

\begin{document}
\clearpage
\title{\huge {\color{black} Artificial Neural Networks-Based Machine Learning for Wireless Networks: A Tutorial}}
%
\author{
{Mingzhe Chen\IEEEauthorrefmark{1}$^,$\IEEEauthorrefmark{2}$^,$\IEEEauthorrefmark{3}}, {Ursula Challita\IEEEauthorrefmark{4}}, Walid Saad\IEEEauthorrefmark{5}, Changchuan Yin\IEEEauthorrefmark{1}, and M\'erouane Debbah \IEEEauthorrefmark{6}\vspace*{0em}\\
\authorblockA{\small \IEEEauthorrefmark{1}Beijing Laboratory of Advanced Information Network, Beijing University of Posts and Telecommunications, Beijing, China 100876,\\ Email: \protect\url{chenmingzhe@bupt.edu.cn} and \protect\url{ccyin@bupt.edu.cn.} \\
\IEEEauthorrefmark{2} The Future Network of Intelligence Institute, The Chinese University of Hong Kong, Shenzhen, China.\\
\IEEEauthorrefmark{3} Department of Electrical Engineering, Princeton University, Princeton, NJ, USA.\\
\IEEEauthorrefmark{4} School of Informatics, The University of Edinburgh, Edinburgh, UK. Email: \protect{ursula.challita@ed.ac.uk.}\\
\IEEEauthorrefmark{5}Wireless@VT, Bradley Department of Electrical and Computer Engineering, Virginia Tech, Blacksburg, VA, USA, Email: \protect\url{walids@vt.edu.}\\
\IEEEauthorrefmark{6}\small Mathematical and Algorithmic Sciences Lab, Huawei France R \& D, Paris, France, \\Email: merouane.debbah@huawei.com.
  }
%
%
%
%

\maketitle
\thispagestyle{empty}
\vspace{0cm}
\begin{abstract}
In order to effectively provide ultra reliable low latency communications and pervasive connectivity for Internet of Things (IoT) devices, next-generation wireless networks can leverage intelligent, data-driven functions enabled by the integration of machine learning notions across the wireless core and edge infrastructure. \color{black}
{\color{black}In this context, this paper provides a 
comprehensive tutorial that overviews how artificial neural networks (ANNs)-based machine learning algorithms can be employed for solving various wireless networking problems.}
 For this purpose, we first present a detailed overview of a number of key types of ANNs that include recurrent, spiking, and deep neural networks, that are pertinent to wireless networking applications. For each type of ANN, we present the basic architecture as well as specific examples that are particularly important for wireless network design. Such examples include echo state networks, liquid state machine, and long short term memory. And then, we provide an in-depth overview on the variety of wireless communication problems that can be addressed using ANNs, ranging from communication using unmanned aerial vehicles to virtual reality applications over wireless networks and edge computing and caching. For each individual application, we present the main motivation for using ANNs along with the associated challenges while we also provide a detailed example for a use case scenario and outline future works that can be addressed using ANNs. In a nutshell, this article constitutes the first holistic tutorial on the development of ANN-based machine learning techniques tailored to the needs of future wireless networks.
\vspace{-0.1cm}

\end{abstract}
\section{Introduction}
The wireless networking landscape is undergoing a major revolution. The smartphone-centric networks of yesteryears are gradually morphing into an \emph{Internet of Things (IoT)} ecosystem{\cite{luong2016data,dawy2017toward,park2016learning}}
that integrates a heterogeneous mix of wireless-enabled devices ranging from smartphones, to drones, connected vehicles, wearables, sensors, and virtual reality {\color{black}devices}. This unprecedented transformation will not only drive an exponential growth in wireless traffic in the foreseeable future, but it will also lead to the emergence of new and untested wireless service use cases, that substantially differ from conventional multimedia or voice-based services{\cite{boccardi2014five}}. For instance, beyond the need for high data rates -- which has been the main driver of the wireless network evolution in the past decade -- {\color{black}next-generation wireless networks will also have to deliver ultra-reliable, low-latency communication \cite{boccardi2014five} and{\cite{3gpp.TR.36.881}}, that is adaptive and in real-time to the dynamics of the IoT users and the IoT's physical environment.}
 For example, drones and connected vehicles \cite{7470933} will place autonomy at the heart of the IoT. This, in turn, will necessitate the deployment of ultra-reliable wireless links that can provide real-time, low-latency control for such autonomous systems{\cite{zeng2018joint,8469055,8320780}}. Meanwhile, in tomorrow's wireless networks, {\color{black}large volumes of data will be collected}, periodically and in real-time, across a massive number of sensing and wearable devices that monitor physical environments. Such massive short-packet transmissions will lead to a substantial traffic over the wireless uplink, which has traditionally been much less congested than the downlink{\cite{7529226}}. This same wireless network must also support cloud-based gaming \cite{gopal2016emerging}, immersive virtual reality services{\cite{3gpp.26.928}}, real-time HD streaming, and conventional multimedia services. This ultimately creates a radically different networking environment whose novel applications and their diverse quality-of-service (QoS) and reliability requirements mandate a fundamental change in the way in which wireless networks are modeled, analyzed, designed, and optimized.

The need to cope with this ongoing and rapid evolution of wireless services has led to a considerable body of research that investigates what the optimal cellular network architecture will be within the context of the emerging fifth generation (5G) wireless networks (e.g., see \cite{andrews2014will} {\color{black}and the references} therein). While the main ingredients for 5G -- such as dense small cell deployments, millimeter wave  (mmWave) communications, and device-to-device (D2D) communications -- have been identified, integrating them into a truly harmonious wireless system that can meet the IoT challenges requires instilling \emph{intelligent functions across both {\color{black}the edge and the core} of the network}. These intelligent functions must be able to adaptively exploit the wireless system resources and the generated data, in order to optimize the network operations and guarantee, in real-time, the QoS needs of emerging wireless and IoT services. {\color{black} Such mobile edge and core intelligence can potentially be realized by integrating fundamental notions of \emph{machine learning (ML)} \cite{segaran2007programming}, in particular, \emph{artificial neural network (ANN)-based ML approaches}, across the wireless infrastructure and the end-user devices. ANNs\cite{yegnanarayana2009artificial} are a computational nonlinear machine learning framework can be used for supervised learning, unsupervised learning \cite{822529}, semi-supervised learning \cite{bennett1999semi}, and reinforcement learning \cite{mnih2015human}, in various wireless networking scenarios. Hereinafter, ML is used to refer to ANN-based ML.}

 
{\color{black} \subsection{Role of ANNs in Wireless Networks}
 ML tools are undoubtedly one of the most important tools for endowing wireless networks with intelligent functions, as evidenced by the wide adoption of ML in 
 a myriad of applications domains \cite{andrieu2003introduction,freeman2000learning,sebastiani2002machine,collobert2008unified,pang2002thumbs,bishop2006pattern}. 
In the context of wireless networks, ML will enable any wireless device \emph{to actively and intelligently monitor its environment by learning and predicting the evolution of the various environmental features} (e.g., wireless channel dynamics, traffic patterns, network composition, content requests, user context, etc.) and \emph{proactively taking actions that maximize the chances of success for some predefined goal, which, in a wireless system, pertains to some sought after quality-of-service.} 
 ML enables the network infrastructure to learn from the wireless networking environment and take adaptive network optimization actions.
 In consequence, ML is expected to play several roles in the next-generation of wireless networks \cite{bi2015wireless,VerizonAIML, Ericssonmobilityreport, QualcommAIResearch,ITUml5g}. 
 
 First, the most natural application of { ML} in a wireless system is to exploit \emph{intelligent and predictive data analytics} to enhance situational awareness and the overall network operations {\cite{bi2015wireless}}. In this context, ML will provide the wireless network with the ability to parse through massive amounts of data, generated from multiple sources that range from wireless channel measurements and sensor readings to drones and surveillance images, in order to create a comprehensive operational map of the massive number of devices within the network{\cite{ferber1999multi}}. This map can, in turn, be exploited to optimize various functions, such as fault monitoring and user tracking, across the wireless network.

 Second, beyond its powerful intelligent and predictive data analytics functions, ML will be a major driver of intelligent and data-driven \emph{wireless network optimization}{\cite{ferber1999multi}}. For instance, ML tools will enable the introduction of intelligent resource management tools, that can be used to address a variety of problems ranging from cell association and radio access technology selection to frequency allocation, spectrum management, power control, and intelligent beamforming. In contrast to the conventional distributed optimization techniques, that are often done iteratively in an offline or semi-offline manner{\cite{8187196}}, ML-guided resource management mechanisms will be able to operate in a fully online manner by learning, in real time, the states of the wireless environment and the network's users. Such mechanisms will therefore be able to continuously improve their own performance over time which, in turn, will enable more intelligent and dynamic network decision making. Such ML-driven decision making is essential for much of the envisioned IoT and 5G services, particularly those that require real-time, low latency operation, such as autonomous driving, drone guidance, and industrial control. In fact, if properly designed, {\emph{ML optimization algorithms will provide inherently} \emph{self-organizing, self-healing, and self-optimizing} solutions for a broad range of problems within the context of network optimization and resource management. Such ML-driven self-organizing solutions are particularly apropos for {ultra dense wireless networks} in which classical centralized and distributed optimization approaches can no longer cope with the scale and the heterogeneity of the network.

 Third, beyond its system-level functions, ML can play a key role at the \emph{physical layer} of a wireless network{\cite{o2017introduction}}. As shown in \cite{o2017introduction,o2017learning,o2017deep,liang2017iterative,nachmani2017deep,samuel2017deep}, {ML} tools can be used to redefine the way in which physical layer functions, such as coding and modulation, are designed, at both transmitter and receiver levels, within a generic communication system. Such an ML-driven approach has been shown \cite{o2017introduction,o2017learning,o2017deep,liang2017iterative,nachmani2017deep,samuel2017deep} to have a lot of promise in delivering lower bit error rates and better robustness to the wireless channel impediments.

 Last, but not least, the rapid deployment of highly \emph{user-centric wireless services}, such as virtual reality~\cite{bacstuug2016towards}, in which the gap between the end-user and the network functions is almost minimal, strongly motivates the need for wireless networks that can track and adapt to the human user behavior. In this regard, ML is perhaps the only tool that is capable to learn and mimic human behavior, which will help in creating the wireless network to adapt its functions to its human users, thus creating a truly immersive environment and to maximize the overall quality-of-experience (QoE) of the users.
 
{From the above discussion, we can further narrow down the introduction of ML in wireless networks to imply two key functions: 1) {\color{black}Intelligent and predictive data analytics}, the ability of the wireless network to intelligently process large volumes of data, gathered from its devices, in order to analyze and predict {\color{black}the context of the wireless users and the wireless network's} environmental states thus enabling data-driven network-wide operational decisions, and 2) intelligent/self-organizing network control and optimization, the ability of the wireless network to dynamically learn the wireless environment and intelligently control the wireless network and optimize its resources according to information smartly learned about the wireless environment and users' states.}  


  Clearly, the ML-based system operation is no longer a privilege, but rather a necessity for future wireless networks. ML-driven wireless network designs will pave the way towards an unimaginably rich set of new network functions and wireless services. For instance, even though 5G networks may not be fully ML capable, we envision that the subsequent, sixth generation (6G) \cite{saad2019vision} of wireless cellular networks will surely integrate important tools from ML, as evidenced {\color{black}by the recent} development of intelligent mobile networks proposed by Huawei \cite{HuaweiR} and the ``big innovation house" proposed by Qualcomm \cite{QualcommCEO}. \emph{As such, the question is no longer {if} ML tools are going to be integrated into wireless networks but rather {when} such an integration will happen}. {In fact, the importance of an ML-enabled wireless network has already been motivated by a number of recent wireless networking paradigms, such as mobile edge caching, context-aware networking, and mobile edge computing~\cite{hu2015mobile,ahmed2016survey,sardellitti2015joint,nunna2015enabling,lee2017online,mao2016dynamic,chen2016efficient,semiari2015context}, the majority of which use ML techniques for various tasks such as user behavior analysis and predictions so as to determine which contents to cache and how to proactively allocate computing resources}. However, despite their importance, these works have a narrow focus and do not provide any broad, tutorial-like material that can shed light on the challenges and opportunities associated with the use of ML for designing intelligent wireless networks.

{\color{black}\subsection{Previous Works}}
{A number of surveys and tutorials on {ML} applications in wireless networking have been published, for example, {\color{black}\cite{park2016learning}, \cite{o2017introduction}, and \cite{sun2017learning,jiang2017machine,bkassiny2013survey,alsheikh2014machine, demuth2014neural,         8444669,8382166,8373692,7932863,7982603,sun2018application,luong2018applications, you2019ai}.} Nevertheless, these works are limited in a number of ways. First, {\color{black}a majority of the existing works} focuses on a single ML technique (often the basics of deep learning \cite{o2017introduction}, \cite{sun2017learning}, and \cite{8382166,8373692,7932863} or reinforcement learning \cite{luong2018applications}) and, as such, they do not capture the rich spectrum of available ML frameworks. Second, they mostly restrict their scope to a single wireless application such as sensor networks~\cite{alsheikh2014machine}, cognitive radio networks~\cite{bkassiny2013survey}, machine-to-machine (M2M) communication~\cite{park2016learning}, physical layer design~\cite{o2017introduction}, software defined networking~\cite{8444669}, Internet of Things~\cite{8373692}, or~self-organizing networks (SONs)~\cite{7982603}, and, hence, they do not comprehensively cover the broad range of applications that can adopt ML in future networks. Third, a large number of the existing surveys and tutorials, such as \cite{park2016learning}, \cite{jiang2017machine,bkassiny2013survey,alsheikh2014machine,sun2018application}, and \cite{you2019ai}{{\footnote{\color{black} The main difference between our tutorial and \cite{you2019ai} is that the authors in \cite{you2019ai} do not provide a comprehensive tutorial on how a broad range of ANNs can be used for solving the wireless communication problems related to drone-based communications, spectrum management with multiple radio access technologies, wireless virtual reality, mobile edge caching and computing, and the IoT.}}, are highly qualitative and do not provide an in-depth technical and quantitative description on the variety of existing ML tools that are suitable for wireless communications.
 Last, but not least, some surveys discuss the basics of neural networks with applications outside of wireless communications. However, these surveys are largely inaccessible to the wireless community, due to their reliance on examples from rather orthogonal disciplines such as computer vision. Moreover, most of the existing tutorials or surveys do not provide concrete guidelines on how, when, and where to use different artificial neural network (ANN) tools in the context of wireless networks.
Finally, the introductory literature on ML for wireless networks such as in \cite{park2016learning}, \cite{ o2017introduction}, and \cite{sun2017learning,jiang2017machine,bkassiny2013survey,alsheikh2014machine,demuth2014neural,          8444669,8382166,8373692,7982603,7932863,sun2018application,luong2018applications, you2019ai}, is largely sparse and fragmented and provides very scarce details on the role of ANNs, hence, making it difficult to understand the intrinsic details of this broad and far reaching area. {\color{black}Table \ref{ta:existingworks} summarizes the difference between this tutorial and the magazine, tutorial, and survey papers. From Table \ref{ta:existingworks},  we can see that, compared to the existing works such as \cite{park2016learning, o2017introduction, sun2017learning ,jiang2017machine, bkassiny2013survey,alsheikh2014machine,demuth2014neural,8444669,8382166,8373692,7932863,7982603,sun2018application,luong2018applications,you2019ai }, our tutorial provides a more detailed exposition of several types of ANNs that are particularly useful for wireless applications and explains, pedagogically and, in detail, how to develop ANN-based ML solutions to endow intelligent wireless networks and realize the full potential of 5G systems, and beyond.}

 \begin{table*}
 {\color{black}
\centering
  \newcommand{\tabincell}[2]{\begin{tabular}{@{}#1@{}}#2\end{tabular}}
\renewcommand\arraystretch{1}
 \caption{
    \vspace*{-0.1em}Comparison of This Work With Existing Survey and Tutorial Papers. {Here, ``CC", ``CR", ``DT", ``PL", and ``DA"  refer to caching and computing, cognitive radio network, data traffic domain, physical layer domain, and data analytics.  }  }\label{ta:existingworks}\vspace*{-0.6em}
\centering
\begin{tabular}{|c | c| c| c |c |c |c |c|c |c|c |c|c |c|c |c |c |}
\hline
 \multicolumn{1}{|c|}{\multirow{2}{1cm}{\textbf{Existing Works}}} &  \multicolumn{7}{|c|}{\multirow{1}{*}{\textbf{ Key Machine Learning Tools }}} &   \multicolumn{9}{|c|}{\multirow{1}{*}{\textbf{ Key Applications}}}  \\
  \cline{2-17}
 &\textbf{FNN}&\textbf{RNN}&\textbf{DNN}&\textbf{ESN}&\textbf{SNN}&\textbf{DA }&\textbf{RL}&\textbf{UAV}&\textbf{VR}&\textbf{CC}&\textbf{SON} &\textbf{Multi-RAT}&\textbf{IoT}&\textbf{PL}&\textbf{CR}&\textbf{DT} \\
\hline
\cite{park2016learning}&&&  & &&$\surd$&$\surd$&&&&&& $\surd$&&&\\
\cite{o2017introduction}& &  &$\surd$ & &&&&&&&&&&$\surd$&&\\
\cite{sun2017learning}&& & $\surd$ & &&&&&&&&&&&&\\
\cite{jiang2017machine}   &&&&&&$\surd$&$\surd$&&&&&$\surd$&&&$\surd$& \\
\cite{bkassiny2013survey}&&&&&&$\surd$&$\surd$&&&&&&&$\surd$&& \\
\cite{alsheikh2014machine}&& &$\surd$&   && &&&&&&$\surd$&$\surd$&$\surd$&&\\
\cite{demuth2014neural}&$\surd$ && & &&$\surd$&$\surd$&&&&&&&&& \\
\cite{schmidhuber2015deep}& && $\surd$ & &&&&&&&&&&&& \\
\cite{8444669}& & &$\surd$  & &&&&&&&$\surd$ &&&&&\\
\cite{8382166}& $\surd$& & $\surd$ & &&$\surd$ &$\surd$&&&&&&&&&\\
\cite{8373692}& &$\surd$ & $\surd$ & &&$\surd$ &$\surd$&&&&&&$\surd$&&&$\surd$\\
\cite{7932863}& $\surd$&$\surd$ &$\surd$  &&&$\surd$&$\surd$&&&&&&&&&$\surd$ \\
\cite{7982603}&$\surd$ & &$\surd$  & &&$\surd$ &$\surd$&&&&$\surd$&&&&&\\
\cite{sun2018application}&$\surd$ &$\surd$ &$\surd$  & &&$\surd$ &$\surd$&&&$\surd$&&$\surd$&&&&$\surd$\\
\cite{luong2018applications}&$\surd$ &$\surd$ &$\surd$  & &&$\surd$ &$\surd$&&&$\surd$&&$\surd$&&$\surd$&&$\surd$\\
\cite{you2019ai}&$\surd$ &&$\surd$  & &&$\surd$ &$\surd$&&&&$\surd$&$\surd$&&$\surd$&&\\
Our tutorial&$\surd$ &$\surd$ &$\surd$  &$\surd$ &$\surd$&$\surd$ &$\surd$&$\surd$&$\surd$&$\surd$&&$\surd$&$\surd$&&&\\
\hline
\end{tabular}
}
\end{table*}

{\color{black}\subsection{Contributions}}
{\color{black} The main contribution of this paper is, thus, to provide a tutorial on the topic of ANN-based ML for wireless network design}
  The overarching goal is to give {\color{black}a tutorial} on the emerging research contributions, from ANNs and wireless communications,
that address the major opportunities and challenges in developing ANN-based ML frameworks for understanding and designing intelligent wireless systems. {To the best of our knowledge, this is the first tutorial that gathers the state-of-the-art and emerging research contributions related to the use of ANNs 
for addressing a set of communication problems in beyond 5G wireless networks.  } Our main contributions include:
\begin{itemize}
\item We provide a comprehensive treatment of \emph{artificial neural networks}, 
with an emphasis on how such tools can be used to create a new breed of ML-enabled wireless networks.
\item {\color{black}After providing a brief introduction to the basics of ML, we provide a more detailed exposition of ANNs that are particularly useful for wireless applications, such as recurrent, spiking, and deep neural networks. For each type, we provide an introduction on their basic architectures and a specific use-case example. Other ANNs that can be used for wireless applications are also briefly mentioned where appropriate.}
\item Then, we discuss a broad range of wireless applications that can make use of ANN. These applications include drone-based communications, spectrum management with multiple radio access technologies, wireless virtual reality, mobile edge caching and computing, and {the IoT system}, among others. For each application, we first outline the main rationale for applying ANNs while pinpointing illustrative scenarios. Then, we expose the challenges and opportunities brought forward by the use of ANNs in the specific wireless application. We complement this discussion with a detailed example drawn from the state-of-the-art and, then, we conclude by shedding light on the potential future works within each specific area.
\end{itemize}


  \begin{figure*}
  \begin{center}
   \vspace{0cm}
    \includegraphics[width=14cm]{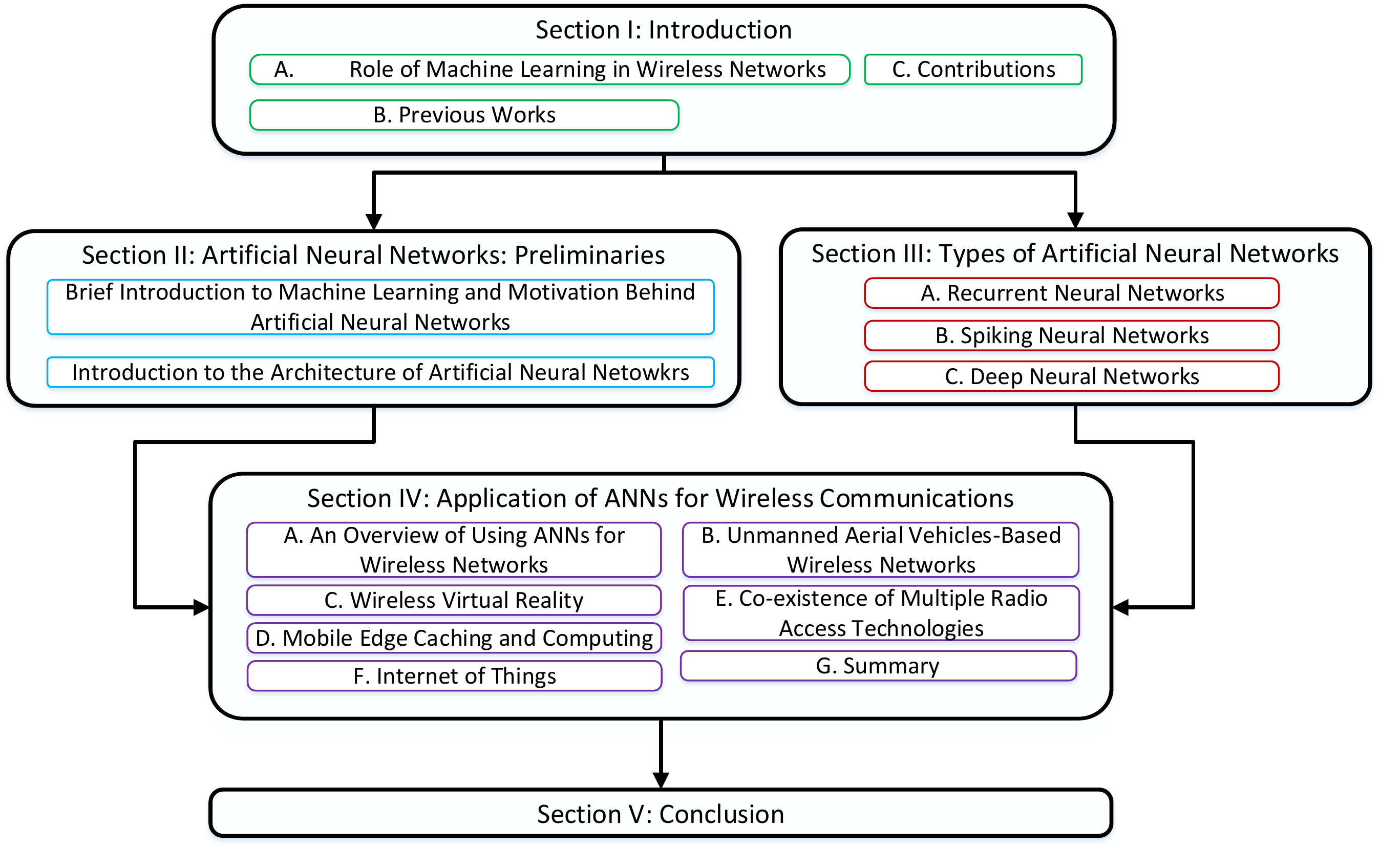}
    \vspace{-0.3cm}
 {\color{black}   \caption{\label{architecture} Organization of the tutorial.}}
  \end{center}\vspace{-0.8cm}
\end{figure*}

The rest of this tutorial is organized as follows {(Fig. \ref{architecture})}. In Section II, we introduce the basics of ANNs. Section III presents several key types of ANNs such as recurrent neural networks {(RNNs)}, spiking neural networks {(SNNs)}, and deep neural networks {(DNNs)}. In Section IV, we discuss the use of ANNs for wireless communication and the corresponding challenges and opportunities.   Finally, conclusions are drawn in Section V.

\section{Artificial Neural Networks: Preliminaries}\label{se:ANNP}
ML was born from pattern recognition and it is essentially based on {\color{black} the premise that intelligent machines should be able to} learn from and adapt to their environment through experience \cite{andrieu2003introduction,freeman2000learning,sebastiani2002machine,collobert2008unified,pang2002thumbs,bishop2006pattern}. Due to {the ever} growing volumes of generated data -- across critical infrastructures, communication networks, and smart cities --  and the need for intelligent data analytics, the use of ML algorithms has become ubiquitous \cite{MLW} across many sectors, such as in financial services, health care, technology, and entertainment. Using ML algorithms to build models that uncover connections and predict dynamic system or human behavior, system operators can make intelligent decisions without any human intervention. For example, in a wireless system such as the IoT, ML tools can be used for intelligent data analytics and edge intelligence. ML tasks often depend on the nature of their training data. In ML, \emph{training} is the process that teaches the machining learning framework to achieve a specific goal, such as for speech recognition. In other words, training enables the ML framework to discover potential relationships between the input data and the output data of this machine learning framework. There exist, in general, four key classes of learning approaches \cite{alpaydin2014introduction}: a) supervised learning, b) unsupervised learning, c) semi-supervised learning, and d) reinforcement learning.

Supervised learning algorithms are trained using labeled data{\cite{alpaydin2014introduction}}. When dealing with labeled data, both the input data and its desired output data are known to the system. Supervised learning is commonly used in applications that have enough historical data. In contrast, the training of unsupervised learning tasks is done without labeled data{\cite{alpaydin2014introduction}}. The goal of unsupervised learning is to explore the data and infer some structure directly from the unlabeled data. Semi-supervised learning is used for the same applications as supervised learning but it uses both labeled and unlabeled data for training{\cite{alpaydin2014introduction}}. This type of learning can be used with methods such as classification, regression, and prediction. Semi-supervised learning is useful when the cost of a fully-labeled training process is relatively high. In contrast to the previously discussed learning methods that need to be trained with historical data, RL is trained by the data collected from implementation of the RL {\cite{alpaydin2014introduction}}. The goal of RL is to learn an environment and find the best strategies for a given agent, in different environments. The RL algorithms are particularly interesting in the context of wireless network optimization \cite{5978427}. To perform supervised, unsupervised, semi-supervised, or RL learning tasks, several  frameworks have been developed. 
Among those frameworks, ANNs \cite{demuth2014neural} are arguably the most important, as they are able to mimic human intelligence.

ANNs are inspired by the structure and functional aspects of biological neural networks, that can learn from complicated or imprecise data{\cite{demuth2014neural}}. 
Within the context of wireless communications, as it will be clearer from the later sections, ANNs can be used to investigate and predict network and user behavior so as to provide user information for solving diverse wireless networking problems such as cell association, spectrum management, computational resource allocation, and cached content replacement. Moreover, recent developments of smart devices and mobile applications have significantly increased the level at which human users interact with mobile systems. A trained ANN can be thought of as an ``expert'' in dealing with human-related data. Therefore, using ANNs to extract information from the user environment can provide a wireless network with the ability to predict the users' future behaviors and, hence, to design an optimal strategy to improve the resulting QoS and reliability.
 
There are various types of ANNs (see {Fig. \ref{machinelearning}):
 \begin{itemize}
 \item \emph{Modular neural networks}: A modular neural network (MNN) is composed of several independent ANNs and an intermediary. In an MNN, each ANN is used to complete one subtask of the entire task that an MNN wants to perform. An intermediary is used to process the output of each independent ANN and generate the output of an MNN.
 
\item \emph{Recurrent neural networks}: RNNs are ANN architectures that allow neuron connections from a neuron in one layer to neurons in previous layers. According to different activation functions and connection methods for the neurons in an RNN, RNNs can be used to define several different architectures: a) stochastic neural networks, b) bidirectional neural networks (BNNs), c) fully recurrent neural network (FRNN), d) neural Turing machines (NTMs), e) long short-term memories (LSTMs), e) echo state networks (ESNs), f) simple recurrent neural networks (SRNNs), and g) gated recurrent units (GRUs). 

\item \emph{Generative adversarial networks}: Generative adversarial networks (GANs) consist of two neural networks. One neural network is used to learn a map from a latent space to a particular data distribution, while another neural network is used to discriminate between the true data distribution and the distribution mapped by the neural network.

\item \emph{Deep neural networks}:  All the ANNs that have multiple hidden layers are known as DNNs. 

\item \emph{Spiking neural networks}: The spiking neural networks consist of spiking neurons that accurately mimic the biological neural networks.

\item \emph{Feedforward neural networks}: In a feedforward neural network (FNN), each neuron has incoming connections only from the previous layer and outgoing connections only to the next layer. FNNs can be used to define more advanced architectures such as: a) extreme learning machines (ELMs), b) convolutional neural networks (CNNs), c) time delay neural networks (TDNNs), d) autoencoders, e) probabilistic neural networks (PNUs), and e) radial basis functions (RBFs).

\item \emph{Physical neural networks}: In a physical neural network (PNN), an electrically adjustable resistance material is used to emulate the function of a neural activation.

\end{itemize}
}
Each type of ANN is suitable for a particular learning task. {For instance, RNNs are effective in dealing with time-dependent data while SNNs are effective in dealing with continuous data.}
{It should be noted that most of the data collected by wireless networks is time-dependent and continuous. In particular, in wireless networks, the user context and behavior, the wireless signals, and the wireless channel conditions are all time-dependent and continuous. RNNs and SNNs are effective in dealing with such collected data. They can exploit this data for various purposes, such as network control and user behavior predictions. }  {However, since RNNs or SNNs can record only a limited size of historical data, they may not be able to solve all of the wireless communication problems. To solve complex wireless problems that cannot be solved by shallow RNNs and SNNs, one can use DNNs which have a high memory capacity for data analytics and can separate the complex problem that needs to be learned into a composition of several simpler problems thus making the learning process effective. In consequence, in Section~III, we specifically introduce RNNs, SNNs, and DNNs that are most suited for wireless network use cases. } 

  \vspace{-0cm}
\begin{figure*}[!t]
  \begin{center}
   \vspace{0cm}
    \includegraphics[width=18cm]{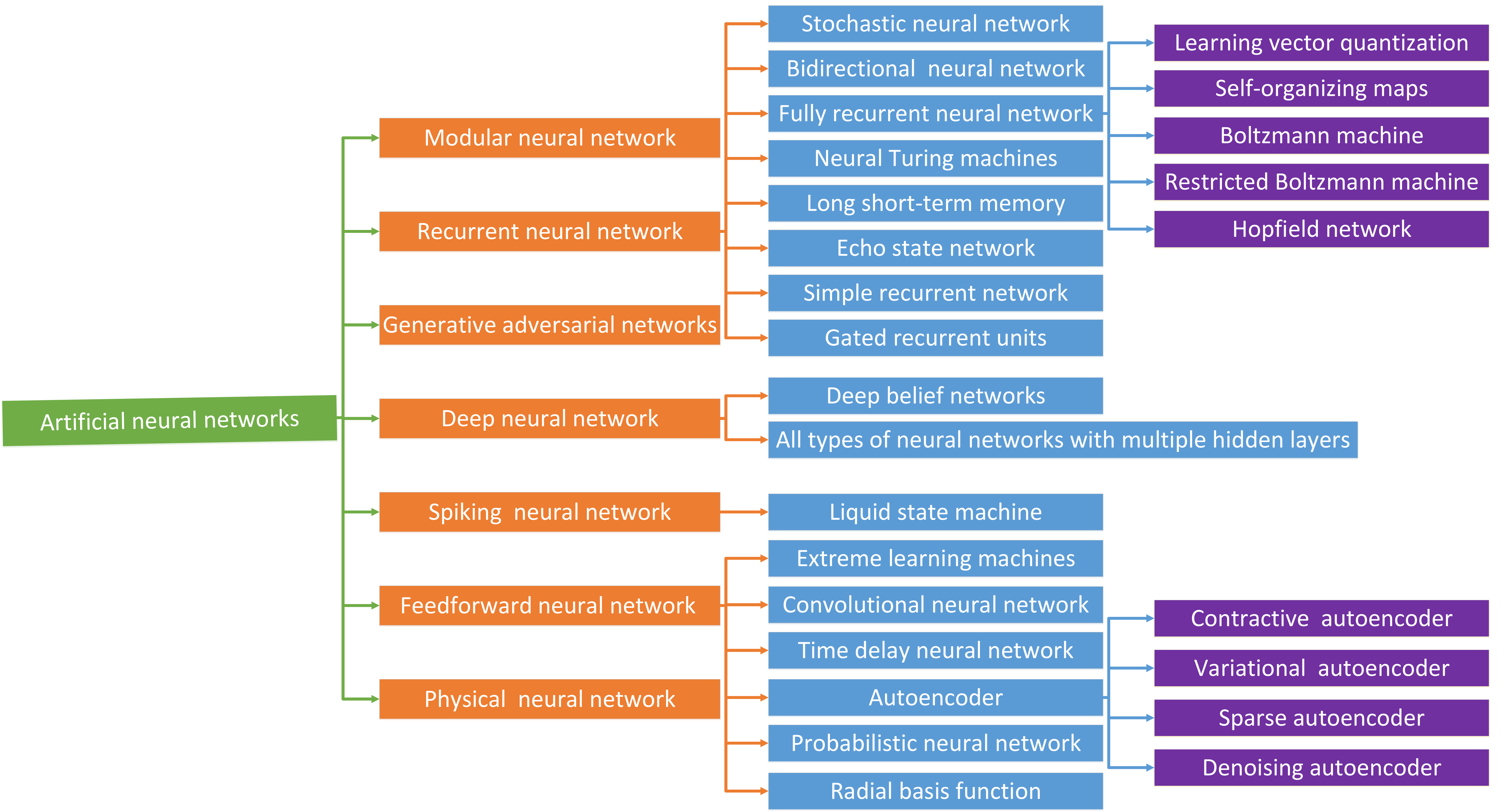}
    \vspace{-0.1cm}
    \caption{\label{machinelearning}{Summary of artificial neural networks.}}
  \end{center}\vspace{-0.1cm}
\end{figure*}

  \section{Types of Artificial Neural Networks}\label{se:ANN}
In this section, we specifically discuss three types of ANNs: RNNs, SNNs, and DNNs, that have a promising potential for wireless network design, as will become clear in Section~IV. For each kind of ANN, we briefly introduce its architecture, advantages, and properties. Then, we present specific example architectures. 
 \subsection{Recurrent Neural Networks} \label{section:RNN}
  \subsubsection{\textbf{Architecture of Recurrent Neural Networks}}
  \vspace{-0cm}
\begin{figure}[!t]
  \begin{center}
    \includegraphics[width=7cm]{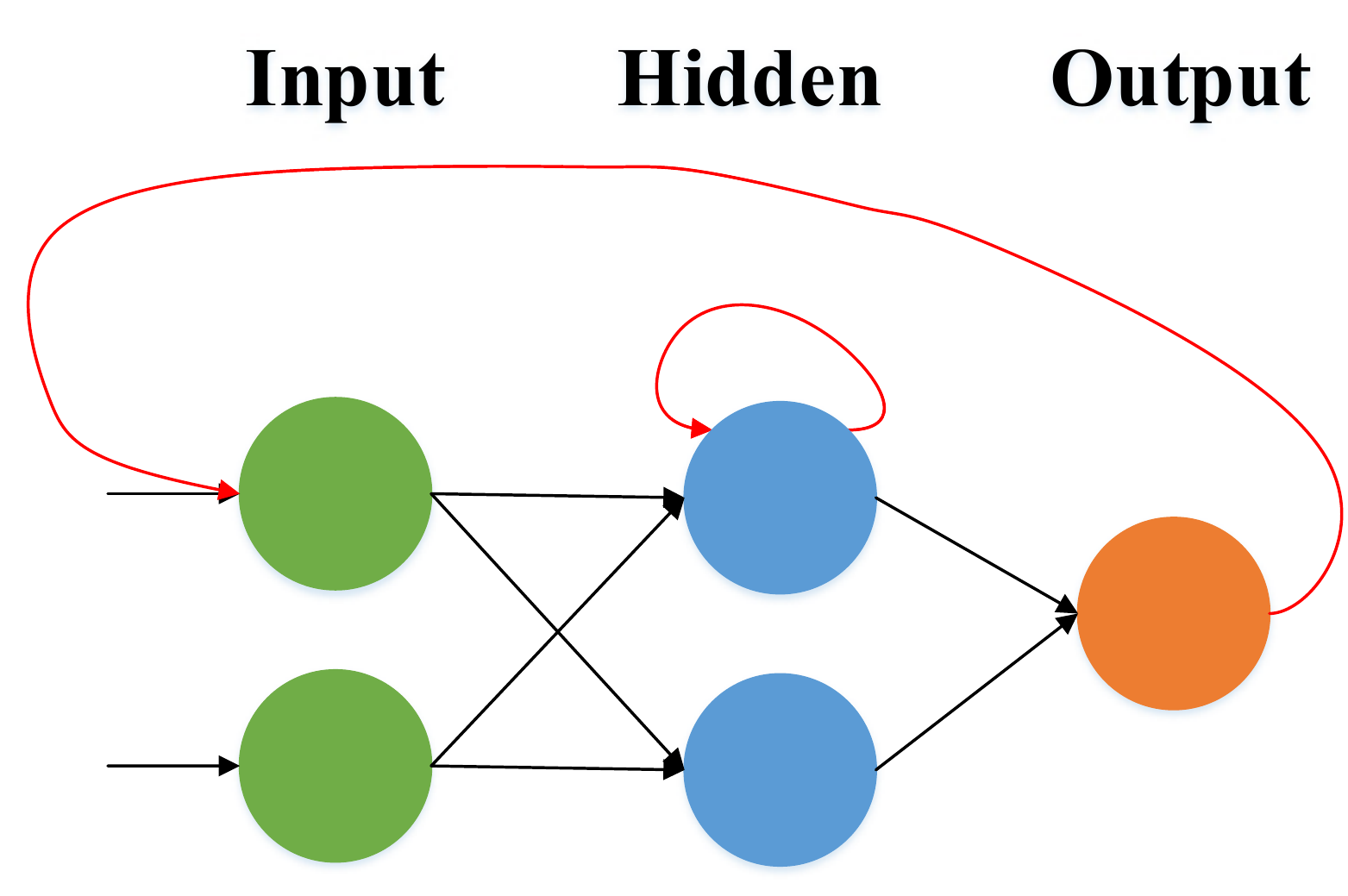}
    \vspace{-0.3cm}
    \caption{\label{RNN} Recurrent neural network architecture.}
  \end{center}\vspace{-0.8cm}
\end{figure}

  \vspace{-0cm}
\begin{figure*}[!t]
  \begin{center}
   \vspace{0cm}
    \includegraphics[width=16cm]{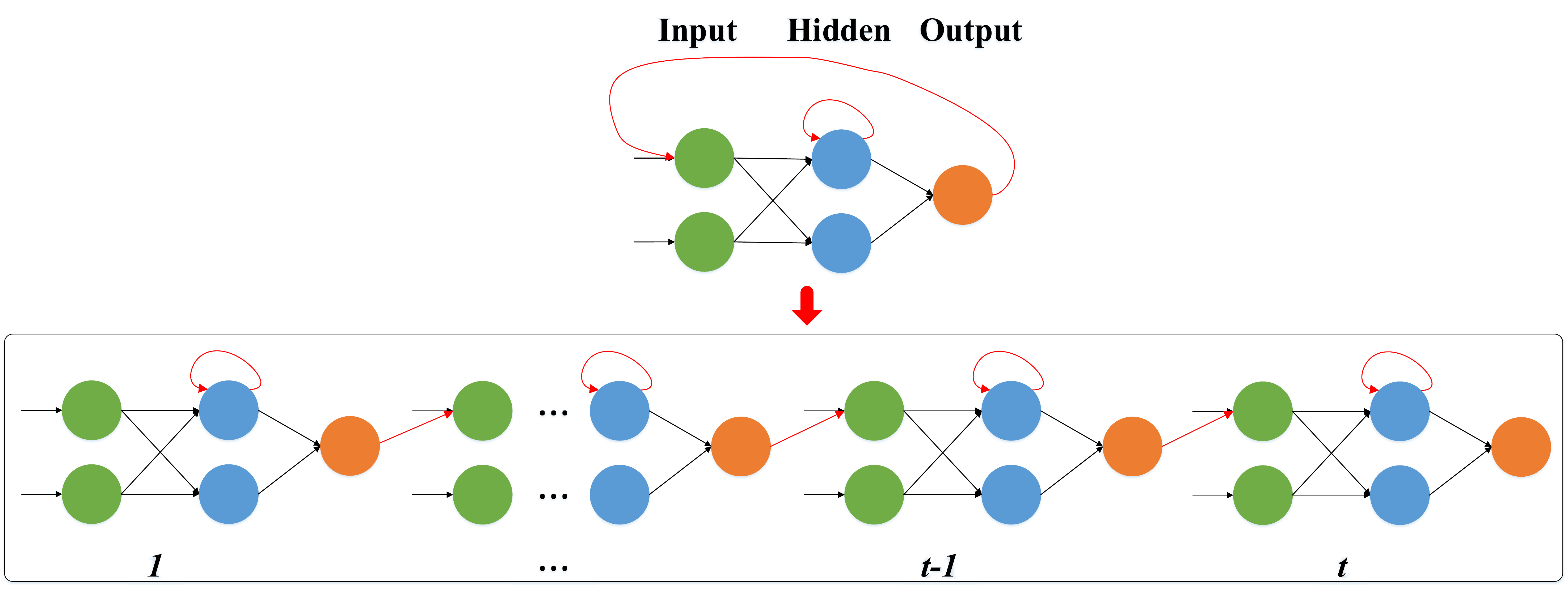}
    \vspace{-0.3cm}
    \caption{\label{exRNN}Architecture of an unfolded recurrent neural network.}
  \end{center}\vspace{-0.8cm}
\end{figure*}

In a traditional ANN, it is assumed that all the inputs or all the outputs are independent from each other. However, for many tasks, the inputs (outputs) are related. For example, for predicting the mobility patterns of wireless devices, the input data, that is the users' locations, are certainly related. To this end,
\emph{recurrent neural networks} \cite{mandic2001recurrent}, which are ANN architectures that allow neuron connections from a neuron in one layer to neurons in previous layers \cite{mandic2001recurrent}, as shown in Fig. \ref{RNN}, have been introduced. This seemingly simple change enables the output of a neural network to depend, not only on the current input, but also on the historical input, as shown in Fig. \ref{exRNN}. This allows RNNs to make use of sequential information and exploit dynamic temporal behaviors such as those faced in mobility prediction or speech recognition. For example, an RNN can be used to predict the mobility patterns of  mobile devices and wireless users. These patterns are related to the historical locations that the wireless users have visited. This task cannot be done in one step without combing historical locations from previous steps. 
Therefore, the ANNs whose output depends only on the current input, such as FNNs, cannot perform such highly time-dependent tasks.
A summary of the key advantages and disadvantages of RNNs for wireless applications is presented in Table~\ref{ta:2}.
 Note that, in theory, RNNs can make use of historical information in arbitrarily long sequences, but in practice they are limited to only a subset of the historical information \cite{lukovsevivcius2009reservoir}. 
For training RNNs, the most commonly used algorithms include the \emph{backpropagation through time (BPTT) algorithm}\cite{werbos1990backpropagation}. 
However, RNNs require more time to train compared to traditional ANNs (e.g., FNNs) since each value of the activation function depends on the series data recorded in RNNs. To reduce the training complexity of RNNs, one promising solution is to develop an RNN that needs to only train the output weight matrix. Next, we specifically introduce this type of RNNs, named echo state networks (ESNs) \cite{APractical}.

 \begin{table*}
\centering
  \newcommand{\tabincell}[2]{\begin{tabular}{@{}#1@{}}#2.4\end{tabular}}
\renewcommand\arraystretch{1}
 \caption{
    \vspace*{-0.05em}Summary of the Advantages and Disadvantages of ANNs for Wireless Applications }\label{ta:2}\vspace*{-0.5em}
\centering
\begin{tabular}{|c|c|l|l|}
\hline
 \multicolumn{1}{|c|}{\multirow{2}{*}{\textbf{}}}&\multicolumn{1}{|c|}{\multirow{2}{*}{\textbf{Typical type of input data}}}   &  \multicolumn{1}{|c|}{\multirow{2}{*}{\textbf{ Advantages}}} &  \multicolumn{1}{|c|}{\multirow{2}{*}{\textbf{ Drawbacks}}} \\ 
 &&&\\
\hline
\multirow{6}{*}{\textbf{RNNs}}& \multirow{6}{*}{{Time-dependent data}} &\multirow{1}{6cm}{$\bullet$ Effectiveness in processing time-related data\\~~~such as wireless traffic} &\multirow{2}{6cm}{$\bullet$ Training complexity due to the loop connections\\~~~between neurons} \\
&&&\\
&&\multirow{2}{6cm}{$\bullet$ Ability to capture dynamic temporal behaviors\\ ~~~~such as content requests or device mobility} & \multirow{1}{*}{$\bullet$ Limited memory to record historical data}\\ 
&&&\\
&&\multirow{1}{6cm}{$\bullet$ Ability to make use of sequential information\\~~~~such as sequential symbols received by a user} & \\
&&&\\
\hline

\multirow{6}{*}{\textbf{SNNs}}& \multirow{6}{*}{{Continuous data}} &\multirow{1}{6cm}{$\bullet$ Effectiveness in processing continuous data\\~~~such as amplitudes of wireless signals} &\multirow{1}{6cm}{$\bullet$ Training complexity due to dynamic neurons} \\
&&&$\bullet$ Specific training method is needed for each type of SNN\\
&&$\bullet$ Large memory available for data collection&$\bullet$ Need to sample the states of neurons \\ 
&&\multirow{1}{6cm}{$\bullet$ Ability to cope with rapidly changing, dynamic\\~~~network behavior (e.g., dynamic traffic)} & \\ 
&&&\\
&&$\bullet$ Ability to perform multiple learning tasks& \\ 
\hline
\multirow{5}{*}{\textbf{DNNs}}& \multirow{5}{*}{{High-dimensional data}}  &$\bullet$\multirow{3}{6cm}{ Inherent ability to find low-dimensional representations (features) of high-dimensional data such as images and wireless traffic pattern} &\multirow{1}{6cm}{$\bullet$ Hard to tune for practical applications} \\
&&&$\bullet$ Large training dataset is required\\
&&&$\bullet$ Computationally intensive to train\\
&&$\bullet$ Better learning capability compared to shallow ANNs &  \\ 
&&$\bullet$ Effective in learning very complex functions&\\
\hline
\end{tabular}
 \vspace{-0.2cm}

\end{table*}

 \subsubsection{\textbf{Example RNN -- Echo State Networks}}\label{se:ESN}
ESNs are known to be a highly practical type of RNNs due to their effective approach for training \cite{Short}. In fact, ESNs reinvigorated interest in RNNs \cite{Minimum} by making them accessible to wider audiences due to their apparent simplicity. In an ESN, the input weight matrix and the hidden weight matrix are randomly generated without any specific training. Therefore, ESN needs to only train the output weight matrix.  
ESNs can, in theory, approximate any arbitrary nonlinear dynamical system with any arbitrary precision, they have an inherent temporal processing capability, and are therefore a very powerful enhancement of the linear blackbox modeling techniques in nonlinear domain. 
Due to the ESN's appealing properties such as training simplicity and ability to record historical information, it has been widely applied for supervised learning tasks, 
RL tasks, 
classification, and regression. In wireless networks, ESNs have been applied for various natural applications, such as content prediction, resource management, and mobility pattern estimation, as it will be clear in Section IV. Next, the specific architecture and training methods for ESNs are introduced.

\noindent $\bullet$ \emph{Architecture of an Echo State Network:} ESNs use an RNN architecture with only one hidden layer\footnote{Deep generalizations of ESNs also exist \cite{gallicchio2016deep}}. {\color{black} We define that the input vector of an ESN as ${\boldsymbol{x}_{t}}=\left[{x}_{t,1},\ldots,{x}_{t,N_\textrm{in}} \right]^{\rm T}$ and the output vector of an ESN as ${\boldsymbol{y}_{t}}=\left[{y}_{t,1},\ldots,{y}_{t,N_\textrm{out}} \right]^{\rm T}$. 
An ESN model consists of the input weight matrix $\boldsymbol{W}_\textrm{in} \in {\mathbb{R}^{N \times N_\textrm{in}}} $, the recurrent weight matrix ${\boldsymbol{W} \in {\mathbb{R}^{N \times \left(N+1\right)}} }$, the leaking rate $\alpha$, and the output weight matrix {$\boldsymbol{W}_\textrm{out} \in {\mathbb{R}^{N_\textrm{out} \times \left(1+N+N_\textrm{in} \right)}} $}, where $N$ is the number of neurons in the hidden layer.  
The leaking rate $\alpha$ must be chosen to match the speed of the dynamics of the hidden states {${\boldsymbol{s}_{t}}=\left[{s}_{t,1},\ldots,{s}_{t,N} \right]^{\rm T}$, where ${s}_{t,i}$ represents the state of neuron $i$ at time $t$,} and output $\boldsymbol{y}_t$. 
To allow ESNs to store historical information, the hidden state $\boldsymbol{s}_t$ should satisfy the so-called \emph{echo state property}, which means that the hidden state $\boldsymbol{s}_t$  should be uniquely defined by the fading history of the input $\boldsymbol{x}_0, \boldsymbol{x}_1, \ldots, \boldsymbol{x}_t$. This is in contrast to traditional ANNs, such as FNNs, that need to adjust the weight values of the neurons in the hidden layers, ESNs only need to guarantee the echo state property.
 Typically, in order to guarantee the {echo state property} of an ESN, the spectral radius of $\boldsymbol{W}$ should be smaller than 1. The setting of other ESN components to guarantee the {echo state property} and to optimize ESN performance can be found in \cite{APractical}.}

Having described the main components of ESNs, we now describe the activation value of each neuron. Even though the input and the hidden weight matrices are fixed (randomly), all the neurons of an ESN will have their own activation values (hidden state). As opposed to the classical RNNs in which the hidden state depends only on the current input, in ESNs, the hidden state will be given by:
\begin{equation}\label{eq:hs}
{{\boldsymbol{\tilde s}}_{t}} ={\mathop{f}\nolimits}\!\left( {\boldsymbol{W}{\left[1; \boldsymbol{s}_{t - 1}\right]} + \boldsymbol{W}_\textrm{in}{\boldsymbol{x}_{t}}} \right),
\end{equation}
\begin{equation}
{\boldsymbol{ s}}_{t}=\left(1-\alpha\right)\boldsymbol{ s}_{t-1}  +\alpha {{\boldsymbol{\tilde s}}_{t}},
\end{equation}
where $f\!\left(x\right)=\frac{{{e^x} - {e^{ - x}}}}{{{e^x} + {e^{ - x}}}}$ and $\left[ { \cdot ; \cdot }  \right]$ represents a vertical vector (or matrix) concatenation. The model is also sometimes used without the leaky integration, which is a special case for $\alpha=1$ yielding ${{\boldsymbol{\tilde s}}_{t}}={{\boldsymbol{ s}}_{t}} $. From (\ref{eq:hs}), we can see that the scaling of $\boldsymbol{W}_\textrm{in}$ and $\boldsymbol{W}$ determines the proportion of how much the current state $\boldsymbol{s}_t$ depends on the current input $\boldsymbol{x}_t$ and how much on the previous state $\boldsymbol{s}_{t-1}$. Here, a feedback connection from $\boldsymbol{y}_{t-1}$ to $\boldsymbol{s}_t$ can be applied to the ESNs, defined as a weight matrix {$\boldsymbol{W}_\textrm{fb} \in {\mathbb{R}^{N \times N_\textrm{out}}} $}. Hence, (\ref{eq:hs}) can be rewritten as ${{\boldsymbol{\tilde s}}_{t}} ={\mathop{f}\nolimits}\!\left( {\boldsymbol{W}{\left[1; \boldsymbol{s}_{t - 1}\right]} + \boldsymbol{W}_\textrm{in}{\boldsymbol{x}_{t}}+\boldsymbol{W}_\textrm{fb}\boldsymbol{y}_{t-1} } \right)$.

Based on the hidden state ${\boldsymbol{ s}}_{t}$, the output signal of the ESN can be given by:
\begin{equation}\label{eq:ESNy}
y_t= \boldsymbol{W}_\textrm{out}\left[1;{\boldsymbol{ s}}_{t};\boldsymbol{x}_t\right].
\end{equation}
Here, an additional nonlinearity can be applied to (\ref{eq:ESNy}), i.e., $\boldsymbol{y}_t= \textrm{tanh}\left( \boldsymbol{W}_\textrm{out}\left[1;{\boldsymbol{ s}}_{t};\boldsymbol{x}_t\right]\right)$. \vspace{0.5 em}

\noindent $\bullet$ \emph{Training in Echo State Networks}: The training process in ESNs seeks to minimize the mean square error (MSE) between the targeted output and the actual output. When this MSE is minimized, the actual output will be the target output which can be given by $\boldsymbol{y}_t^\textrm{D}= \boldsymbol{W}_\textrm{out}\left[1;{\boldsymbol{ s}}_{t};\boldsymbol{x}_t\right]$ where $\boldsymbol{y}_t^\textrm{D}$ is the targeted output. Therefore, the training goal is to find an optimal $\boldsymbol{W}_\textrm{out}$ such that $\boldsymbol{W}_\textrm{out}\left[1;{\boldsymbol{ s}}_{t};\boldsymbol{x}_t\right]=\boldsymbol{y}_t^\textrm{D}$. In contrast to conventional RNNs that require gradient-based learning algorithms to adjust all the inputs, the hidden, and the output weight matrices, ESNs only need to train the output weight matrix with simple training methods such as ridge regression. The most universal and stable solution to this problem is via the so-called \emph{ridge regression} approach, also known as \emph{regression with Tikhonov regularization} \cite{bishop2008training}, which is given by:
\begin{equation}\label{eq:W}
\boldsymbol{W}_\textrm{out}=\boldsymbol{y}_t^\textrm{D}\left[1;{\boldsymbol{ s}}_{t};\boldsymbol{x}_t\right]^{\rm T}\left(\left[1;{\boldsymbol{ s}}_{t};\boldsymbol{x}_t\right]\left[1;{\boldsymbol{ s}}_{t};\boldsymbol{x}_t\right]^{\rm T}+\theta \boldsymbol{I} \right)^{-1},
\end{equation}
where $ \boldsymbol{I} $ is an identity matrix and $\theta$ is a regularization coefficient which should be selected individually for a concrete reservoir based on validation data. When $\theta=0$, the ridge regression will become a generalization of a regular linear regression. {However, ridge regression is an offline training method for ESNs. In fact, ESNs can be also trained by using online methods such as the \emph{least mean squares} (LMS) algorithm \cite{farhang2013adaptive}, or the recursive least squares (RLS) algorithms \cite{jaeger2004harnessing}.

  \subsection{Spiking Neural Networks}
  Another important type of ANNs is the so-called \emph{spiking neural networks}. In contrast to FNNs and RNNs that simply use a single value to denote the activations of neurons, SNNs use a more accurate model of biological neural networks to denote the activations of neurons. 
  In the following, we first briefly introduce the architecture of SNNs. Then, we give an example for SNNs, the so-called liquid state machine.
  \subsubsection{\textbf{Architecture of a Spiking Neural Network}}
The architecture of SNNs is similar to the neurons in the biological neural networks. Therefore, we first discuss how the neurons operate in a real-world biological neural network. Then, we discuss the model of neurons in SNNs.

In biological neural networks, the neurons use spikes to communicate with each other. The incoming signals alter the voltage of a neuron and when the voltage exceeds a threshold value, the neuron sends out an action potential which
is a short ($1$~ms) and sudden increase in voltage that is created in the cell body or soma. Due to the form and the nature of this process, 
we refer to it as a \emph{spike} or a pulse. 
For SNNs, the use of such spikes can significantly improve the dynamics of the network. {Therefore, SNNs} can model a central nervous system and study the operation of biological neural circuits.
Since the neurons in SNNs are modeled based on the spike, SNNs have two major advantages over traditional ANNs: fast real-time decoding of signals and high information carriage capacity by adding a temporal dimension. Therefore, an SNN can use fewer neurons to accomplish the same task compared to traditional ANNs and it can also be used for real-time computations on continuous streams of data, which means that both the inputs and outputs of an SNN are streams of data in continuous time. However, the training of SNNs is more challenging (and potentially more time-consuming) than that of traditional ANNs due to their complex spiking neural models. A summary of the key advantages and disadvantages of SNNs for wireless applications is presented in Table~\ref{ta:2}. To reduce the training complexity of SNNs and keep the dynamics of spiking neurons, one promising solution is to develop a spiking neuron network that  needs to only train the output weight matrix, like ESNs in RNNs. Next, we specifically present this type of SNNs, named liquid state machine.

\subsubsection{\textbf{Example SNN - Liquid State Machine}}

  \vspace{-0cm}
\begin{figure}[!t]
  \begin{center}
   \vspace{0cm}
    \includegraphics[width=8cm]{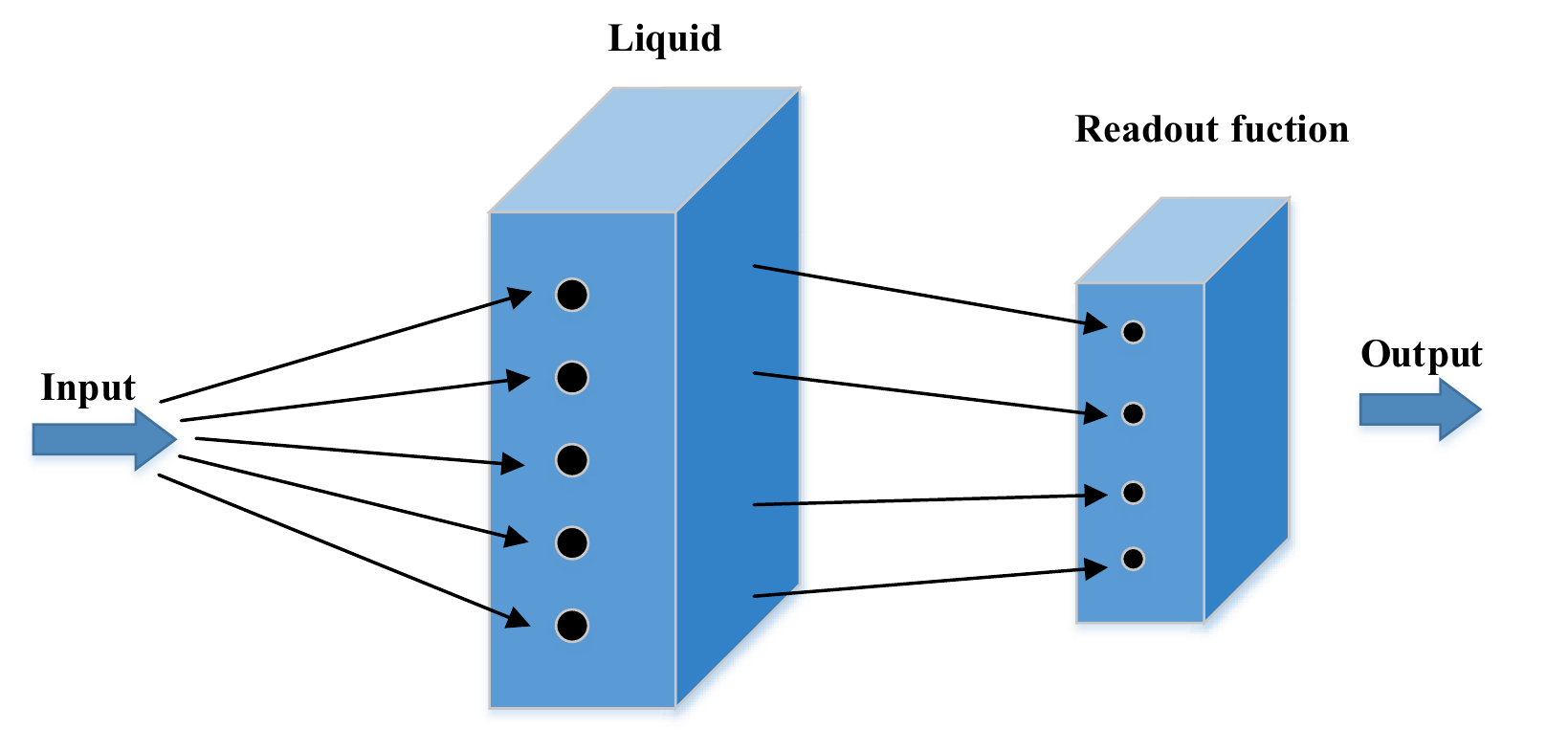}
    \vspace{-0.3cm}
    \caption{\label{LSM}Architecture of a LSM.}
  \end{center}\vspace{-0.8cm}
\end{figure}

The architecture of an LSM consists of only two components: the liquid and the readout function, as shown in Fig. \ref{LSM}. Here, the \emph{liquid} represents a spiking neural network with {leaky-integrate-and-fire (LIF)} model neurons and \emph{the readout} function is {a number of} FNNs. 
For an LSM, the connections between the neurons in the liquid is randomly generated, allowing LSM to possess a recurrent nature that turns the time-varying input into a spatio-temporal pattern.
In contrast to the general SNNs that need to adjust the weight values of all neurons, LSMs need to only train the comparatively simple FNN of the readout function. In particular, simple training methods for FNNs such as the feedforward propagation algorithm can be used for training SNNs to minimize the errors between the desired output signal and the actual output signal, which enables LSM to be widely applied for practical applications such as \cite{verstraeten2005isolated} and \cite{maass2010liquid}. Due to the LSM's spiking neurons, it can perform ML tasks on continuous data like general SNNs but, it is also possible to train it using effective and simple algorithms. In wireless networks, this can be suitable for signal detection and nonlinear audio prediction. Next, we specifically introduce the LSM architecture.

\noindent $\bullet$ \emph{Liquid Model:} In LSM, the liquid is made up of a large number of spiking LIF model neurons, located in a virtual three-dimensional column. The liquid has two important functions in the classification of time-series data. First, its fading memory is responsible for collecting and integrating the input signal over time. Each one of the neurons in the liquid keeps its own state, which gives the liquid a strong fading memory. The activity in the network and the actual firing of the neurons can also last for a while after the signal has ended, which can be viewed as another form of memory. Second, in the liquid of an LSM, the different input signals are separated, allowing for the readout to classify them. This separation is hypothesized to happen by increasing the dimensionality of the signal. For example, if the input signal has 20 input channels, this is transformed into 135 ($3\times3\times15 $) signals and states of neurons in the liquid. For every pair of input signal and liquid neuron, there is a certain chance of being connected, e.g., $30\%$ in \cite{maass2002real}. The connections between the neurons are allocated in a stochastic manner (e.g., see \cite[Appendix B]{maass2002real}). All neurons in a liquid will connect to the readout functions.

\noindent $\bullet$ \emph{Readout Model:} 
 The readout of an LSM consists of one or more FNNs that use the activity state of the liquid to approximate a specific function. The purpose of the readout is to build the relationship between the dynamics of the spiking neurons and the desired output signals. The inputs of the readout networks are called readout-moments. These are snapshots of the liquid activity taken at a regular interval. 
 Whatever measure is used, the readout represents the state of the liquid at some point in time. In general, {in LSM, FNNs are used as the readout function. FNNs will use the liquid dynamics (i.e., spikes) as their input and the desired output signals as their output.} Then, the readout function can be trained using traditional training methods used for FNNs, mainly backpropagation. 
  Once the readout function has been trained, the LSM can be used to perform the corresponding tasks.


     \subsection{Deep Neural Networks}\label{se:DNN}
    \vspace{-0cm}
\begin{figure}[!t]
  \begin{center}
   \vspace{0cm}
    \includegraphics[width=8cm]{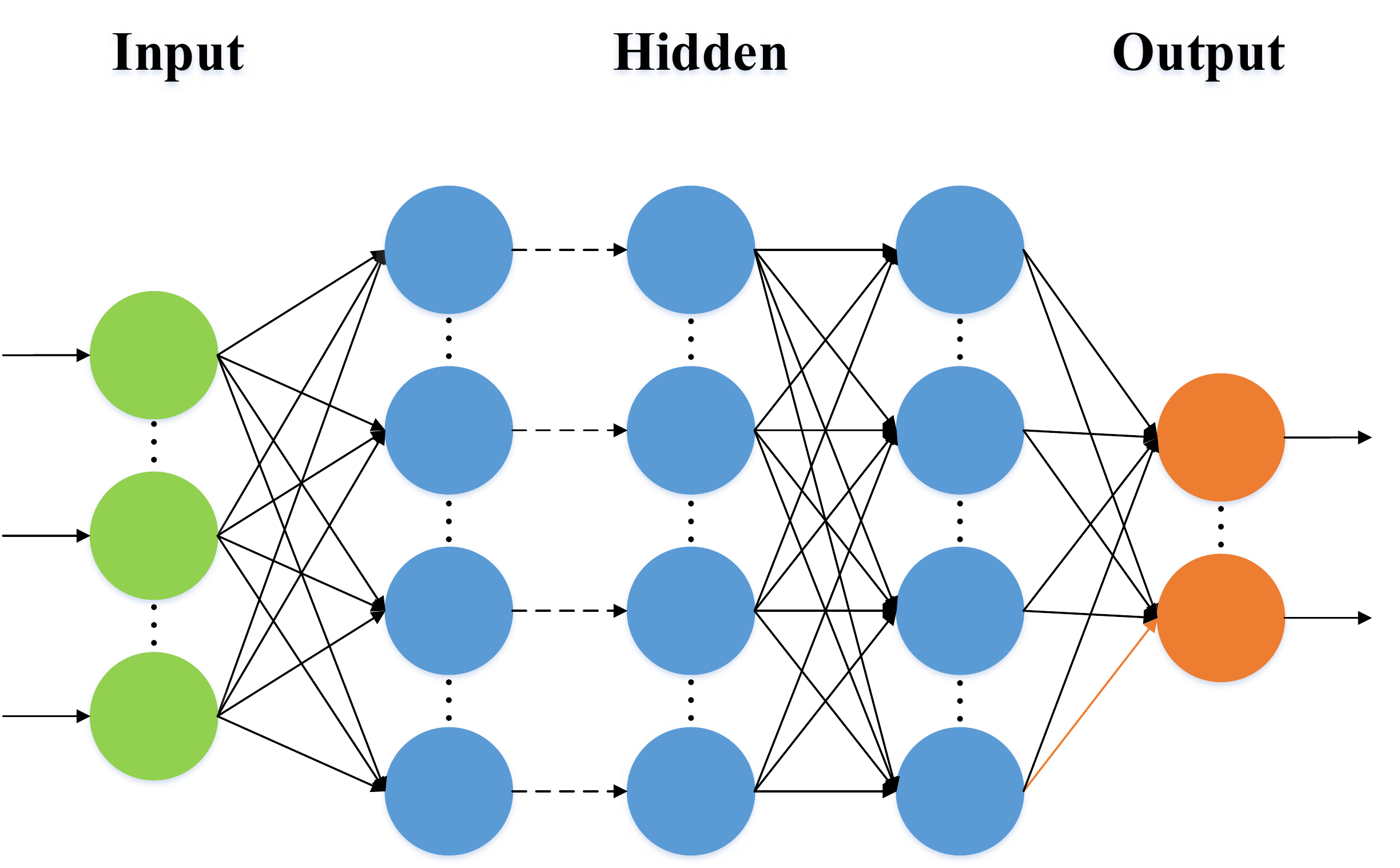}
    \vspace{-0.3cm}
    \caption{\label{DNN}Architecture of a DNN.}
  \end{center}\vspace{-0.8cm}
\end{figure}

    Thus far, all of the discussed ANNs, including ESNs and LSMs, have assumed a single hidden layer. Such an architecture is typically referred to as a shallow ANN. In contrast, a \emph{deep neural network} is an ANN with multiple hidden layers between the input and the output layers~\cite{deep_learning}, as shown in Fig. \ref{DNN}. Therefore, a DNN models high-level abstractions in data through multiple nonlinear transformations to learn multiple levels of representation and abstraction~\cite{deep_learning}. Several types of DNNs exist such as deep CNNs, deep ESNs, deep LSMs, and  LSTM~\cite{deep_learning}. The main reasons that have enabled a paradigm shift from conventional, shallow ANNs, towards DNN, include recent advances in computing capacity due to the emergence of capable processing units, the wide availability of data for DNN training, and the emergence of effective DNN training algorithms \cite{relu}.
As opposed to shallow ANNs that have only one hidden layer, a DNN having multiple layers is more beneficial due to the following reasons:

\begin{itemize}
  \item \emph{Number of neurons:} Generally, a shallow ANN would require a lot more neurons than a DNN for the same level of performance. In fact, the number of units in a shallow ANN grows exponentially with the complexity of the task.
  \item \emph{Task learning:} While the shallow ANNs can be effective to solve small-scale problems, they can be ineffective when dealing with more complex {problems, such as wireless environment mapping.} In fact, the main issue is that shallow ANNs are very good at memorization, but not so good at generalization. As such, DNNs are more suitable for many real-world tasks which often involve complex problems that are solved by decomposing the function that needs to be learned into several simpler functions so as to improve the efficiency of the learning process.
\end{itemize}
It is worth noting that, although DNNs have a large capacity to model a high degree of nonlinearity in the input data, a central challenge is that of overfitting.  
In DNNs, overfitting becomes particularly acute due to the presence of a very large number of parameters. To overcome this issue, several advanced regularization approaches, such as dataset augmentation and weight decay \cite{regularization} have been proposed. These methods modify the learning algorithm so that the test error is reduced at the expense of increased training errors. 
 A summary of key advantages and disadvantages of DNNs for wireless applications are presented in Table~\ref{ta:2}.

Next, we elaborate more on LSTM, a special kind of DNN that is capable of storing information for long periods of time {by using an identity activation function for the memory cell. This, in turn, makes LSTM suitable for various wireless communication problems such as channel selection.  

\subsubsection{\textbf{Example DNN - Long Short Term Memory}}

    \vspace{-0cm}
\begin{figure*}[!t]
  \begin{center}
   \vspace{0cm}
    \includegraphics[width=11cm]{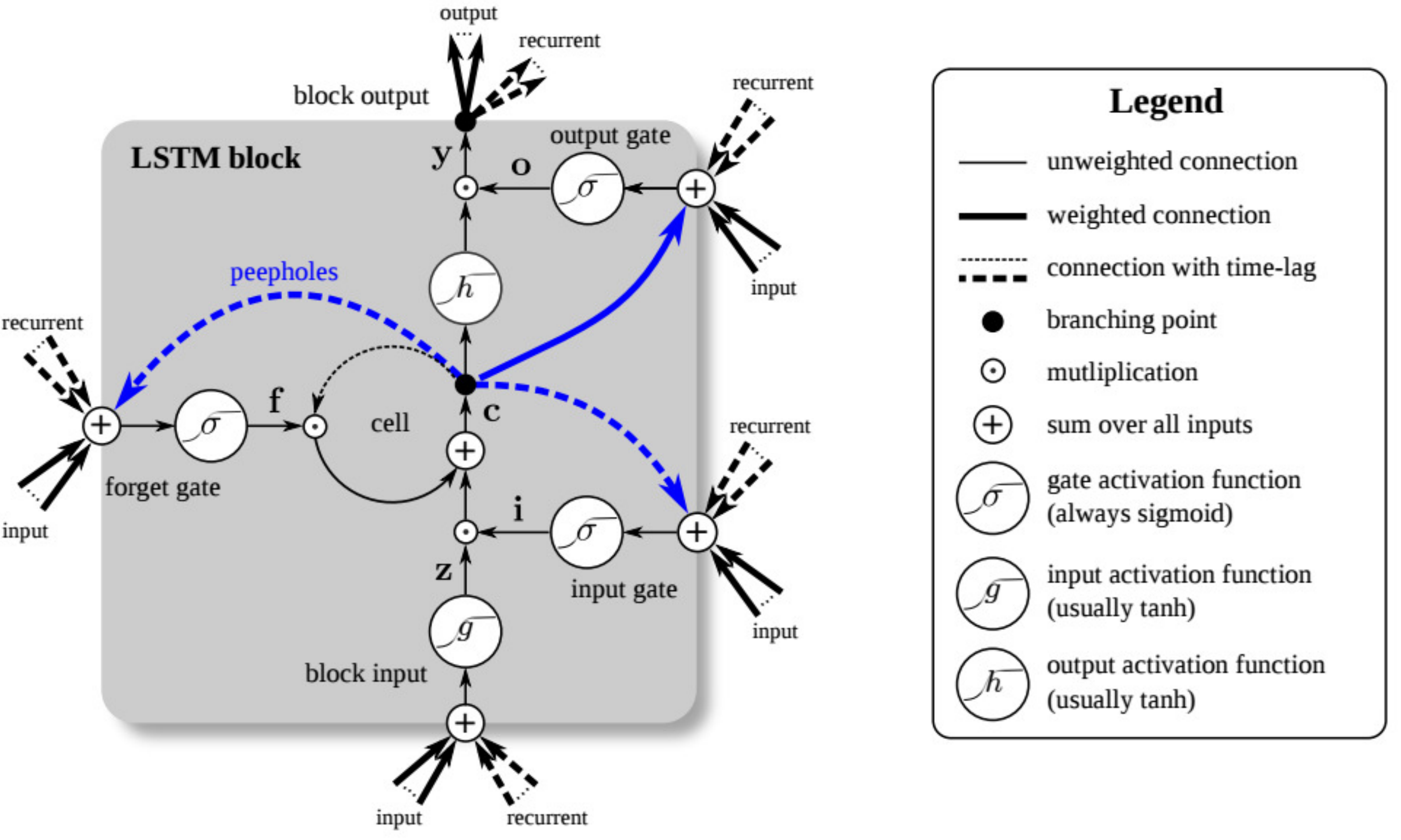}
    \vspace{-0.3cm}
    {\caption{\label{LSTM}Architecture of an LSTM as shown in \cite{ABeginner}.}}
  \end{center}\vspace{-0.6cm}
\end{figure*}
LSTMs {that typically consist of three hidden layers} are a special kind of ``deep learning'' RNNs that are capable of storing information for either long or short periods of time. In particular, the activations of an LSTM network correspond to short-term memory, while the weights correspond to long-term memory. Therefore, if the activations can preserve information over long periods of time, then this makes them long-term short-term memory. Although both ESNs and LSTMs are good at modeling time series data, LSTM cells have the capability of dealing with long term dependencies. An LSTM contains LSTM units each of which having a cell with a state $c_t$ at time $t$. Access to this memory unit, as shown in Fig.\ref{LSTM}, for reading or modifying information is controlled via three gates:
\begin{itemize}
  \item \emph{Input gate ($i_t$):} controls whether the input is passed on to the memory cell or ignored.
  \item \emph{Output gate ($o_t$):} controls whether the current activation vector of the memory cell is passed on to the output layer or not.
  \item \emph{Forget gate ($f_t$):} controls whether the activation vector of the memory cell is reset to zero or maintained.
\end{itemize}

\begin{table}[t!]\footnotesize
\setlength{\belowcaptionskip}{0pt}
\setlength{\abovedisplayskip}{3pt}
\newcommand{\tabincell}[2]{\begin{tabular}{@{}#1@{}}#1.6\end{tabular}}
 \setlength{\abovecaptionskip}{2pt}
  \caption{
    \vspace*{-0.2em}Various Behaviors of an LSTM Cell}\label{table_lstm}\vspace*{0em}
\centering
\tabcolsep=0.11cm 
\scalebox{0.99}{
\begin{tabular}{|c|c|c|}
\hline
\textbf{Input gate} & \textbf{Forget gate} & \textbf{Behavior}\\
\hline
0 & 1 & remember the previous value \\
\hline
1 & 1 & add to the previous value \\
\hline
0 & 0 & erase the value \\
\hline
1 & 0 & overwrite the value \\
\hline
\end{tabular}
}
\vspace{-0.34cm}
\end{table}

Therefore, an LSTM cell makes decisions about what to store, and when to allow reads, writes, and erasures, via gates that open and close. At each time step $t$, an LSTM receives inputs from two external sources, the current frame $x_t$ and the previous hidden states of all LSTM units in the same layer $h_{t-1}$, at each of the four terminals (the three gates and the input). These inputs get summed up, along with the bias factors $b_f$, $b_i$, $b_o$, and $b_c$. The gates are activated by passing their total input through the logistic functions. Table~\ref{table_lstm} summarizes the various behaviors an LSTM cell can achieve depending on the values of the input and the forget gates. Moreover, the update steps of a layer of LSTM units are summarized in the following equations:
\begin{equation}
g_t=f_g(\boldsymbol{W}_f\boldsymbol{x}_t+\boldsymbol{U}_f\boldsymbol{s}_{t-1}+\boldsymbol{b}_f),
\end{equation}
\begin{equation}
i_t=f_g(\boldsymbol{W}_i\boldsymbol{x}_t+\boldsymbol{U}_i\boldsymbol{s}_{t-1}+\boldsymbol{b}_i),
\end{equation}
\begin{equation}
o_t=f_g(\boldsymbol{W}_o\boldsymbol{x}_t+\boldsymbol{U}_o\boldsymbol{s}_{t-1}+\boldsymbol{b}_o),
\end{equation}
\begin{equation}
c_t=g_t\odot c_{t-1} + i_t \odot f_c(\boldsymbol{W}_c\boldsymbol{x}_t+\boldsymbol{U}_c\boldsymbol{h}_{t-1}+\boldsymbol{b}_c),
\end{equation}
\begin{equation}
\boldsymbol{s}_t=o_t\odot f_h(c_t),
\end{equation}
\noindent where $g_t$, $i_t$, and $o_t$ are the forget, the input, and the output gate vectors at time $t$, respectively. $x_t$ is the input vector, $h_t$ is the hidden/output vector, and $c_t$ is the cell state vector (i.e., internal memory) at time $t$. $\boldsymbol{W}_f$ and $\boldsymbol{U}_f$ are the weight and transition matrices of the forget gate, respectively.  $\boldsymbol{W}_i$ and $\boldsymbol{U}_i$ are the weight and transition matrices of the input gate, respectively. $\boldsymbol{W}_o$ and $\boldsymbol{U}_o$ are the weight and transition matrices of the output gate, respectively. $\boldsymbol{W}_c$ and $\boldsymbol{U}_c$ are the weight and transition matrices of the cell state, respectively.  
$f_g$, $f_c$, and $f_h$ are the activation functions, corresponding respectively to the sigmoid and the tanh functions. $\odot$ denotes the Hadamard product. Compared to a standard RNN, LSTM uses additive memory updates and separates the memory $c$ from the hidden state $\boldsymbol{s}$, which interacts with the environment when making predictions. To train an LSTM network, the stochastic gradient descent algorithm can be used.


{Finally, another important type of DNNs is the so-called \emph{convolutional neural networks} that were recently proposed for analyzing visual imagery \cite{krizhevsky2012imagenet}.
CNNs are essentially a class of deep, FNNs. In CNNs, the hidden layers have neurons arranged in three dimensions: width, height, and depth.
These hidden layers are either convolutional, pooling, or fully connected, and, hence, if one hidden layer is convolutional (pooling/fully connected), then it is called convolutional (pooling/fully connected) layer. The convolutional layers apply a convolution operation to the input, passing the result to the next layer. 
The pooling layers are mainly used to simplify the information from the convolutional layer while fully connected layers connect every neuron in one layer to every neuron in another layer.
As opposed to LSTM, that are good at temporal modeling, CNNs are appropriate at reducing frequency variations which therefore makes them suitable for applications that deal with spatial data such as interference identification in wireless networks~\cite{CNN_wireless}.  
Moreover, CNNs can be combined with LSTM, resulting in a CNN LSTM architecture that can be used for sequence prediction problems with spatial inputs, like images or videos~\cite{CNN_LSTM}.}

 In summary, different types of ANNs will have different architectures, activation functions, connection methods, and data storage capacities.  
   Each specific type of ANNs is suitable for dealing with a particular type of data. For example, RNNs are good at dealing with time-related data while SNNs are good at dealing with continuous data. Moreover, each type of ANNs has its own advantages and disadvantages in terms of learning tasks, specific tasks such as time-related tasks or space-related tasks, training data size, training time, and data storage space. 
 Given all of their advantages, ANNs are ripe to be exploited in a diverse spectrum of applications in wireless networking, as discussed in the following section.


  \section{Applications of Neural Networks in Wireless Communications}
  In this section, we first overview the motivation behind developing ANN solutions for wireless communications and networking problems.
   Then, we introduce the use of ANNs for various wireless applications.
    In particular, we discuss how to use ANNs for unmanned aerial vehicles (UAVs), wireless virtual reality (VR), mobile edge caching and computing, multiple radio access technologies, and the IoT.

  \subsection{Artificially Intelligent Wireless Networks using ANNs: An Overview}

 Recently, ANNs have started to attract significant attention in the context of wireless communications and networking \cite{boccardi2014five,bi2015wireless} and \cite{o2017introduction}, since the development of smart devices and mobile applications has significantly increased the autonomy of a wireless network, as well as the level at which human users interact with the wireless communication system.
 Moreover, the development of mobile edge computing and caching technologies makes it possible for base stations to store and analyze the behavior of the users of a wireless network. In addition, the emergence of the Internet of Things motivates the use of ANNs to improve the way in which wireless data is processed, collected, and used for various sensing and autonomy purposes. 


   In essence, within the wireless communication domains, ANNs have been proposed for two major applications. First, they can be used for \emph{prediction, inference, {\color{black}and the intelligent and predictive data analytics purposes}}. Within this application domain, the ANN-based ML algorithms enable the wireless network to learn from the datasets generated by its users, environment, and network devices. For instance, ANNs can be used to analyze and predict the wireless users' mobility patterns and content requests {therefore allowing the BSs to optimize the use of their resources, such as frequency, time, or the files that will be cached across the network.} 
Moreover, predictions and inference will be a primary enabler of the emerging IoT and smart cities paradigms. Within an IoT or within a smart city ecosystem, sensors will generate massive volumes of data that can be used by the wireless network to optimize its resources usage, understand its network operation, monitor failures, or simply deliver smart services, such as intelligent transportation. In this regard, the use of ANNs for optimized predictions is imperative. In fact, ANNs will equip the network with the capability to process massive volumes of data and to parse useful information out of this data, as a pre-cursor to delivering smart city services. For example, road traffic data gathered from IoT sensors can be processed using ANN tools to predict the road traffic status at various locations in the city. This can then be used by the wireless network that connects road traffic signals, apparatus, and autonomous/connected vehicles to inform the vehicles of the traffic state and to potentially re-route some traffic to respond to the current state of the system. 
  Furthermore, ANNs can be beneficial for integrating different data from multiple sensors thus facilitating more interesting and complex wireless communication applications. In particular, ANNs can identify nonintuitive features largely from cross-sensor correlations which can result in a more accurate estimation of a wireless network's conditions and an efficient allocation of the available resources. 
Finally, the wireless network can use ANNs to learn about faults, infrastructure failure, and other disruptive events, so as to improve its resilience to such events.



 Second, a key application of ANNs in wireless networks is for enabling \emph{self-organizing network operation} by instilling ANN-based ML at the edge of the network, as well as across its various components (e.g., base stations and end-user devices). Such edge intelligence is a key enabler of self-organizing solutions for resource management, user association, and data offloading. In this context, ANNs can serve as RL tools \cite{lin1993reinforcement} that
{can be used by a wireless network's devices to learn the wireless environment and to make intelligent decisions.} 
An ANN-based RL algorithm also can be used to learn the users' information such as their locations and data rate, and determine the UAV's path based on the learned information. Traditional learning algorithms, such as Q-learning, that use tables or matrices to record historical data, do not scale well for dense wireless networks. On the other hand, ANNs recently use a nonlinear function approximation method to find the relationship using historical information. {Therefore, ANN-based RL algorithms can learn complex relationships between wireless users and their networking environments to provide solutions for the notoriously challenging problems of network performance optimization and resource management.}
  

ANNs can be simultaneously employed for both prediction and intelligent/self-organizing operation, for scenarios in which two functions are largely interdependent.
 For instance, data can help in decision making, while decision making can generate new data. For example, when considering virtual reality applications over wireless networks, one can use ANNs to predict the behavior of users, such as head movement and content requests. These predictions can help an ANN-based RL algorithm to allocate computational and spectral resources to the users hence improving their QoS.
 Next, we discuss specific applications that use ANNs for wireless communications.



  \subsection{Wireless Communications and Networking with Unmanned Aerial Vehicles}
  \subsubsection{\textbf{UAVs for Wireless Communications}}

   \begin{figure}[!t]
  \begin{center}
   \vspace{0cm}
    \includegraphics[width=8cm]{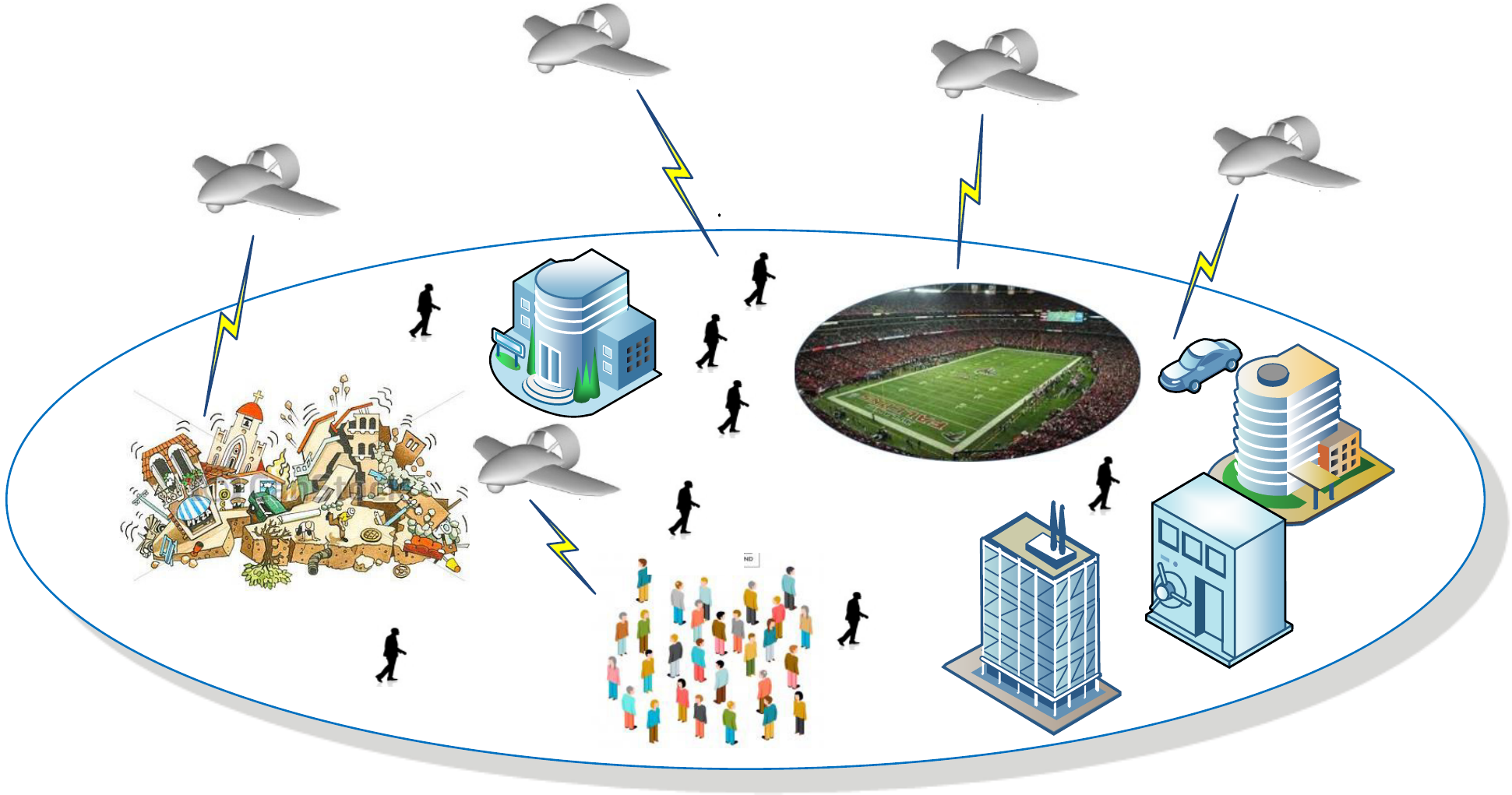}
    \vspace{-0.3cm}
    \caption{\label{UAVfigure}{UAV-enabled wireless networks. In this figure, UAVs can be used as BSs to serve users in hotspot areas due to special events such as a sport game or a disaster scenarios. }}
  \end{center}\vspace{-0.8cm}
\end{figure}
\vspace{-0cm}

  Providing connectivity from the sky to ground wireless users is an emerging trend in wireless networking \cite{zeng2016wireless} (Fig. \ref{UAVfigure}).
Compared to terrestrial communications, a wireless system with low-altitude UAVs is faster to deploy, more flexibly reconfigured, and likely to experience better communication channels due to the presence of short-range, line-of-sight (LoS) links. 
The use of highly mobile and energy-constrained UAVs for wireless communications also introduces many new challenges{\color{black}\cite{zeng2016wireless}}, such as the need of network modeling, backhaul (fronthaul) limitations for UAV-to-UAV communication when the UAVs act as flying BSs, optimal deployment, air-to-ground channel modeling, energy efficiency, path planning, and security. In particular, compared to the deployment of terrestrial BSs that are static, mostly long-term, and two-dimensional, the deployment of UAVs is flexible, short-term, and three-dimensional. Therefore, there is a need to investigate the optimal deployment of UAVs for coverage extension and capacity improvement.
 Moreover, UAVs can be used for data collection, delivery, and transmitting telematics. Hence, there is a need to develop intelligent self-organizing control algorithms to optimize the flying path of UAVs. In addition, the scarcity of the wireless spectrum, that is already heavily used for terrestrial networks, is also a big challenge for UAV-based wireless communication. Due to the UAVs' channel characteristics (less blockage and high probability for LoS link), the use of mmWave spectrum bands and visible light \cite{8715400} will be a promising solution for UAV-based communication. {Therefore, one can consider resource management problems in the context of mmW-equipped UAVs, given their potential benefits for air-to-ground communications.}
  Finally, one must consider the problems of resource allocation, interference management, and routing when the UAVs act as users. 


\subsubsection{\textbf{Neural Networks for UAV-Based Wireless Communication}}

Due to the flying nature of UAVs, they can track the users' behavior and collect information related to the users and the UAVs within any distance, at any time or any place, which provides an ideal setting for implementing ANN techniques. ANNs have two major use cases for UAV-based wireless communication. First, using ANN-centric RL algorithms, UAVs can be operated in a self-organizing manner. For instance, using ANNs as a RL, UAVs can dynamically adjust their locations, flying directions, resource allocation decisions, and path planning to serve their ground users and adapt to the users' dynamic environment.
{Second, UAVs can be used to map the ground environment as well as the wireless environment itself to collect data and take advantage of ANN algorithms to exploit the collected data and perform data analytics to predict the ground users' behavior.} 
For example, ANNs can exploit the collected mobility data to predict the users' mobility patterns. Based on the behavioral patterns of the users, battery-limited UAVs can determine their optimal locations and design an optimal flying path to service ground users. Meanwhile, using ANNs enables more advanced UAV applications such as environment identification.  
Clearly, within a wireless environment, most of the data of interest, such as that pertaining to the human behavior, UAV movement, and data collected from wireless devices, will be time related. For instance, certain users will often go to the same office for work at the same time during weekdays. ANNs can effectively deal with time-dependent data which makes them a natural choice for the applications of UAV-based wireless communication.

  Using ANNs for UAVs faces many challenges, such as the limited flight time to collect data, the limited power and the computational resources for training ANNs, as well as the data errors due to the air-to-ground channel. First, the limited battery life and the limited computational power of UAVs can significantly constrain the use of ANNs. This stems from the fact that ANNs require a non-negligible amount of time and computational resources for training. For instance, UAVs must consider a tradeoff between the energy used for training ANNs and that used for other applications such as servicing users. Moreover,
due to their flight time constraints \cite{mozaffari2017wireless}, UAVs can only collect data within a limited time period.
 In consequence, UAVs may not have enough collected data for training ANNs. In addition,  the air-to-ground channels of UAVs will be significantly affected by the weather, the environment, and their movement. Therefore, the collected data can include errors that may affect the accuracy of the outcomes of the ANNs.

 The existing literature has studied a number of problems related to using ANNs for UAVs {\cite{8432464,7451189,8353152,nodland2013neural,braga2016image,cui2018multi,chen2017caching,chen2018liquid}.
 In \cite{8432464}, the authors used a deep RL algorithm to efficiently control the coverage and connectivity of UAVs. The authors in \cite{7451189} studied the use of ANNs for UAV assignment to meet the high traffic demands of ground users. 
 The work in \cite{8353152} investigated the use of ANNs for UAV detection.   In \cite{nodland2013neural}, the authors studied the use of ANNs for trajectory tracking of UAVs. The work in \cite{braga2016image} proposed a multilayer perceptron based learning algorithm that uses aerial images and aerial geo-referenced images to estimate the positions of UAVs. In \cite{cui2018multi}, an ESN based RL algorithm is proposed for resource allocation in UAV based networks. {\color{black}In \cite{chen2018liquid}, we proposed an RL algorithm that uses LSM for resource allocation in UAV-based LTE over an unlicensed band (LTE-U) network. }} 
 For UAV-based wireless communications, ANNs can be also used for many applications such as path planning  \cite{liu2019optimized}, as mentioned previously.
Next, we explain a specific ANN  application for UAV-based wireless communication.

\subsubsection{\textbf{Example}} An elegant and interesting use of ANNs for UAV-based communication systems is presented in \cite{chen2017caching} for the study of the proactive deployment of cache-enabled UAVs. The model in \cite{chen2017caching} considers the downlink of a wireless cloud radio access network (CRAN) servicing a set of mobile users via terrestrial remote radio heads and flying cache-enabled UAVs. The terrestrial remote radio heads   (RRHs) transmit over the cellular band and are connected to the cloud's pool of the baseband units (BBUs) via capacity-constrained fronthaul links. Since each user has its own QoE requirement, the capacity-constrained fronthaul links will directly limit the data rate of the users that request content from the cloud. Therefore, the cache-enabled UAVs are introduced to service the mobile users along with terrestrial RRHs.
Each cache-enabled UAV can store a limited number of popular content that the users request. By caching the predicted content, the transmission delay from the content server to the UAVs can be significantly reduced as each UAV can directly transmit its stored content to the users.


A realistic model for periodic, daily, and pedestrian mobility patterns is considered according to which each user will regularly visit a certain location of interest. The QoE of each user is formally defined as function of each user's data rate, delay, and device type. The impact of the device type on the QoE is captured by the screen size of each device. The screen size will also affect the QoE perception of the user, especially for video-oriented applications.  The goal of \cite{chen2017caching} is to find an effective deployment of cache-enabled UAVs to satisfy the QoE requirements of each user while minimizing the transmit powers of the UAVs. This problem involves predicting, for each user, the content request distribution and the periodic locations, finding the optimal contents to cache at the UAVs, determining the users' associations, as well as adjusting the locations and transmit power of the UAVs. ANNs can be used to solve the prediction tasks due to their effectiveness in dealing with time-varying data (e.g., mobility data). Moreover, ANNs can extract the relationships between the user locations and the users' context information such as gender, occupation, and age. {In addition, ANN-based RL algorithms can find the relationship between the UAVs' location and the data rate of each user, enabling UAVs to find the locations that maximize the users' data rates.}

A prediction algorithm using the framework of ESN with conceptors is developed to find the users' content request distributions and their mobility patterns. The predictions of the users' content request distribution and their mobility patterns are then used to find the user-UAV association, optimal locations of the UAVs and content caching at the UAVs. Since the data of the users' behaviors such as mobility and content request are time-related, an ESN-based approach, as previously discussed in Subsection~\ref{se:ESN}, can quickly learn the mobility pattern and content request distributions without requiring significant training data. Conceptors, defined in \cite{Jaeger2014Controlling}, enable an ESN to perform a large number of predictions of mobility and content request patterns. Moreover, new patterns can be added to the reservoir of the ESN without interfering with the previously acquired ones. The architecture of the conceptor ESN-based prediction approach is based on the ESN model specified in Subsection~\ref{se:ESN}. For the content request distribution prediction, the cloud's BBUs must implement one conceptor ESN algorithm for each user. The input is defined as each user's context that includes gender, occupation, age, and device type. The output is the prediction of a user's content request distribution. The generation of the reservoir is done as explained in Subsection~\ref{se:ESN}. The conceptor is defined as a matrix that is used to control the learning of an ESN. For predicting mobility patterns, the input of the ESN-based algorithm is defined as the user's context and current location. The output is the prediction of a user's location in the next time slots. Ridge regression is used to train the ESNs. The conceptor is also defined as a matrix used to control the learning of an ESN. During the learning stage, the conceptor will record the learned mobility patterns and content request distribution patterns. When the conceptor ESN-based algorithm encounters a new input pattern, it will first determine whether this pattern has been learned. If this new pattern has been previously learned, the conceptor will instruct the ESN to directly ignore it. This can allow the ESN to save some of its memory only for the unlearned patterns.

 Based on the users' mobility pattern prediction, the BBUs can determine the user association using a $K$-mean clustering approach. By implementing a $K$-mean clustering approach, the users that are close to each other are grouped into one cluster. In consequence, each UAV services one cluster and the user-UAV association is determined. Then, based on the UAV association and each user's content request distribution, the optimal contents to cache at each UAV and the optimal UAVs' locations can be found. When the altitude of a UAV is much higher (lower) than the size of its corresponding coverage, the optimal location of the UAV can be found \cite[Theorems 2 and 3]{chen2017caching}. For  more generic cases, it can be found by the ESN-based RL algorithm \cite{Chen2016Echo}.


 \begin{figure*}[!t]
  \begin{center}
   \vspace{0cm}
    \includegraphics[width=14cm]{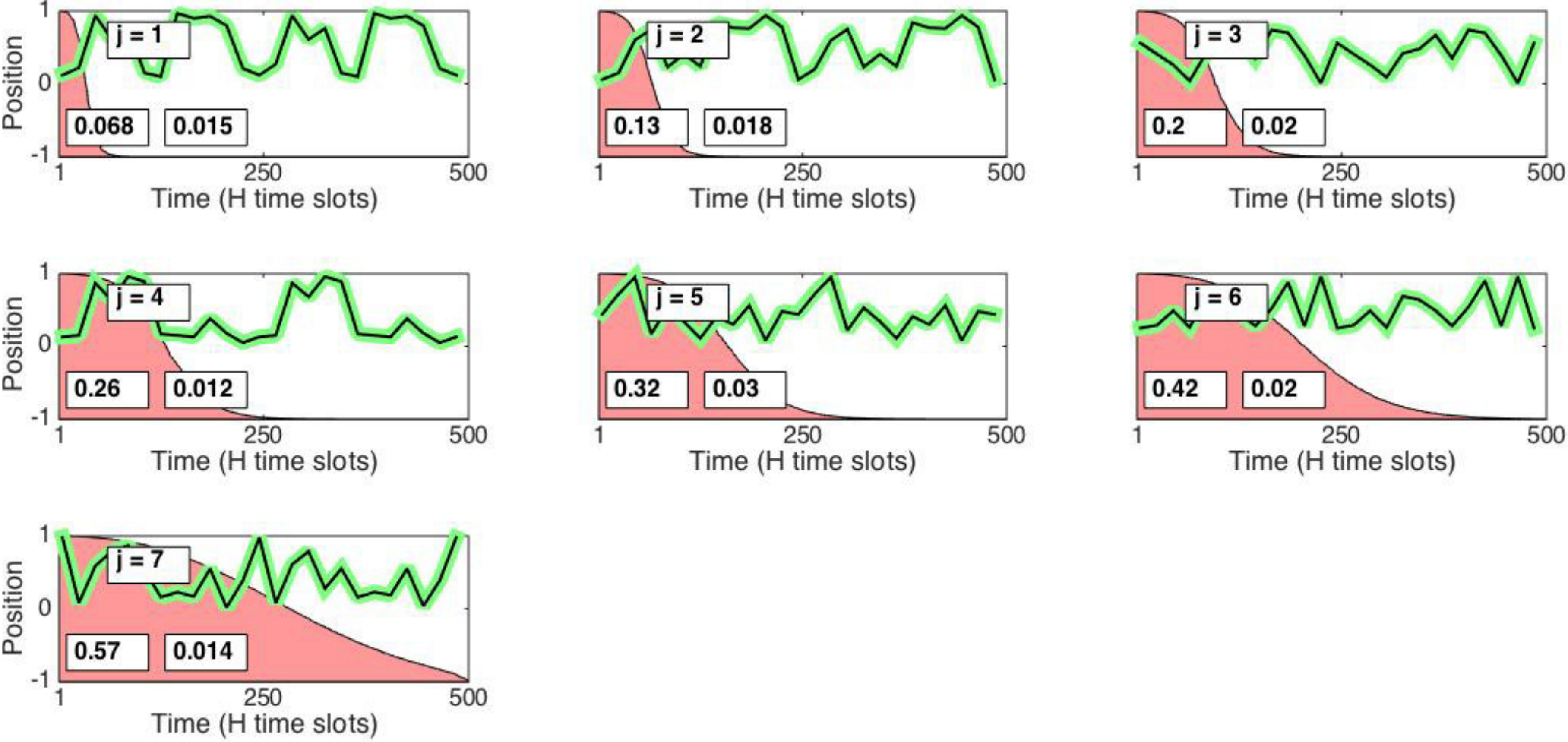}
    \vspace{-0.3cm}
    \caption{\label{figure1} Mobility patterns predictions of conceptor ESN algorithm \cite{chen2017caching}. In this figure, the green curve represents the conceptor ESN prediction, the black curve is the real positions, top rectangle $j$ is the index of the mobility pattern learned by ESN, the legend on the bottom left shows the total reservoir memory used by ESN and the legend on the bottom right shows the normalized root mean square error of each mobility pattern prediction.}
  \end{center}\vspace{-0.6cm}
\end{figure*}

In Fig. \ref{figure1}, based on \cite{chen2017caching}, we show how the memory of the conceptor ESN reservoir changes as the number of mobility patterns that were learned varies. The used mobility data is gathered from \emph{Beijing University of Posts and Telecommunications} by recording the students' locations during each day. In Fig. \ref{figure1}, one mobility pattern represents the users' trajectory in one day and the colored region is the memory used by the ESN. Fig. \ref{figure1} shows that the usage of the memory increases as the number of the learned mobility patterns increases.
 Fig. \ref{figure1} also shows that the conceptor ESN uses less memory for learning mobility pattern 2 compared to pattern 6. In fact, compared to pattern 6, mobility pattern 2 has more similarities to mobility pattern 1, and, hence, the conceptor ESN requires less memory to learn pattern 2. This is because the proposed approach can be used to only learn the difference  between the learned mobility patterns and the new ones rather than to learn the entirety of every new pattern.
 \begin{figure}[!t]
  \begin{center}
   \vspace{0cm}
    \includegraphics[width=7cm]{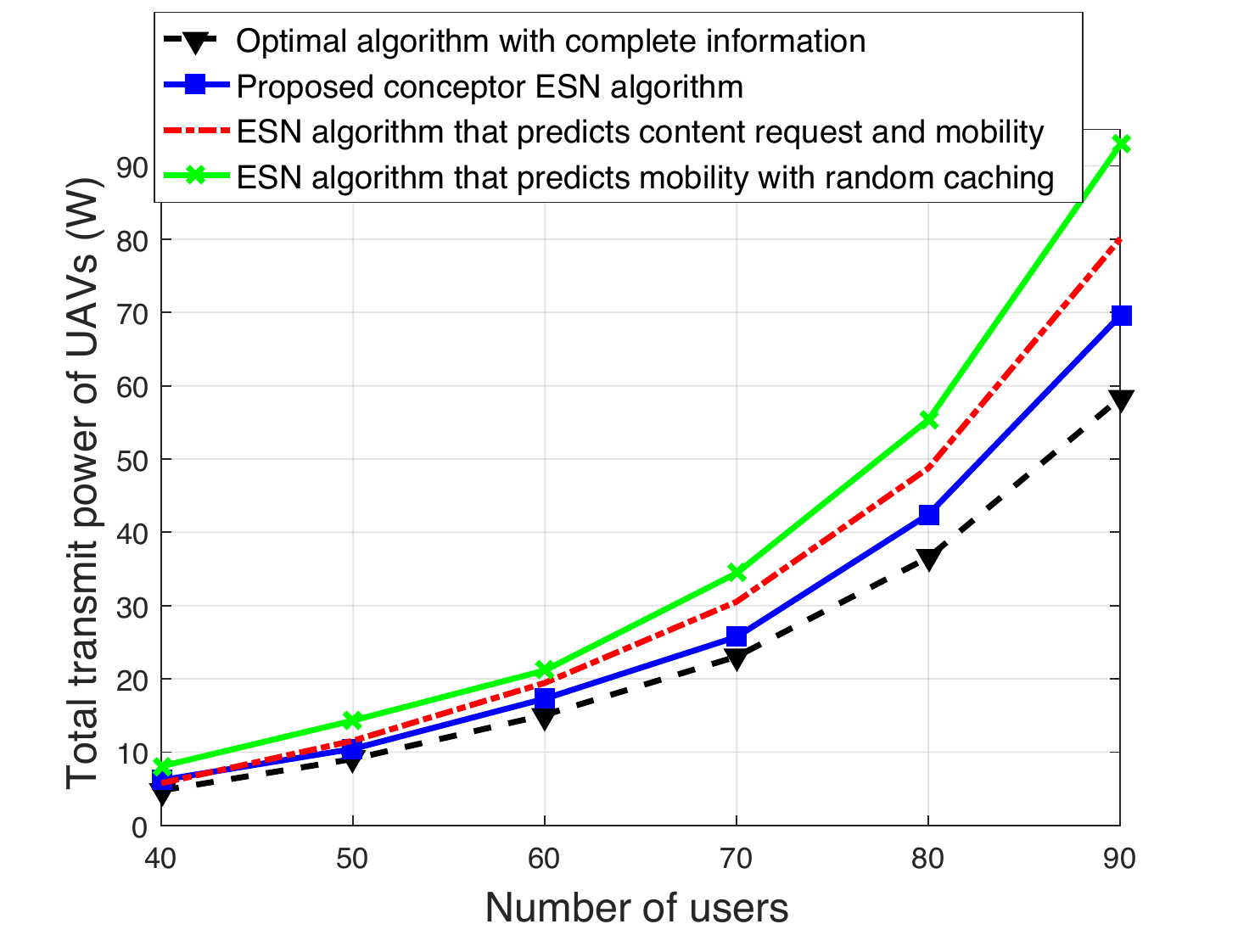}
    \vspace{-0.3cm}
    \caption{\label{figure2} Simulation result showing the transmit power as the number of users varies  \cite{chen2017caching}.}
  \end{center}\vspace{-0.6cm}
\end{figure}
\vspace{-0cm}

Fig. \ref{figure2} shows how the {total transmit power} of the UAVs changes as the number of users varies. From Fig. \ref{figure2}, we can observe that the {total UAV transmit power} resulting from all the algorithms increases with the number of users. This is due to the fact that the number of users associated with the RRHs and the capacity of the wireless fronthaul links are limited. Therefore, the UAVs must increase their transmit power to satisfy the QoE requirement of each user. From Fig. \ref{figure2}, we can also see that the conceptor based ESN approach can reduce the {total transmit power} of the UAVs by about {16.7\%} compared to the ESN algorithm used to predict the content request and the mobility for a network with 70 users. This is because the conceptor ESN, that separates the users' behavior into multiple patterns and uses the conceptor to learn these patterns, can predict the users' behavior more accurately compared to the ESN algorithm.


{Resource allocation problems in UAV-based wireless networks can also be addressed using LSMs, as explained in \cite{chen2018liquid}. In particular, in \cite{chen2018liquid}, an LSM-based RL algorithm is used for resource and cache management in LTE over unlicensed (LTE-U) UAV networks. The LSM-based RL algorithm in \cite{chen2018liquid} can find the appropriate policies for user association and resource allocation as well as the contents to cache at UAVs, as the users' content requests change dynamically. This is due to the fact that an LSM can record the dynamic user content requests as well as the policies of the user association, resource allocation, and content caching due to its large memory (compared to ESN). Based on the recorded information, the LSM algorithm can build a relationship between content requests, user association, resource allocation and caching content.}

\subsubsection{\textbf{Lessons learned}}{\color{black}
From this example, we have demonstrated {\color{black}that the conceptor ESN} can be used for effective data analytics in wireless networks that integrate UAV base stations, particularly, for mobility pattern and content request distribution predictions (and UAV-level caching). The ML angle in this application stems from the fact that predictions are used for intelligently determining the user association, optimal caching, and optimal UAV locations. 
The key lessons learned here include: 
\begin{itemize}
\item The advantage of the conceptor ESN for UAV-based networks is that it provided the network with an ability to proactively determine the deployment of UAVs and the optimal content stored at UAVs. Since UAVs are flexible in their deployment (unlike terrestrial base stations), such a proactive approach is desirable. The analysis in \cite{chen2017caching} also revealed that the use of a conceptor in the ESN scheme allows it to separate a user's weekly mobility into several patterns and use various non-linear systems for predictions thus improving accuracy. Moreover, the conceptor ESNs enable the cloud to add new patterns to the ESN without interfering with previously acquired ones and, hence, they can improve the usage of an ESN's memory (i.e., its capacity to store past data). 


\item The conceptor ESN algorithm that we presented in this section is able to perform its predictions over a long period of time. In this case,  the conceptor ESN can be trained in a completely offline manner and its training process can be implemented at the cloud, thus leveraging its computational power. Once trained at the cloud, the UAVs can then directly use the cloud-trained conceptor ESNs for predictions and deployment. Thus, this results energy savings which is particularly important for resource-limited UAVs.
Another reason to train conceptor ESNs at the cloud is that the cloud is better positioned in the network to collect mobility information. Due to this implementation, one can neglect the overhead for the training of the conceptor ESNs.

\item From this work, we have observed that, for mobility prediction, a shallow conceptor ESN learning algorithm can achieve the same prediction accuracy compared to a deep learning algorithm (e.g., similar to the one that will be introduced in the multi-RAT application of Subsection IV-E). This is mainly due to the fact that the future locations of each user depend only on a small number of the locations that the user has previously visited. In consequence, a shallow conceptor ESN is sufficient to record these visited locations and perform reasonable predictions. 
\item One disadvantage of using a conceptor ESN learning algorithm for intelligent and predictive data analytics is that the conceptor will increase the training complexity of each ESN. This is due to the fact that, during the training process, the conceptor needs to identify the input data of a given ESN and also needs to find appropriate memory space of the ESN for data recording. This further motivates the need to train the conceptor ESNs at the cloud so as to save the UAV energy.
\end{itemize}{\color{black}Note that, observations in the third and fourth bullets above can be generalized to other shallow RNNs. }}

\subsubsection{\textbf{Future Works}}
Clearly, ANNs are an important tool for addressing key challenges in UAV-based communication networks. In fact, different types of ANNs can be suitable for various UAV applications. For instance, given their effectiveness in dealing with time-dependent data, RNNs can be used for predicting user locations and traffic demands. This allows UAVs to optimize their location based on the dynamics of the network.
DNN-based RL algorithms can be used to determine the time duration that the UAVs need to service the ground users and how to service the ground users (e.g., stop or fly to service the users). Since DNNs have the ability to store large amount of data, DNN-based RL algorithms can also be used to store the data related to the users' historical context and, then, predict each ground user's locations, content requests, and latency requirement. {Based on these predictions, the UAVs can find their optimal trajectory and, as a result, determine which area to serve at any given time.}
 In addition, SNNs can be used for modeling the air-to ground channel, in general, and over mmWave frequencies, in particular. This is because SNNs are good at dealing with continuous data and the wireless channel is time-varing and continuous \cite{challita2016chance}. For instance, UAVs can use SNNs to analyze the data that they can collect from the radio environment, such as the
received signal strength, UAVs' positions, and users' positions, and then generate an air-to-ground channel model to fit the collected data. Finally, SNNs are a good choice for the prediction of UE UAVs' trajectories. Then, the networks can select the appropriate BSs to service UE UAVs.  
A summary of key problems that can be solved by using ANNs for UAV-based communications is presented in Table \ref{ta:UAV} along with the challenges and future works.

\begin{table*}
\centering
  \newcommand{\tabincell}[2]{\begin{tabular}{@{}#1@{}}#2\end{tabular}}
\renewcommand\arraystretch{1}
 \caption{
    \vspace*{-0.1em}Summary of the use of ANNs for Specific Application }\label{ta:UAV}\vspace*{-0.6em}
\centering
\begin{tabular}{|c|l|l|l|}
\hline
 \multicolumn{1}{|c|}{\multirow{2}{*}{\textbf{Applications}}} &  \multicolumn{1}{|c|}{\multirow{2}{*}{\textbf{ Existing Works }}} &   \multicolumn{1}{|c|}{\multirow{2}{*}{\textbf{ Challenges}}}   &   \multicolumn{1}{|c|}{\multirow{2}{*}{\textbf{ Future Works and Suggested Solutions }}}  \\
 &&&\\
\hline
\multirow{6}{*}{\textbf{UAV}} & {\color{black}  $\bullet$ UAV control~\cite{8432464,nodland2013neural}}&\multirow{2}{4cm}{$\bullet$ Limited power and computation \\~~~for training ANNs}&   $\bullet$ UAV path planning $ \Rightarrow $ RNN-based RL algorithm \\
& {\color{black}  $\bullet$ Position estimation \cite{braga2016image}}& & $\bullet$ Resource management $ \Rightarrow $ DNN-based RL algorithm\\
& {\color{black}  $\bullet$ UAV detection~\cite{8353152}}& \multirow{1}{*}{$\bullet$ Limited time for data collection} & $\bullet$ Channel modeling for air-to-ground $ \Rightarrow $ SNN-based algorithm\\
& {\color{black}  \multirow{1}{4cm}{$\bullet$ Deployment and caching~\cite{7451189},\\~~~\cite{cui2018multi}, and \cite{chen2017caching}} }&$\bullet$ Errors in training data&  $\bullet$  Handover for UE UAVs $ \Rightarrow $ RNN-based algorithm\\
& &&$\bullet$ Design multi-hop aerial network $ \Rightarrow $ CNN-based algorithm\\
&&&$\bullet$ UE UAV trajectory prediction $ \Rightarrow $ SNN-based algorithm\\
\hline
\multirow{7}{*}{\textbf{VR}} & $\bullet$ Resource allocation~\cite{VRTL,VROWNchen}         &\multirow{1}{4cm}{$\bullet$ Errors in collected data}&   $\bullet$ VR users' movement $ \Rightarrow $ RNNs prediction algorithm\\
&$\bullet$  Head movement prediction \cite{chen2018federated} &$\bullet$ Limited computational resources  & $\bullet$  Content correlation $ \Rightarrow $  CNN-based algorithm \\
&$\bullet$ Gaze prediction~\cite{koulieris2016gaze}& \multirow{1}{4cm}{$\bullet$ Limited time for training ANNs} & $\bullet$ VR video coding and decoding $ \Rightarrow $ CNN-based algorithm \\
&\multirow{1}{4.8cm}{$\bullet$ Content caching and transmission~\cite{chen2018echo}}  &&  $\bullet$ Correction of inaccurate VR images $ \Rightarrow $ CNN-based algorithm \\
&&&$\bullet$  Viewing video prediction $ \Rightarrow $  SNN-based algorithm \\
& &&\multirow{1}{7cm}{$\bullet$  Joint wireless and VR user environment prediction \\~~~$ \Rightarrow $ RNNs prediction algorithm}\\
&&&\\
& &&\multirow{1}{7cm}{$\bullet$ Manage computational resources and video formats\\~~~$ \Rightarrow $ DNN-based RL algorithm}\\ 
&&&\\
\hline

\multirow{6}{1.6cm}{\textbf{Caching and\\~Computing}} & {\color{black}  $\bullet$ Architecture for caching \cite{zeydan2016big}} &\multirow{1}{4cm}{$\bullet$ Data cleaning}&   $\bullet$ Analysis of content correlation $ \Rightarrow $ CNN-based RL algorithm\\

& {\color{black}  $\bullet$ Cache replacement  \cite{cobb2008web,8478380,8513863}}&$\bullet$ Content classification  & $\bullet$ Content transmission methods $ \Rightarrow $ RNN-based RL algorithm \\

& {\color{black}  \multirow{1}{4cm}{$\bullet$ Content popularity prediction\\~~~\cite{Bigdata} and \cite{7873292}}}& \multirow{1}{4cm}{$\bullet$ Limited storage of ANNs for \\~~~recording all types of contents} & $\bullet$ Clustering of users and tasks $ \Rightarrow $ CNN-based algorithm \\
&&&  $\bullet$ Computational demand prediction $ \Rightarrow $ SNN-based algorithm \\
& {\color{black}  \multirow{1}{4cm}{$\bullet$ Content request distribution\\~~~prediction~\cite{chen2017caching} and \cite{chen2017echo}}}  &&  $\bullet$ Computing time prediction $ \Rightarrow $ SNN-based algorithm\\
&&&$\bullet$ Computational resource allocation $ \Rightarrow $ RNN-based RL approach \\ 
&&&$\bullet$ Computational caching $ \Rightarrow $ RNN-based RL algorithm \\

\hline

\multirow{6}{1.5cm}{\textbf{ Multi-RAT}} & {\color{black}  $\bullet$ Resource management \cite{fuzzy},\cite{Chen2016Echo} }&\multirow{1}{4cm}{$\bullet$ Channel selection}&   $\bullet$ Detection of LoS links $ \Rightarrow $ CNN-based algorithm\\

& {\color{black}  $\bullet$ RAT selection \cite{8353153} }&  $\bullet$  Mobility predictions    & $\bullet$ Antenna tilting $ \Rightarrow $ DNN-based RL algorithm \\

&\multirow{2}{3.5cm}{$\bullet$ Transmission technology\\~~~classification \cite{MLP_RAT}}& \multirow{1}{4cm}{$\bullet$ Channel load estimation  } & $\bullet$ Channel estimation $ \Rightarrow $ SNN-based algorithm \\
&  & $\bullet$  Load balancing &  $\bullet$ Handover among multi-RAT BSs $ \Rightarrow $ RNN-based algorithm \\
&$\bullet$ Multi-radio packet scheduling \cite{hopfield} & &  $\bullet$ MmWave links for multi-RAT $ \Rightarrow $ DNN-based algorithm\\
& {\color{black}  $\bullet$ Mode selection~\cite{mehdi,8468000}}& &  $\bullet$ MmWave channel modeling $ \Rightarrow $ SNN-based algorithm\\
\hline

\multirow{8}{0.7cm}{\textbf{IoT}} &$\bullet$ Model IoT as ANNs \cite{kaminski2017neural} &\multirow{2}{4cm}{$\bullet$ Massive amounts of data and\\~~~large number of devices}&   $\bullet$  Data compression and recovery $ \Rightarrow $ CNN-based algorithm \\

&$\bullet$ Failure detection \cite{naidu1990use},\cite{ning2011future} &  & $\bullet$ Resource management $ \Rightarrow $ RNN-based RL algorithm \\

&\multirow{1}{*}{$\bullet$ User activities classification \cite{alam2016analysis}}& \multirow{2}{4cm}{$\bullet$  Limited computation and energy\\~~~resources } & $\bullet$ User identification $ \Rightarrow $ DNN-based algorithm \\

&  \multirow{1}{*}{$\bullet$ Tracking accuracy improvement \cite{luo2016laguerre}}   &  & $\bullet$ IoT devices management $ \Rightarrow $ SNN-based algorithm \\
 &$\bullet$ Image detection \cite{du2017reconfigurable}&$\bullet$ Errors in collected data&$\bullet$ Data relationship extraction $ \Rightarrow $  RNN-based RL algorithm\\
 &{\color{black}  $\bullet$ Data sampling~\cite{8496746}}&$\bullet$ Real-time training for ANNs& \multirow{1}{6cm}{$\bullet$ Modeling autonomous M2M communication\\~~~$ \Rightarrow $ FNN and SNN based algorithm} \\
 & {\color{black}  $\bullet$ Entity state prediction~\cite{8502822}} &&\\
  & {\color{black}  $\bullet$ Target surveillance~\cite{8463616}} &&\\
\hline

\end{tabular}
 \vspace{-0.2cm}

\end{table*}

  \subsection{Wireless Virtual Reality}

     \begin{figure}[!t]
  \begin{center}
   \vspace{0cm}
    \includegraphics[width=7cm]{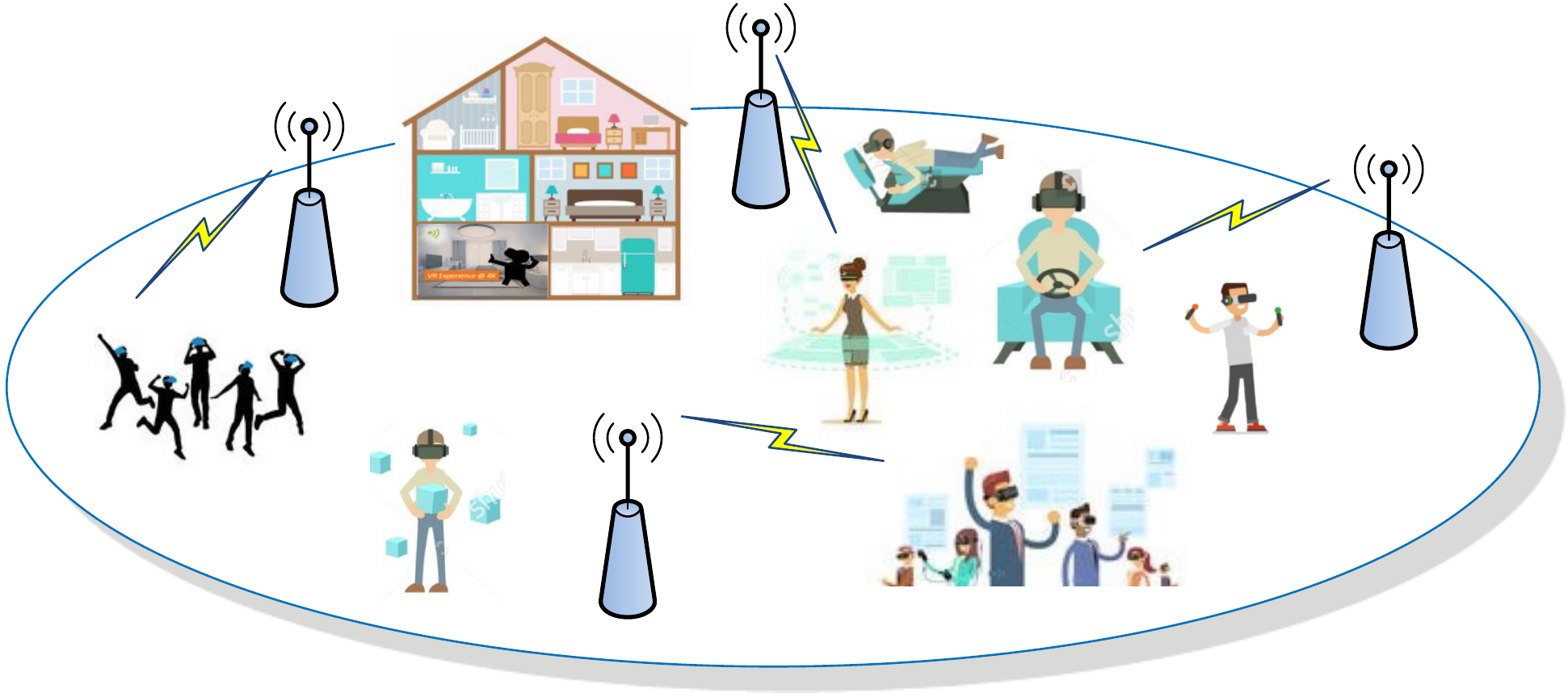}
    \vspace{-0.3cm}
    \caption{\label{VRfigure}{{\color{black}Wireless VR networks. In this figure, BSs that are acted as VR controllers generate and transmit VR videos to VR users according to the tracking information collected from VR users.}}}
  \end{center}\vspace{-0.8cm}
\end{figure}
\vspace{-0cm}

   \subsubsection{\textbf{Virtual Reality over Wireless Networks}} 
{\color{black}Recently, the wireless industry such as Qualcomm \cite{qualcommVR} and Nokia \cite{Elbamby180107587}, has rated VR as one of the most important applications in 5G and beyond networks. Moreover, 3GPP is standardizing wireless VR, called extended reality (XR) \cite{3gpp.26.928}. In addition, several industrial players such as HTC Vive \cite{wirelessHTCVIVE}, and Oculus \cite{wirelessOculus}, and Intel \cite{wirelessIntel} are all developing wireless VR devices that can operate over wireless cellular networks. These recent developments motivate us to analyze wireless VR as a key use case of ANNs in future wireless networks.} 

When a VR device is operated over a wireless link, the users must send the tracking information that includes the users' locations and orientations to the BSs and, then, the BSs will use the tracking information to construct $360^\circ$ images and send these images to the users. Therefore, for wireless VR applications, the uplink and downlink transmissions must be jointly considered. Moreover, in contrast to traditional video that consists of $120^\circ$ images, a VR video consists of high-resolution $360^\circ$ vision with three-dimensional surround stereo. This new type of VR video requires a much higher data rate than that of traditional mobile video. In addition, as the VR images are constructed according to the the users' movement such as their head or eye movement, the tracking accuracy of the VR system will directly affect the user experience.
In summary, the challenges of operating VR devices over wireless networks \cite{bacstuug2016towards} include tracking accuracy, low delay, high data rate, user experience modeling, effective image compression as well as VR content and tracking information transmission over wireless links. 

   \subsubsection{\textbf{Neural Networks for Wireless Virtual Reality}} The use of ANNs is a promising solution for a number of problems related to wireless VR. This is due to the fact that, compared to other applications such as UAV or caching, VR applications depend more on the users' environment and their behavior vis-a-vis the VR environment. 
{\color{black}In a wireless VR network, the head and eye movements will significantly affect resource management and network control. This is a very new challenge for wireless networks.} 
For instance,
 ANNs are effective at identifying and predicting the users' movements and their actions. Based on the predictions of the users' environment, actions, and movements, the BSs can improve the generation of the VR images and optimize the resource management for wireless VR users.
 ANNs have two major applications for wireless VR.
First, ANNs can be used to predict the users' movement as well as their future interactions with the VR environment. For example, a user displays only the visible portion of a $360^\circ$ video and, hence, transmitting the entire $360^\circ$ video frame can waste the capacity-limited bandwidth. Since all the images are constructed based on the users' movements, using ANNs, one can predict the users' movement and, hence, enable the wireless BSs to generate only the portion of the VR image that a user wants to display. Moreover, the predictions of the users' movement can also improve the tracking accuracy of the VR sensors. In particular, the BSs will jointly consider the users' movement predicted by ANNs and the users' movements collected by VR sensors to determine the users' movements.

Second, ANNs can be used to develop self-organizing algorithms to dynamically control and manage the wireless VR network thus addressing problems such as dynamic resource management. In particular, ANNs can be used for adaptively optimizing the resource allocation and adjusting the quality and format of the VR images according to the cellular network environment.

 Using ANNs for VR faces many challenges. First, in wireless VR networks, the data collected from the users may contain errors that are unknown to the BSs. In consequence, the BSs may need to use erroneous data to train the ANNs and, hence, the prediction accuracy of the ANN will be significantly affected. Second, due to the large data size of each $360^\circ$ VR image, the BSs must spend a large amount of computational resources to process VR images. Meanwhile, the training of ANNs will also require a large amount of computational resources. Thus, how to effectively allocate the computational resources for processing VR images and training ANNs is an important challenge. In addition, the VR applications require ultra-low latency while the training of ANNs can be time-consuming. Hence, how to effectively train ANNs in a limited time is an important question for wireless VR.
 In this regard, training ANNs in an offline manner or using ANNs that converge quickly can be two promising solutions for speeding up the training process of ANNs.

The existing literature has studied a number of problems related to using ANNs for VR such as in \cite{chen2018federated,koulieris2016gaze,VRTL,VROWNchen,chen2018echo}. 
   The work in \cite{chen2018federated} proposed an ESN based distributed learning algorithm to predict the users' head movement in VR applications. 
   In \cite{koulieris2016gaze}, a decision forest learning algorithm is proposed for gaze prediction. 
   The work in \cite{VRTL} developed a neural network based transfer learning algorithm for data correlation aware resource allocation.
   $360^\circ$ content caching and transmission is optimized in \cite{chen2018echo} using an ESN and SSN based deep RL algorithm. 
    {Table \ref{ta:EX} summarizes the type of ANNs and learning algorithms used for each existing work in virtual reality networks.} In essence, the existing VR literature such as \cite{chen2018federated,koulieris2016gaze,VROWNchen,VRTL,chen2018echo} has used ANNs to solve a number of VR problems such as hand gestures recognition, interactive shape changes, video conversion, head movement prediction, and resource allocation. However, with the exception of our works in \cite{VROWNchen} and \cite{chen2018federated}, all of the other works that use ANNs for VR applications are focused on wired VR. Therefore, they do not consider the challenges of wireless VR such as
scarce spectrum resources, limited data rates, and how to transmit the tracking data accurately and reliably. In fact, ANNs can be used for wireless VR to solve the problems such as users movement prediction, spectrum management, and VR image generation.
 Next, a specific ANNs' application for VR over wireless network is introduced.

\subsubsection{\textbf{Example}} One key application of using ANNs for wireless VR systems is presented in \cite{VROWNchen} for the study of resource allocation in cellular networks that support VR users. In this model, the BSs act as the VR control centers that collect the tracking information from the VR users over the cellular uplink and then send the generated images (based on the tracking information) and accompanying surround stereo audio to the VR users over the downlink. Therefore, this resource allocation problem in wireless VR must jointly consider both the uplink and the downlink transmissions. To capture the VR users' QoS in a cellular network, the model in \cite{VROWNchen} jointly accounts for VR tracking accuracy, processing delay, and transmission delay. The tracking accuracy is defined as the difference between the tracking vector transmitted wirelessly from the VR headset to the BS and the accurate tracking vector obtained from the users' force feedback. The tracking vector represents the users' positions and orientations. The transmission delay consists of the uplink transmission delay and the downlink  transmission delay. The uplink transmission delay represents the time that a BS uses to receive the tracking information while the downlink transmission delay is the time that a BS uses to transmit the VR contents. The processing delay is defined as the time that a BS spends to correct the VR image from the image constructed based on the inaccurate tracking vector to the image constructed according to the accurate tracking vector. In \cite{VROWNchen},  the relationship between the delay and the tracking is not necessarily linear nor independent and, thus, multi-attribute utility theory \cite{abbas2010constructing} is used to construct a utility function assigns a unique value to each tracking and delay components of the VR QoS.

The goal of \cite{VROWNchen} is to develop an effective resource block allocation scheme to maximize the users' utility function that captures the VR QoS. This maximization jointly considers the coupled problems of user association, uplink resource allocation, and downlink resource allocation. Moreover, the VR QoS of each BS depends not only on its resource allocation scheme but also on the resource allocation decisions of other BSs. Consequently, the use of centralized optimization for such a complex problem is largely intractable and yields significant overhead. In addition, for VR resource allocation problems, we must jointly consider both uplink and downlink resource allocation, and, thus, the number of actions will be much larger than conventional scenarios that consider only uplink or downlink resource allocation.
Thus, as the number of actions significantly increases, each BS may not be able to collect all the information needed to calculate the utility function.

To overcome these challenges, an ANN-based RL algorithm can be used for self-organizing VR resource allocation. In particular, an ANN-based RL algorithm can find the relationship between the user association, resource allocation, and the user data rates, and, then, it can, directly select the optimal resource allocation scheme after the training process.
For the downlink and uplink resource allocation problem in \cite{VROWNchen}, an ANN-based RL algorithm can use less exploration time to build the relationship between the actions and their corresponding utilities and then optimize resource allocation.     

To simplify the generation and training process of an ANN-based RL algorithm, an ESN-based RL algorithm is selected for VR resource allocation. The ESN-based learning algorithm enables each BS to predict the value of VR QoS resulting from each resource allocation scheme without having to traverse all the resource allocation schemes. The architecture of the ESN-based self-organizing approach is based on the ESN model specified in Subsection~\ref{se:ESN}. To use ESNs for RL, each row of the ESN's output weight matrix is defined as one action. Here, one action represents one type of resource allocation. The input of each ESN is the current action selection strategies of all BSs. The generation of the ESN model follows Subsection \ref{se:ESN}. The output is the estimated utility value. In the learning process, at each time slot, each BS will implement one action according to the current action selection strategy. After the BSs perform their selected actions, they can get the actual utility values. Based on the actual utility values and the utility values estimated by ESN, each BS can adjust the values of the output weight matrix of an ESN according to (\ref{eq:W}). As time elapses, the ESN can accurately estimate the utility values for each BS and can find the relationship between the resource allocation schemes and the utility values. Based on this relationship, each BS can find the optimal action selection strategy that maximizes the average VR QoS for its users.

 \begin{figure}[!t]
  \begin{center}
   \vspace{0cm}
    \includegraphics[width=7cm]{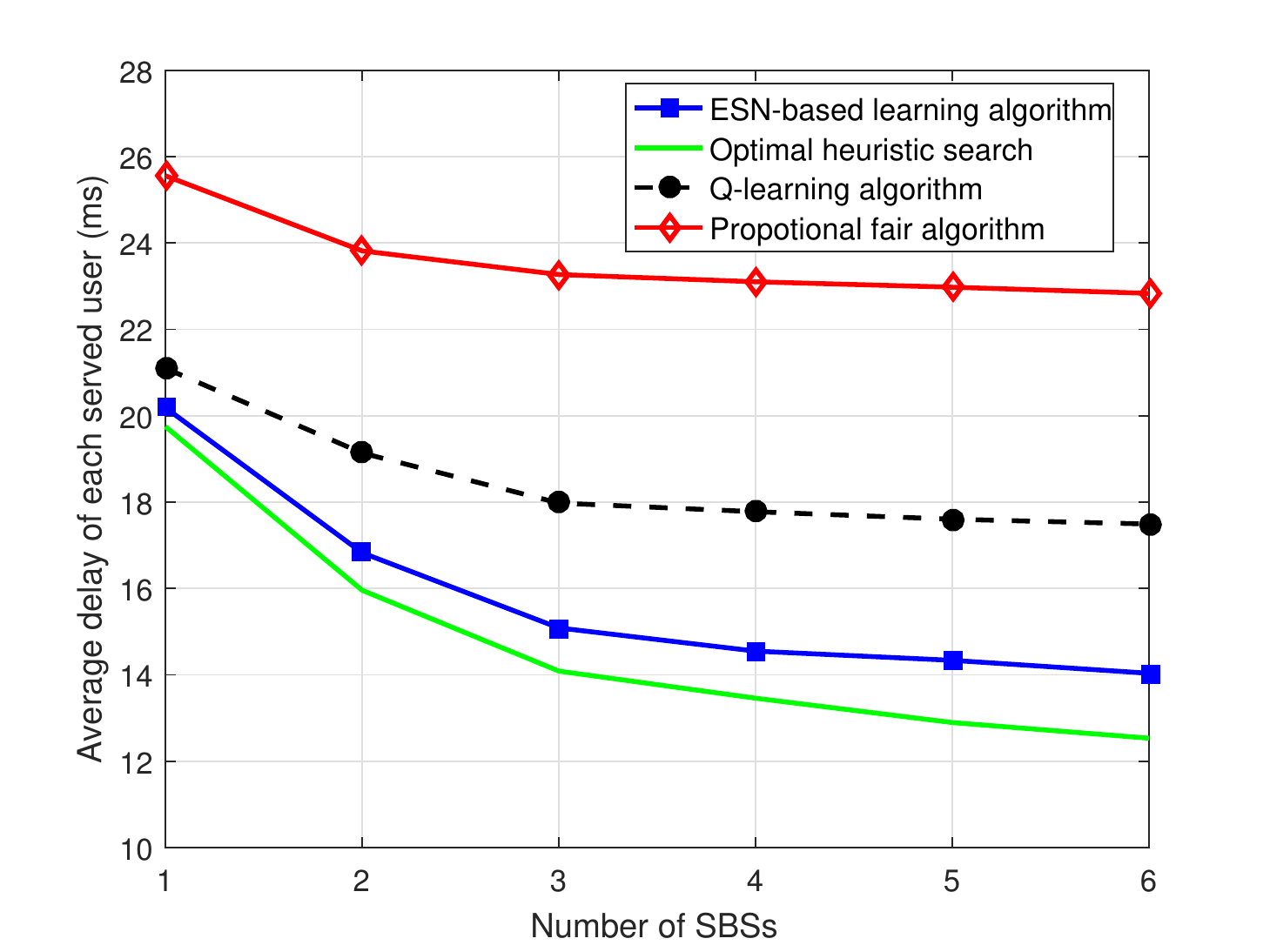}
    \vspace{-0.3cm}
    \caption{\label{figure11}{{\color{black}Delay for each served user vs. the number of BSs \cite{VROWNchen}.}}}
  \end{center}\vspace{-0.7cm}
\end{figure}
\vspace{-0cm}


Fig. \ref{figure11} shows how average delay of each user varies as the number of BSs changes. From Fig. \ref{figure11}, we can see that, as the number of BSs increases, the transmission delay for each served user increases.
This is due to the fact that, as the number of BSs increases, the number of users located in each BS's coverage decreases and, hence, the average delay increases. However, as the number of BSs increases, the delay increase becomes slower due to the additional interference. {This stems from the fact that, as the number of BSs continues to increase, the number of the users associated with each BS decreases and more spectrum will be allocated to each user. Hence, the delay of each user will continue to decrease. However, as the number of the BSs increases, the increasing interference will limit the reduction in the delay.}
Fig. \ref{figure11} also shows that the ESN-based algorithm achieves up to a {19.6\%} gain in terms of average delay compared to the Q-learning algorithm for the case with 6 BSs. Fig. \ref{figure11} also shows that the ESN-based approach allows the wireless VR transmission to meet the VR delay requirement that includes both the transmission and processing delay (typically 20 ms \cite{WhatVRMichael}). These gains stem from the adaptive nature of ESNs. 


From this example, we illustrated the use of ESN as an RL algorithm for self-organizing resource allocation in wireless VR. An ESN-based RL algorithm enables each BS to allocate downlink and uplink spectrum resource in a self-organizing manner that adjusts the resource allocation according to the dynamical environment. Moreover, an ESN-based RL algorithm can use an approximation method to find the relationship between each BS's actions and its corresponding utility values, and, hence, an ESN-based RL algorithm can speed up the training process. Simulation results show that an ESN-based RL algorithm enables each BS to achieve the delay requirement of VR transmission.

{\color{black}\subsubsection{\textbf{Lessons learned}} Clearly, we have demonstrated that ESNs can be an effective tool for resource management in a wireless VR network that needs to jointly consider the uplink and downlink resource block allocation. Some key outcomes learned from this application include the following:
\begin{itemize}
\item In non-wireless applications such as speech recognition, ESNs are used for data analytics.  In this VR application, ESNs are used as a reinforcement learning algorithm for downlink and uplink resource block management. The advantage of the ESN based RL algorithm is that it provided the network with an ability to predict the value of the VR QoS that results from each action (instead of relying on a Q-table to record the observed utility values as done in Q-learning) and, hence, it can {find the optimal action selection strategy that can maximize the individual (per SBS) VR QoS utilities}  
 without having to traverse all actions. As a result, ESN-based RL is suitable for wireless VR resource management problems in which both uplink and downlink resources must be managed jointly, thus increasing the search space for the wireless VR QoS optimization problem, compared to standard wireless resource management problems. This was a novel use case of ESNs that is motivated by the underlying wireless system, rather than by the need to process some data as done in computer vision.

\item Compared to most of the existing DNN-based RL algorithms that cannot analytically guarantee convergence to a final equilibrium or optimization solution, in this application, we have proved that ESN-based RL algorithms will finally converge to the expected VR QoS utilities if the learning parameters are appropriately set. 

 
 \item Due to the limited memory capacity of each ESN, the application of an ESN-based RL algorithm depends on the complexity of the underlying wireless problems. ESN-based RL algorithms can be used to solve the optimization problem with a moderate number of optimized variables while DNN-based algorithms can be used to solve more complex optimization problems. In this work, the ESN-based RL algorithms can achieve the same performance for resource block allocation as DNN-based RL algorithms. However, the time needed for training DNNs such as LSTMs will be much higher than the time needed for training ESNs. In consequence, one must choose an appropriate ANN architecture for RL depending on the complexity of the wireless optimization problems. In the wireless VR application, it could be more suitable to use a shallow ANN in the RL algorithm for problems such as channel selection and user association, while DNN-based RL algorithms are more suitable for {power allocation. This is due to the fact that, in power allocation problems, the optimized variables are continuous and, thus, the number of actions needed for RL will be much larger than those used in other problems (e.g., user association).}
\end{itemize}}
{\color{black}Here, we note that, the above lesson learned can be generalized to other shallow ANNs. }

\subsubsection{\textbf{Future Works}}

Clearly, ANNs are a promising tool to address challenges in wireless VR applications. In fact, the above application of ANNs for spectrum resource allocation can be easily extended to manage other types of resources such as computational resources, and video formats. Moreover, SNNs can be used for the prediction of the viewing VR video which is the VR video displayed at the headset of one user. Then, the network can reduce the data size of each transmitted VR video and pre-transmit each viewing VR video to the users. This is because SNNs are good at processing the rapidly changing, dynamic VR videos. Furthermore, RNNs can be used to predict and detect the VR users' movement such as eye movement and head movement and their interactions with the environment. Then, the network can pre-construct VR images based on these predictions which can reduce the time spent to construct the VR images. The user-VR system interactions are all time-dependent and, hence, RNNs are a good choice for performing such tasks. Note that, the prediction of the users' movement will directly affect the VR images that are sent to the users at each time slot and, hence, the learning algorithm must complete the training process during a short time period. In consequence, we should use RNNs that are easy to train for the prediction of the users' movement. Finally, CNNs can be used for VR video compression and recovery so as to reduce the data size of each transmitted VR video and improve the QoS for each VR user. This is because CNNs are good at storing large amount of data in spatial domain and {learn the features of VR images.}
 A summary of key problems that can be solved by using ANNs in wireless VR system is presented in Table \ref{ta:UAV} along with the challenges and future works.

 \subsection{Mobile Edge Caching and Computing}

       \begin{figure}[!t]
  \begin{center}
   \vspace{0cm}
    \includegraphics[width=7cm]{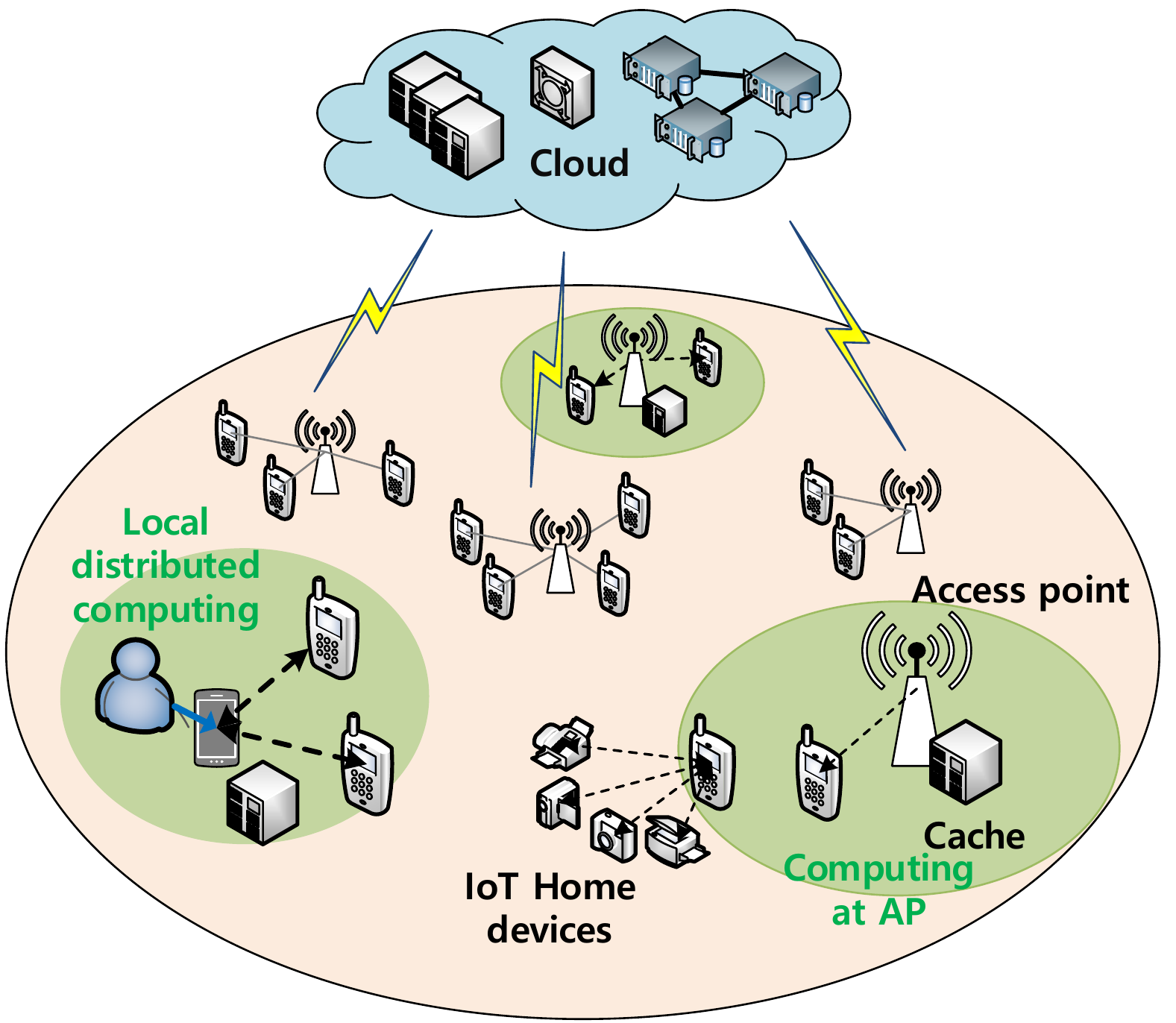}
    \vspace{-0.3cm}
    \caption{\label{CCfigure}{Mobile edge caching and computing wireless networks.}}
  \end{center}\vspace{-0.7cm}
\end{figure}
\vspace{-0cm}

   \subsubsection{\textbf{Mobile Edge Caching and Computing}}
   Caching at the edge of the wireless networks, as shown in Fig. \ref{CCfigure}, enables the network devices (BSs and end-user devices) to store the most popular content to reduce the data traffic (content transmission), delay, and bandwidth usage, as well as to improve the energy efficiency and the utilization of the users' context and social information{\color{black} \cite{6871674}}.
Recently, it has become possible to jointly consider cache placement and content delivery, using coded caching{\color{black}\cite{7537173}}. Coded caching enables network devices to create multicasting opportunities for specific content, via coded multicast transmissions, thus significantly improving the bandwidth efficiency \cite{fadlallah2017coding}.
However, designing effective caching strategies for wireless networks faces many challenges{\color{black}\cite{6871674}} such as solving optimized cache placement, cache update, and content popularity analytics problems.

In addition to caching, the wireless network's edge devices can be used for performing effective and low-latency computations using the emerging paradigm of \emph{mobile edge computing} \cite{mao2017survey}. The basic premise of mobile edge computing is to exploit local resources for computational purposes (e.g., for VR image generation or for sensor data processing), in order to avoid high-latency transmission to remote cloud servers. Mobile edge computing, which includes related concepts such as fog computing \cite{MugenRecent}, can decrease the overall computational latency by reducing the reliance on the remote cloud while effectively offloading computational resources across multiple local and remote devices. The key challenge in mobile edge computing is to optimally allocate computational tasks across both the edge devices (e.g., fog nodes) and the remote data servers, in a way to optimize latency.
Finally, it is worth noting that some recent works \cite{elbamby2017proactive} have jointly combined caching and computing. In this case, caching is used to store the most popular and basic computational tasks. Based on the caching results, the network will have to determine the optimal computational resource allocation to globally minimize latency.
However, optimizing mobile edge computing faces many challenges such as computing placement, computational resource allocation, computing tasks assignment, end-to-end latency minimization, and minimization of the energy consumption for the devices.

  \subsubsection{\textbf{Neural Networks for Mobile Edge Caching and Computing}} ANNs can play a central role in the design of the new mobile edge caching and computing mechanisms. For instance, the problems of optimal cache placement and cache update are all dependent on the predictions of the users' behaviors such as the users' content request problems. For example, the cache placement depends on the users' locations while the cache update depends on the frequency with which a user requests a certain content. Since human behavior can be predicted by ANNs, ANNs are a promising solution for effective mobile edge caching and computing.

In essence, ANNs can play a vital role in three major applications for mobile edge caching and computing.
  First, ANNs can be used for prediction and inference purposes. For example, ANNs can be used to predict the users' content request distributions and content request frequency. The content request distribution and content request frequency can be used to determine which content to store at the end-user devices or BSs.
Furthermore, ANNs can also be used to find social information from the collected data. In particular, ANNs can learn the users' interests, activities, and interactions. By exploiting the correlation between the users' data, their social interests, and their common interests, the accuracy of predicting future events such as the users' geographic locations, the next visited cells, and the requested contents can be dramatically improved \cite{bastug2014living}. For example, ANNs can be used to predict the users' interests. The users that have the same interests are highly likely to request the same content. Therefore, the system operator can cluster the users that have the same interests and store the popular contents they may request. Similarly, ANNs can be used to predict the computational requirements of tasks which in turn enables the network devices to schedule the computational resources in advance thus minimizing latency.

Second, ANNs can be used as an effective clustering algorithm to classify the users based on their activities such as content request, which enables the system operator to determine which contents to store at a storage unit and, thus, improve the usage of cached contents. 
For instance, the content requests of users can change over time while the cached content will be updated for a long time (i.g., one day) and, hence, the system operator must determine which content to cache by reviewing all the collected content requests. 
ANNs, such as CNNs, can be used to store the content request information and classify the large amount of content requests for cache update. In fact, predictions and clustering are interrelated and, therefore, ANNs can be used for both applications simultaneously. For example, ANNs can first be used to predict the users' content request distributions, and, then, ANNs can be used to classify the users that have similar content request distributions. Meanwhile, ANN-based clustering algorithms can be used to classify the computing tasks. Then, the computing tasks that are clustered into a group can be assigned to a certain computing center. In this case, each computing center will process one type of computing tasks thus reducing the computational time. Finally, ANNs can also be used for intelligently scheduling the computing tasks to different computing centers. In particular, ANNs can be used as an RL algorithm to learn each computing center's state such as its computational load, and then, allocate computing tasks based on the learned information to reduce the computational time.


Using ANNs for mobile edge caching and computing faces many challenges.
Data cleaning is an essential part of the data analysis process for mobile edge processing. For example, to predict the users' content requests, the data processing system should be capable of reading and extracting useful data from huge and disparate data sources. For example, one user's content request depends on this user's age, job, and locations.
In fact, the data cleaning process usually takes more time than the learning process. For instance, the type and volume of content that users may request can be in the order of millions and, hence, the data processing system should select appropriate content to analyze and predict the users' content request behaviors. For caching, the most important use of ANNs is to predict the users' content requests which directly determines the caching update. However, each user may request a large volume of content types such as video, music, and news, each of which having different formats and resolutions. Hence, for each user, the total number of the requested content items will be significantly large. However, the memory of an ANN is limited and, hence, each ANN can record only a limited number of requested contents. In consequence, an ANN must be able to select the most important content for content request prediction so as to help the network operator determine which content to store at mobile edge cache. Similarly, for computing tasks predictions, the limited-memory ANNs can only store a finite number of the computing tasks and, hence, they must select suitable computing tasks to store and predict. Moreover, as opposed to mobile edge caching that requires a long period of time to update the cached contents, mobile edge computing needs to process the tasks as soon as possible. Therefore, the ANNs used for mobile edge computing must complete their training process in a short period time.

  
 The existing literature has studied a number of problems related to the use of ANNs for caching {\color{black}\cite{chen2017caching,zeydan2016big,cobb2008web,chen2017echo}, and \cite{Bigdata,7873292,8478380,8513863}.} The authors in \cite{zeydan2016big} proposed a big data-enabled architecture to investigate proactive content caching in 5G wireless networks. {\color{black}In \cite{cobb2008web,8478380,8513863}, ANNs are used to determine the cache replacement and content delivery.} 
 The authors in \cite{Bigdata} developed a data extraction method using the Hadoop platform to predict content popularity. {\color{black}In \cite{7873292}, an extreme-learning machine neural network is used to predict content popularity.}
 The works in \cite{chen2017caching} and \cite{chen2017echo} developed an ESN-based learning algorithm to predict the users' mobility patterns and content request distributions.  In general, existing works such as in{\color{black}\cite{chen2017caching,zeydan2016big,cobb2008web,chen2017echo}, and \cite{Bigdata,7873292,8478380,8513863}} have used ANNs to solve the caching problems such as cache replacement, content popularity prediction, and content request distribution prediction.
   For mobile edge computing, in general, there is no existing work that uses ANNs to solve these relevant problems. 
   Next, we explain a specific ANNs' application for mobile edge caching.


  \subsubsection{\textbf{Example}}One illustrative application for the use of ANNs for mobile edge caching is presented in \cite{chen2017echo} which studies the problem of proactive caching in CRANs.
 In this model, the users are served by the RRHs which are connected to the cloud pool of the BBUs via capacity-constrained wired fronthaul links. The RRHs and the users are all equipped with storage units that can be used to store the most popular content that the users request. The RRHs which have the same content request distributions are grouped into a virtual cluster and serve their users using zero-forcing method. The content request distribution for a particular user represents the probabilities with which the user requests different content. Virtual clusters are connected to the content servers via capacity-constrained wired backhaul links. Since the backhaul (fronthaul) links are wired, we assume that the total transmission rate of the backhaul (fronthaul) links is equally allocated to the content that must be transmitted over the backhaul (fronthaul) links. Each user has a periodic mobility pattern and regularly visits a certain location. Since cache-enabled RRHs and BBUs can store the requested content, this content can be transmitted over four possible links: a) content server-BBUs-RRH-user, b) cloud cache-BBUs-RRH-user, c) RRH cache-RRH-user, and d) remote RRH cache-remote RRH-BBUs-RRH-user. The notion of \emph{effective capacity}\footnote{The effective capacity is a link-layer channel model that can be used to measure a content transmission over multiple hops. In particular, 
 the effective capacity can be used to measure a content transmission from the BBUs to the RRHs, then from RRHs to the users.}  \cite{Effectivecapacity} was used to capture the maximum content transmission rate of a channel {under a certain QoS requirement.} 
The effective capacity of each content transmission depends on the link that is used to transmit the content and the actual link capacity between the user and the associated RRHs.

The goal of \cite{chen2017echo} is to develop an effective framework for content caching and RRH clustering in an effort to reduce the network's interference and to offload the traffic of the backhaul and of the fronthaul based on the predictions of the users' content request distributions and mobility patterns. To achieve this goal, a QoS and delay optimization problem is formulated, whose objective is to maximize the long-term sum effective capacity of all users. This optimization problem involves the prediction of the content request distribution and of the periodic location for each user, and the finding of the optimal content to cache at the BBUs and at the RRHs. To predict the content request distribution and mobility patterns for each user, an ESN-based learning algorithm is used, similarly to the one described in Subsection \ref{se:ESN}. For each user, the BBUs must implement one ESN algorithm for content request distribution prediction and another ESN algorithm for mobility pattern prediction.

For the content request distribution prediction, the input of the developed ESN is a user's context which includes content request time, week, gender, occupation, age, and device type. The output is the predicted content request distribution. The ESN model consists of the input weight matrix, the output weight matrix, and the recurrent weight matrix (see Subsection \ref{se:ESN}). A linear gradient descent approach
is used to train the output weight matrix.
For mobility pattern prediction, the input of the developed ESN is the current location of each user and the output is the vector of locations that a user is predicted to visit for the next steps. In contrast to the recurrent matrix that is a sparse matrix and generated randomly, the recurrent matrix of the ESN used for mobility prediction contains only $W$ non-zero elements, where $W$ is the dimension of the recurrent matrix. This simplified recurrent matrix can speed up the training process of the ESNs. An offline manner using ridge regression is used to train the output weight matrix.

Based on the users' content request distribution and locations, the cloud can estimate the users' RRH association, determine each RRH's content request distribution, and, then, cluster the RRHs into several groups. Finally, the content that must be cached at the cloud and at the RRHs can be determined.
%
The analysis result proved that the ESN-based algorithm will reach an optimal solution to the content caching problem.


 \begin{figure}[!t]
  \begin{center}
   \vspace{0cm}
    \includegraphics[width=7cm]{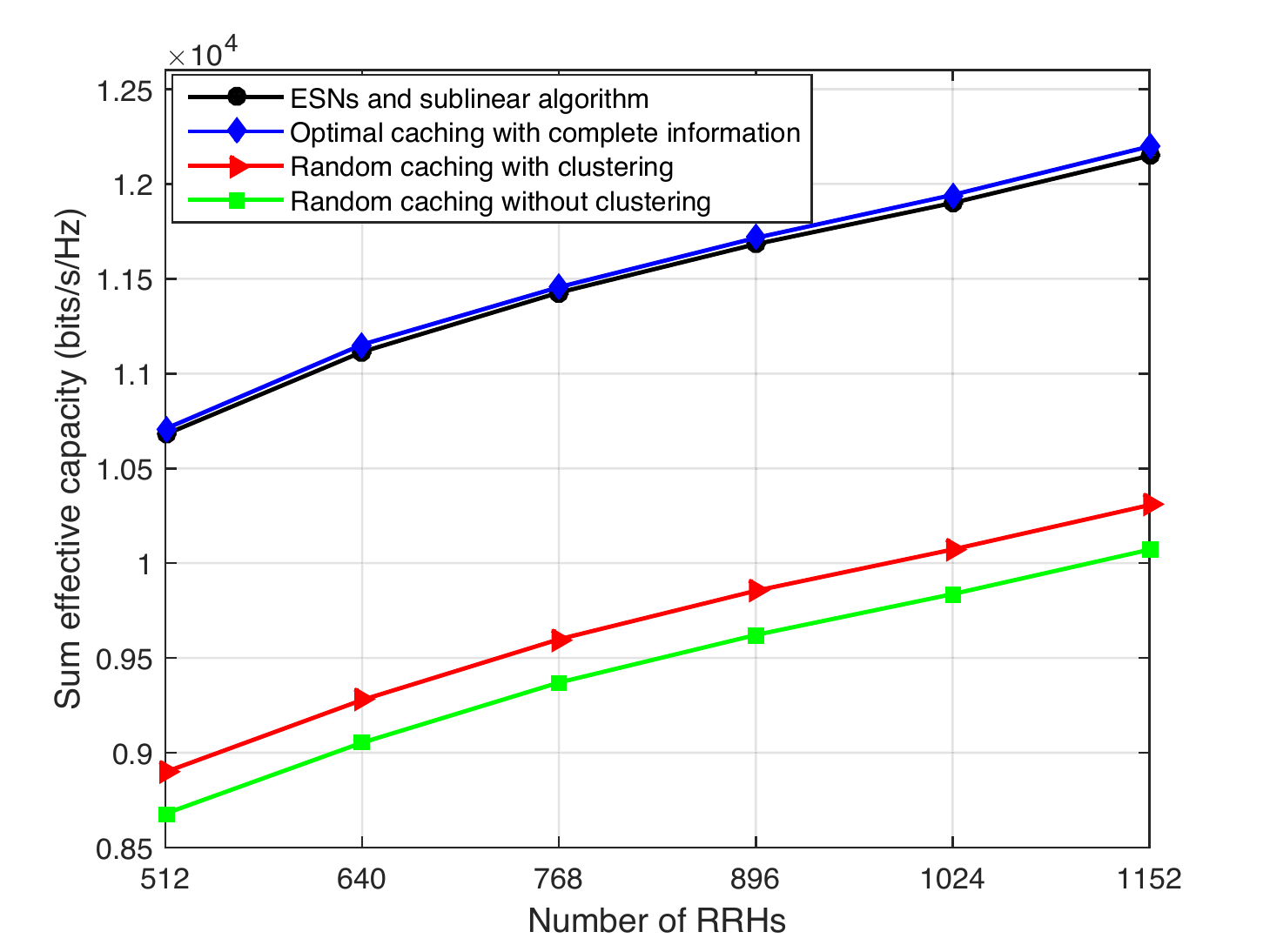}
    \vspace{-0.3cm}
    \caption{\label{figure10cache}{Sum effective capacity as function of the number of RRHs \cite{chen2017echo}.}}
  \end{center}\vspace{-0.7cm}
\end{figure}
\vspace{-0cm}

Fig. \ref{figure10cache} shows how the sum of the effective capacities of all the users in a period changes with the number of RRHs. As the number of the RRHs increases, the effective capacities of all the algorithms increase as the users become closer to their RRHs. The ESN approach can yield up to 21.6\% and 24.4\% of improvements in the effective capacity compared to random caching with clustering and random caching without clustering, respectively, for a network with 512 RRHs. This stems from the fact that the ESN-based algorithm can effectively use the predictions of the ESNs to determine which content to cache.


{\color{black}\subsubsection{\textbf{Lessons learned}}{The presented example of the mobile edge caching and computing application demonstrated that ESNs are effective for the prediction of the users' mobility patterns and content request distribution, based on which the cloud can determine the content stored at the cloud and at the RRHs.} Some key outcomes learned from this application include:
\begin{itemize}


\item Even though analyzing the memory capacity of an ESN is generally challenging, in this application, we were able to derive the memory capacity for an ESN that uses a linear activation function. Based on this analysis, {we can accurately set the size of the matrices and the memory capacity of each ESN that can precisely predict the users' mobility and content request distributions.} Here, we need to note that, as the memory capacity increases, the training complexity of an ESN will significantly increase. In this context, for mobility prediction in this application, we build an ESN model with minimum memory capacity that can accurately predict the users' mobility patterns and quickly converge. 
In fact, for different prediction tasks, one can adjust the memory capacity of each ESN using the obtained results to enable the ESNs to record all of the information needed for the predictions.

\item This example also showed that ESN-based learning algorithms can be trained to predict only one mobility pattern for each user. {For example, to predict the weekly mobility pattern of each user, an ESN-based learning algorithm cannot separate the mobility pattern in a week into several days and use a specific non-linear system to predict the users' mobility in each day. In fact, as we discussed in the UAV application in Subsection IV-B, using a unique non-linear system to predict the mobility of each user each day can significantly improve the accuracy of weekly mobility pattern prediction.} Learning using ESNs is more appropriate for predicting a single task, rather than for multiple prediction tasks. To overcome this challenge, one can use the conceptor notion that was discussed in Subsection IV-B. {\color{black}Note that, this observation can be generalized to other shallow RNNs and SNNs.}

\item Compared to conceptor ESNs, ESN based learning algorithms have a lower training complexity and faster convergence speed. However, as already mentioned, ESNs cannot separate the users' contexts for multiple mobility pattern predictions which will affect the prediction accuracy. In consequence, one must choose between standard ESN or a conceptor ESN depending on the number of prediction tasks needed and their complexity.
 In Subsection IV-D, the predictions are used to determine the cached content whose prediction is somewhat less challenging compared to other metrics that require more precise predictions such as the UAV locations in Subsection IV-B.
Therefore, we choose the ESN based prediction algorithms for mobility and content request distribution predictions.      



\end{itemize}

}

\subsubsection{\textbf{Future Works}} Clearly, ANNs will be an important tool for solving challenges in mobile edge caching and computing applications, especially for content request prediction and computing tasks prediction. In fact, CNNs, that are good at storing voluminous data in spatial domains, can be used to investigate the content correlation in the spatial domains. Based on the content correlation, each BS can store the contents that are the most related to other contents to improve the caching efficiency and hit ratio. Moreover, RNNs can be used as self-organizing RL algorithms to allocate computational resources. RNNs are suitable here because they can record the utility values resulting from different computational resources allocation schemes as time  elapses. Then, the RNN-based RL algorithms can find the optimal computational resource allocation after several implementations. Meanwhile, in contrast to the user association in cellular network where each user can only associate with one BS, one computing task can be assigned to several computing centers and one computing center can process different computing tasks. Therefore, the problem of computing task assignment is a many-to-many matching problem~\cite{gu2015matching}. RNN-based RL algorithms can also be used to solve the computing task assignment problem due to their advantages in analyzing {historical data pertaining to past assignments of computing tasks.}
 In addition, DNN-based RL algorithms can be used to jointly optimize the cache replacement and the content delivery. To achieve this purpose, each action of the DNN-based RL algorithm must contain one content delivery method as well as one cache update scheme. This is because DNNs are good at storing large amounts of utility values resulting from different content delivery and cache update schemes. Last but not as least, SNNs can be used to predict the dynamic computational resource demands for each user due to their advantages in dealing with highly dynamic data.
A summary of the key problems of using ANNs for mobile edge caching and computing is presented in Table \ref{ta:UAV} along with the challenges and future works.





    \subsection{{Co-existence of Multiple Radio Access Technologies}}
    \subsubsection{\textbf{Co-existence of Multiple Radio Access Technologies}} 
    To cope with the unprecedented increase in mobile data traffic and realize the envisioned 5G services, a significant enhancement of per-user throughput and overall system capacity is required~\cite{5G_vision}. Such an enhancement can be achieved through advanced PHY/MAC/network technologies and efficient methods of spectrum management. In fact, one of the main advancements in the network design for 5G networks relies on the integration of multiple different radio access technologies (RATs){\cite{7815331}}. {\color{black}Multi-RAT based networks encompass several technologies in which spectrum sharing is important. These include cognitive radio networks, LTE-U networks, as well as heterogeneous networks that include both mmWave and sub-6 GHz frequencies.} With the multi-RAT integration, a mobile device can potentially transmit data over multiple radio interfaces such as LTE and WiFi \cite{8011313}, at the same time, thus improving its performance~\cite{multiRAT_1}. Moreover, a multi-RAT network allows fast handover between different RATs and, thus, it provides seamless mobility experience for users. Therefore, the integration of different RATs results in an improvement in the utilization of the available radio resources and, thus, in an increase in the system's capacity. It also guarantees a consistent service experience for different users irrespective of the served RAT and it facilitates the network management.

    Spectrum management is also regarded as another key component of Multi-RAT based networks {\color{black}\cite{5473880}}. Unlike early generations of cellular networks that operate exclusively on the sub-6 GHz (microwave) licensed band, Multi-RAT based networks are expected to transmit over the conventional sub-6 GHz band, the unlicensed spectrum and the 60 GHz mmWave frequency band~\cite{multiRAT_2, omid_2}. We note that, on the other hand, the classical LTE microwave licensed band is reliable, however, limited and hence is a scarce resource. On the other hand, the unlicensed bands can be used to serve best effort traffic only since the operation over this spectrum should account for the presence of other coexisting technologies. 
    Therefore, a multi-mode BS operating over the licensed, unlicensed, and mmWave frequency bands can exploit the different characteristics and availability of the frequency bands thus providing robust and reliable communication links for the end users~\cite{omid_2}. However, to reap the benefits of multi-mode BSs, spectrum sharing is crucial. 
    
    
\subsubsection{\textbf{Neural Networks for Spectrum Management and Multi-RAT}} ANNs are an attractive solution approach for tackling various challenges that arise in multi-RAT scenarios. To leverage the advantages of such multi-RAT networks, ANNs can allow the smart use of different RATs wherein a BS can learn when to transmit on each type of frequency band based on the underlying network conditions. For instance, ANNs may allow multi-mode BSs to steer their traffic flows between the mmWave, the microwave, and the unlicensed band based on the availability of a LoS link, the congestion on the licensed band and the availability of the unlicensed band. Moreover, in LTE-WiFi link aggregation (LWA) scenarios, ANNs allow cellular devices to learn when to operate on each band or utilize both links simultaneously. 

Moreover, ANNs can provide multi-mode BSs with the ability to learn the appropriate resource management procedure over different RATs or spectrum bands in an online manner and, thus, to offer an autonomous and self-organizing operation with no explicit communication among different BSs, once deployed. For instance, ANNs can be trained over large datasets which take into account the variations of the traffic load over several days for scenarios in which the traffic load of WiFi access points (WAPs) can be characterized based on a particular traffic model~\cite{traffic_model}. It should be noted that cellular data traffic networks exhibit statistically fluctuating and periodic demand patterns, especially for applications such as file transfer, video streaming, and browsing~\cite{traffic_model}. ANNs can also accommodate the users' mobility patterns to predict the availability of a LoS link, thus, allowing the transmission over the mmWave band. In particular, they can be trained to learn the antenna tilting angle based on the environment changes in order to guarantee a LoS communication link with the users and, thus, to enable an efficient communication over the mmWave spectrum. Moreover, ANNs may enable multiple BSs to learn how to form multi-hop, mmWave links over backhaul infrastructure, while properly allocating resources across those links in an autonomous manner~\cite{UNF, omid_3}. To cope with the changes in the traffic model and/or the users' mobility pattern, ANNs can be combined with online ML~\cite{online_learning} by properly re-training the weights of the developed learning mechanisms. Multi-mode BSs can, thus, learn the traffic patterns over time and, thus, predict the future channel availability status. With proper network design, ANNs can allow operators to improve their network's performance by reducing the probability of congestion occurrence while ensuring a degree of fairness to the other corresponding technologies in the network.

A proactive resource management of the radio spectrum for multi-mode BSs can also be achieved using ANNs. In a \emph{proactive} approach, rather than reactively responding to incoming demands and serving them when requested, multi-mode BSs can predict traffic patterns and determine future off-peak times on different spectrum bands so that the incoming traffic demand can be properly allocated over a given time window. In an LTE-U system, for instance, a proactive coexistence mechanism may enable future delay-intolerant data demands
to be served within a given prediction window ahead of their actual arrival time thus avoiding the underutilization of the unlicensed spectrum during off-peak hours~\cite{ursulaEW}. This will also lead to an increase in the LTE-U transmission opportunity as well as to a decrease in the collision probability with WAPs and other BSs in the network.

Several existing works have adopted various learning techniques in order to tackle a variety of challenges that arise in multi-RAT networks~\cite{you2019ai,fuzzy, mehdi, MLP_RAT, hopfield, Chen2016Echo,8468000,8353153}. The problem of resource allocation with uplink-downlink decoupling in an LTE-U system has been investigated in~\cite{Chen2016Echo} in which the authors propose a decentralized scheme based on ESNs.  The authors in~\cite{fuzzy} propose a fuzzy-neural system for resource management among different access networks. {\color{black}The work in \cite{8353153} used an ANN-based learning algorithm for channel estimation and channel selection.} The authors in~\cite{MLP_RAT} propose a supervised ANN approach, based on FNNs, for the classification of the users' transmission technology in a multi-RAT system. In~\cite{hopfield}, the authors propose a hopfield neural network scheme for multi-radio packet scheduling. In~\cite{mehdi}, the authors propose a cross-system learning framework in order to optimize the long-term performance of multi-mode BSs, by steering delay-tolerant traffic towards WiFi. {\color{black}The work in \cite{8468000} used a deep RL algorithm for mode selection and resource management in a fog radio access network.
}{\color{black}Other important problems in this domain include root cause analysis issues as the ones are studied in \cite{you2019ai}.}
Nevertheless, these prior works~{\color{black}\cite{you2019ai,fuzzy, mehdi, MLP_RAT, hopfield, Chen2016Echo,8468000,8353153}} consider a reactive approach in which the data requests are first initiated and, then, the resources are allocated based on their corresponding delay tolerance value.
In particular, existing works do not consider the predictable behavior of the traffic and, thus, they do not account for future off-peak times during which data traffic could be distributed among different RATs.


Here, note that, ANNs are suitable for learning the data traffic variations over time and, thus, to predict the future traffic load. In particular, since LSTM cells are capable of storing information for long periods of time, they can learn the long-term dependency within a given sequence. Predictions at a given time step are influenced by the network activations at previous time steps, thus, making LSTMs an attractive solution for proactively allocating the available resources in multi-RAT systems. In what follows, we summarize our work in~\cite{LTE-U_DL}, in which we developed a deep RL scheme, based on LSTM memory cells, for allocating the resources in an LTE-U network over a fixed time window $T$.

\subsubsection{\textbf{Example}}
An interesting application of DNNs in the context of LTE-U and WiFi coexistence is presented in~\cite{LTE-U_DL}. The work in~\cite{LTE-U_DL} considers a network composed of several LTE-U BSs belonging to different LTE operators, several WAPs and a set of unlicensed channels on which LTE-U BSs and WAPs can operate on. The LTE carrier aggregation feature, using which the BSs can aggregate up to five component carriers belonging to the same or different operating frequency bands, is adopted. We consider a time domain divided into multiple time windows of duration $T$, each of which consisting of multiple time epochs $t$. Our objective is to proactively determine the spectrum allocation vector for each BS at $t=0$ over $T$ while guaranteeing long-term equal weighted airtime share with WLAN. In particular, each BS learns its channel selection, carrier aggregation, and fractional spectrum access over $T$ while ensuring long-term airtime fairness with the WLAN and the other LTE-U operators. A contention-based protocol is used for channel access over the unlicensed band. The exponential backoff scheme is adopted for WiFi while the BSs adjust their contention window size (and, thus, the channel access probability) on each of the selected channels based on the network traffic conditions while also guaranteeing a long-term equal weighted fairness with WLAN and other BSs.

The proactive resource allocation scheme in~\cite{LTE-U_DL} is formulated as a noncooperative game in which the players are the BSs. Each BS must choose which channels to transmit on along with the corresponding channel access probabilities at $t = 0$ for each $t$ of the next time window $T$. This, in turn, allows the BSs to determine future off-peak hours of the WLAN on each of the unlicensed channels thus transmitting on the less congested channels. Each BS can therefore maximize its total throughput over the set of selected channels over $T$ while guaranteeing long-term equal weighted fairness with the WLAN and the other BSs. To solve the formulated game (and find the so-called Nash equilibrium solution), a DNN framework based on LSTM cells was used. To allow a sequence-to-sequence mapping, we considered an encoder-decoder model as described in Section~\ref{se:DNN}. In this model, the encoder network maps an input sequence to a vector of a fixed dimensionality and then the decoder network decodes the target sequence from the vector. In this scheme, the input of the encoder is a time series representation of the historical traffic load of the BSs and WAPs on all the unlicensed channels. The learned vector representation is then fed into a multi-layer perceptron (MLP) that summarizes the input vectors into one vector, thus accounting for the dependency among all the input time series vectors. The output of the MLP is then fed into different separate decoders, allowing each BS to reconstruct its predicted action sequence.

To train the proposed network, the REINFORCE algorithm \cite{reinforce} is used to compute the gradient of the expected reward with respect to the policy parameters, and the standard gradient descent optimization algorithm~\cite{policy_gradient} is adopted to allow the model to generate optimal action sequences for input history traffic values. In particular, we considered the RMSprop gradient descent optimization algorithm~\cite{rmsprop}, an adaptive learning rate approach, wherein the learning rate of a particular weight is divided by a running average of the magnitudes of the recent gradients for that weight.

\begin{figure}[t!]
  \begin{center}
  \vspace{-0.cm}
    \includegraphics[width=7cm]{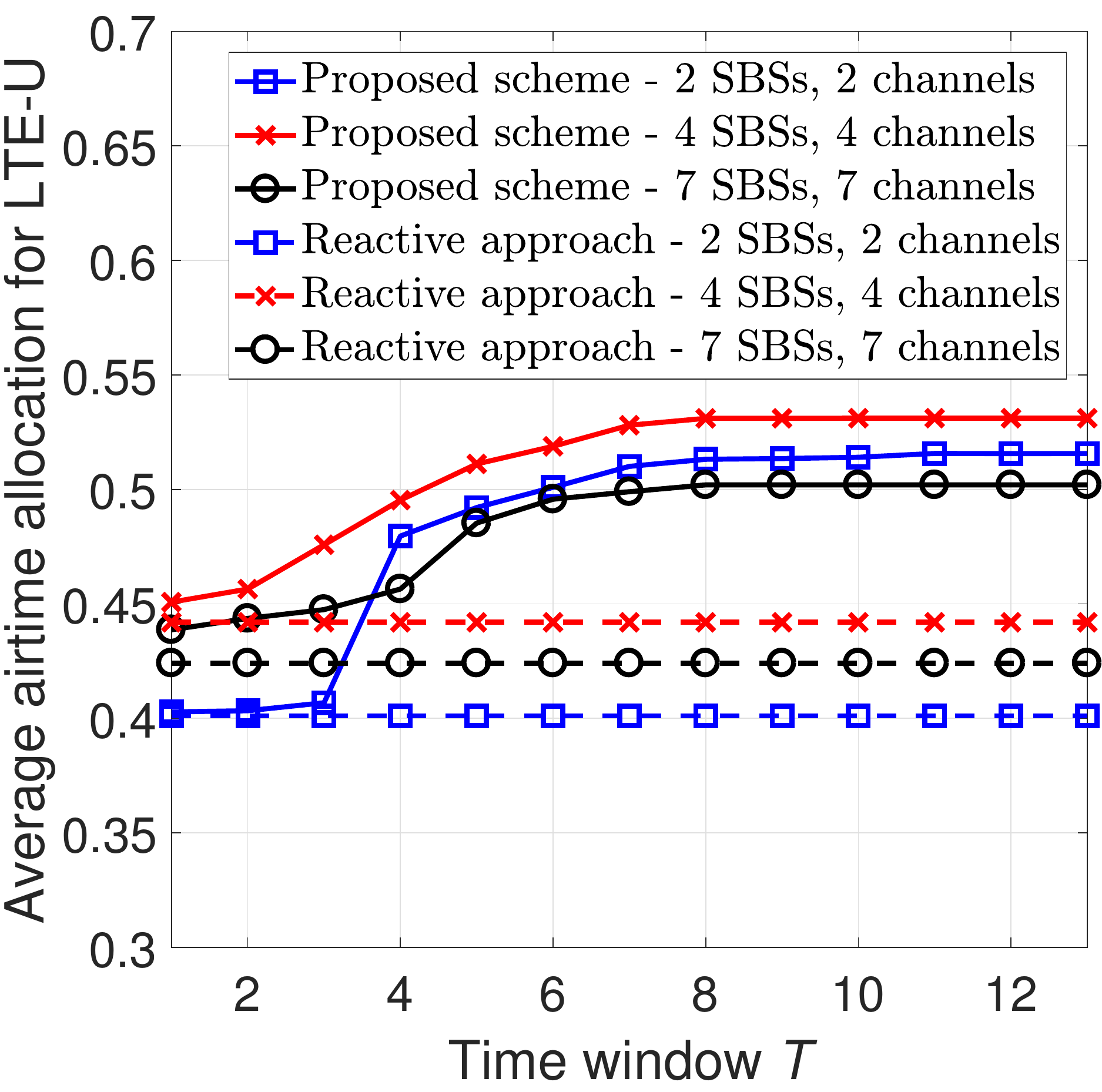}
   \caption{The average throughput gain for LTE-U upon applying a proactive approach (with varying $T$) as compared to a reactive approach~\cite{LTE-U_DL}.}\label{proactive_reactive} 
   \vspace{-0.7cm}
  \end{center}
\end{figure}

The proposed proactive resource allocation scheme was compared with a reactive approach for three different network scenarios. Fig.~\ref{proactive_reactive} shows that for very small values of $T$, the proposed scheme does not yield any significant gains. However, as $T$ increases, the BSs have additional opportunities for shifting part of the traffic into the future and, thus, the gains start to become more pronounced. For example, we can see that, for $4$ BSs and $4$ channels, the proposed proactive scheme achieves an increase of 17\% and 20\% in terms of the average airtime allocation for LTE-U as compared to the reactive approach. Here, note that the gain of the proposed scheme, with respect to the reactive approach, keeps on increasing until it reaches a maximum achievable value, after which it remains almost constant.
{\color{black}\subsubsection{\textbf{Lessons learned}} In the aforementioned application, we have demonstrated that LSTM can be an effective tool for resource management in an LTE-U system that needs to maintain a fair co-existence between WiFi and LTE. The key benefit brought forward by LSTM in this application is that it enabled the cellular system to accurately predict future off-peak hours of WiFi, so as to seize the channels on which to transmit. This, in turn, led to a better co-existence between the two systems, owing to the predictive ability of LSTM that provided the system with the ability to use historical WiFi traffic data to determine future traffic and, thus, make anticipatory resource management decisions. The main lessons learned here include:
\begin{itemize}
\item LSTM has mostly been used for data analytics. In the aforementioned application, the network needed LSTM as a part of a RL algorithm that can determine the solution of a game-theoretic setting, which can be thought of as the solution of a series of optimization problems that are solved at the level of each BS. In this context, LSTM enabled the RL algorithm to estimate future utilities (rather than just observe them from the environment as done in Q-learning) and, hence, be able to seek better optimization problem solutions (equivalent to game-theoretic equilibria). This was a novel use case of LSTM that is motivated by the underlying wireless system, rather than by the need to process some data.
\item Even though proving the optimality properties of the LSTM output itself is difficult, in this application, we have shown that by combining LSTM with a game-theoretic framework, we can ensure that, whenever the RL algorithm converges, it is guaranteed to be at a Nash equilibrium (i.e., as a point at which none of the RL algorithms can find a better outcome). However, guaranteeing convergence analytically is much more challenging than, for example, the ESN-based approaches we used in the VR and the UAV problems, due to the deep nature of LSTM. We do note that our thorough simulations (for many simulation parameters and settings), showed that the algorithm will actually always converge, even though that is not ascertained analytically. An interesting future research to address in this context is to analyze the convergence for LSTM-based RL in an LTE-U context, or more generally, in a multi-RAT resource management context. {\color{black} This difficulty in analyzing the convergence of LSTM can also be encountered when dealing with other types of ANN-based RL schemes.}
\item In this LTE-U scenario, the network operator can train the LSTM in a completely offline manner since all that is needed for this training is to use past observations of WiFi traffic and, it is generally known that, within a geographic area, over long periods of time, the wireless data traffic parameters are more or less consistent. This is a key motivation for using a deep architecture here. 


\item This work has demonstrated that, even though deep learning based on LSTM can provide significant improvements in the predictions of time-stamped sequences of data (here being the time-varying WiFi traffic), in a practical wireless application, one does need to use many layers. In fact, through our simulations, we observed that increasing the number of hidden layers has a very small impact on the achieved performance. This is mainly due to the fact that the WiFi traffic that is used as input to LSTM in this work, is much less time-varying than the datasets that are used in other, non-wireless fields, such as in natural language processing, where multiple layers provide more gains. However, we do note that, in this work, we wanted to predict a future sequence of WiFi traffic data based on a significant history of data and, therefore, using shallow networks like ESN (e.g., as done in the UAV and VR applications) would not have been as effective as using LSTM that has both short and long term memory (as explained in Subsection III-C) and can more effectively handle predictions of future sequences that require significant historical data, as is the case for WiFi traffic. That said, in our simulations, we only needed three hidden layers to reap the benefits of LSTM. 

\item As it is evident from the previous point,  whether or not one adopts a deep architecture or a very advanced type of ANN depends on the type of application that is being addressed. For WiFi traffic prediction, a deep architecture was appropriate. Meanwhile, for prediction of mobility data and user-based content in the VR and UAV applications that were previously discussed, the use of a shallow RNN by itself provided significant gains, even without using a deep architecture. That said, as we will see later in Section IV, in some applications like IoT, one can solve meaningful wireless problems by resorting to very simple ANNs, such as FNNs, without the need for deep architectures or more advanced structures. This is a major contrast to other ML application domains such as computer vision, where oftentimes a complex, deep ANN is needed to obtain meaningful results.

\item One disadvantage of using an ANN within a RL algorithm is that the prediction errors may affect the performance of the outcome. In some sense, within the aforementioned game-theoretic context, the efficiency of the reached equilibrium can be impacted by the prediction errors. While this is true for all of the applications in which we used ANNs as part of a RL algorithm, the effect of the prediction errors may be more pronounced for the LTE-U application because it may lead to the LTE seizing more or less WiFi slots than needed, which can directly impact the operation of the WiFi user. Naturally, this is a more serious drawback than in scenarios where the network is simply  using ANNs to cache data (e.g., as in the previously discussed UAV application) or perform cell association (in which case, if a prediction error occurs, the network can simply resort back to known cell association algorithms).


\end{itemize}}

\subsubsection{\textbf{Future Works}} The above application of ANNs to LTE-U systems can be easily extended to a multi-mode network in which the BSs transmit on the licensed, the unlicensed, and the mmWave spectrum. 
In fact, given their capability of dealing with time series data, RNNs can enhance mobility and handover in highly mobile wireless environments by learning the mobility patterns of users thus decreasing the ping-pong effect among different RATs. For instance, a predictive mobility management framework can address critical handover issues, including frequent handovers, handover failures, and excessive energy consumption for seamless handovers in emerging dense multi-RAT wireless cellular networks. 
ANNs can also predict the QoS requirements, in terms of delay and rate, for the future offered traffic. Moreover, they can predict the transmission links' conditions and, thus, schedule users based on the links' conditions and QoS requirements. Therefore, given the mobility patterns, transmission links' conditions and QoS requirements for each user, BSs can learn how to allocate different users on different bands such that the total network performance, in terms of delay and throughput, is optimized.

An interesting future work of the use of DNNs for mmWave communication is antenna tilting. In particular, DNNs are capable of learning several features of the network environment and thus predicting the optimal tilt angle based on the availability of a LoS link and data rate requirements. This in turn improves the users' throughput thus achieving high data rate. Moreover, LSTMs are capable of learning long time series and thus can allow BSs to predict the link formation for the mmWave backhaul network. In fact, the formation of this backhaul network is highly dependent on the network topology and the traffic conditions. Therefore, given the dynamics of the network, LSTMs enable BSs to dynamically update the formation of the links among each others based on the changes in the network. Moreover, SNNs can be used for mmWave channel modeling since they can process and predict continuous-time data effectively. A summary of key problems that can be solved by using ANNs in multi-RAT system is presented in Table \ref{ta:UAV} along with the challenges and future works.


  \subsection{Internet of Things}
     \subsubsection{\textbf{The Internet of Things}}
   In the foreseeable future, it is envisioned that trillions of machine-type devices such as wearables, sensors, connected vehicles, or mundane objects will be connected to the Internet, forming a massive IoT ecosystem \cite{agiwal2016next}. The IoT will enable machine-type devices to connect with each other over wireless links and operate in a self-organizing manner{\color{black}\cite{7123563}}. Therefore, IoT devices will be able to collect and exchange real-time information to provide smart services.
In this respect, the IoT will allow delivering innovative services and solutions in the realms of smart cities, smart grids, smart homes, and connected vehicles that could provide a significant improvement in people's lives. 
However, the practical deployment of an IoT system still faces many challenges{\color{black}\cite{7123563}} such as data analytics, computation, transmission capabilities, connectivity, end-to-end latency, security \cite{8599014}, and privacy. In particular, how to provide massive device connectivity with stringent latency requirement will be one of the most important challenges. The current centralized communication models and the corresponding technologies may not be able to provide such massive connectivity. Therefore, there is a need for a new communication architecture, such as fog computing models for IoT devices connectivity. Moreover, for each IoT device, energy and computational resources are limited. Hence, how to allocate computational resources and power for all the IoT devices to achieve the data rate and latency requirements is another challenge. 



%



\subsubsection{\textbf{Neural Networks for the Internet of Things}} ANNs can be used to address some of the key challenges within the context of the IoT. 
So far, ANNs have seen four major applications for the IoT. 
First, ANNs enable the IoT system to leverage intelligent data analytics to extract important patterns and relationships from the data sent by the IoT devices. For example, ANNs can be used to discover important correlations among data to improve the data compression and data recovery.  
Second, using ANN-based RL algorithms, IoT devices can operate in a self-organizing manner and adapt their strategies (i.e., channel selection) based on the wireless and users environments. For instance, an IoT device that uses an ANN-based RL algorithm can dynamically select the most suitable frequency band for communication according to the network state. Third, the IoT devices that use ANN-based algorithms can identify and classify the data collected from the IoT sensors. Finally, one of the main goals of the IoT is to improve the life quality of humans and reduce the interaction between human and IoT devices. Thus, ANNs can be used to predict the users behavior to provide advanced information for the IoT devices. For example, ANNs can be used to predict the time that an individual will come home, and, hence, adjust the control strategy for the IoT devices at home.

Using ANNs for IoT faces many challenges. First, in IoT, both energy and computational resources are limited. Therefore, one should consider the tradeoff between the energy and computational needs of training ANNs and the accuracy requirement of a given ANN-based learning algorithm. In particular, the higher the required accuracy, the higher the computational and energy requirements. Second, within an IoT ecosystem, the collected data may have different structure and even contain several errors. Therefore, when data are used to train ANNs, one should consider how to classify the data and deal with the flaws in the data. In other words, the ANNs in IoT must tolerate erroneous data. Third, in the IoT system, ANNs can exploit thousands of types of data for prediction and self-organizing control. For a given task, the data collected from the IoT devices may not all be related to the task. Hence, ANNs must select suitable data for the task. 


The existing literature{\color{black}\cite{kaminski2017neural,naidu1990use,ning2011future,alam2016analysis,luo2016laguerre,du2017reconfigurable,8496746,8502822,8463616}} has studied a number of problems related to using ANNs for IoT. In \cite{kaminski2017neural}, the authors use a framework to treat an IoT network as an ANN to reduce delivery latency. The authors in \cite{naidu1990use} and \cite{ning2011future} used a backpropagation neural network for sensor failure detection in an IoT network. In \cite{alam2016analysis}, eight ML algorithms, including DNNs and FNNs, are tested for human activities classification and robot navigation as well as body postures and movements. In \cite{luo2016laguerre}, the authors used the Laguerre neural network-based approximate dynamic programming scheme to improve the tracking efficiency in an IoT network. The authors in \cite{du2017reconfigurable} develped a streaming hardware accelerator for CNNs to improve the accuracy of image detection in an IoT network. {\color{black} The work in \cite{8496746} used a denoising autoencoder neural network for data sampling in an IoT network.  In \cite{8502822}, a deep belief network is used for entity state prediction. The authors in \cite{8463616} used ANNs for target surveillance.}
In summary, the prior works used ANNs to solve a number of IoT problems such as IoT network modeling, failure detection, human activities classification, and tracking accuracy improvement.
  However, ANNs can also be used to analyze the data correlation for data compression and data recovery, to identify humans, to predict human activities, and to manage the resources of devices. Next, we explain a specific ANNs' application for IoT.

  \subsubsection{\textbf{Example}} One illustrative application for the use of ANNs within the context of the IoT is presented in \cite{kaminski2017neural} which studies how to improve the communication quality by mapping IoT networks to ANNs. The considered IoT network is primarily a wireless sensor network.
 Two objective functions are considered : a) minimizing the overall cost of communication between the devices mapped to the neurons in the input layer and the devices mapped to the neurons in the output layers. Here, the overall cost represents the total transmit power of all devices used to transmit the information signals, and b) minimizing the expected transmission time to deliver the information signals.

 To minimize the total transmit power and the expected transmit time for the IoT, the basic idea of \cite{kaminski2017neural} is to train an ANN so as to approximate the objective functions discussed above and, then, map the IoT network to the ANN. FNNs, are used for this mapping since they transmit the information in only one direction, forward, from the input nodes, through the hidden nodes, and to the output nodes. First, one must identify the devices that want to send signals as well as the devices that will receive signals. The IoT devices that want to send signals are mapped to the neurons in the input layers. The IoT devices that want to receive signals are mapped to the neurons in the output layers. The other IoT devices are mapped to the neurons in the hidden layers. Some of the devices that are mapped to the hidden layers will be used to forward the signals. Then, the FNN is trained in an offline manner to approximate the objective functions. 
 The IoT network devices are mapped into neurons and wireless links into connections between neurons, and, hence, a method is needed to map the trained FNN to the IoT network. Since the computational resources of each IoT device is limited, IoT devices with different computational resources will map to a different number of neurons. For example, an IoT device that has more computational resources can map to a larger number of neurons. Moreover, to ensure the integrity of the mapping model,  each neuron can only map to one of the IoT devices. Given that there are several ways to map the IoT network to the trained FNN, the optimal mapping is formulated as an integer linear program which is then solved using CPLEX. {When the optimal mapping between the IoT network and the trained FNN is found, the optimal connections between the IoT devices are built.} Hence, if the IoT network can find the optimal connections for all devices based on the objective functions, the transmit power and expected transmit time can be reduced. Simulation results show that the mapping algorithm can achieve significant gains in terms of total transmit power and expected transmit time compared to a centralized algorithm. This is because the IoT network uses FNNs to approximate the objective functions and find the optimal device connections.

{\color{black}\subsubsection{\textbf{Lessons learned}}  {This IoT application has shown that FNNs are an effective tool for network mapping in IoTs so as to find the optimal transmission links from the transmitters to the receivers through the relays. We can summarize the main lessons learned here as follows:
}
\begin{itemize}
\item The advantage of FNNs for the studied IoT application is that it enabled the IoT devices to optimally build the transmission links between the receivers and the transmitters so as to reduce the transmission delay without any communications among the IoT devices.
In this application, the wireless network only consists of the receivers, the transmitters, and the relays, and, the data in this wireless network will only be transmitted from the transmitters to the relays, then from the relays to the receivers. 
 The use of FNNs to map this network is appropriate as it allows one to find the optimal transmission links between the transmitters and the receivers, through the relays. This was a novel use case of FNNs that is motivated by the underlying wireless system. 
 
\item FNNs are very simple neural networks with little training overhead, which makes them suitable for implementation in IoT systems in which the devices are resource-constrained.
 

\item One disadvantage of using FNNs for mapping wireless networks is that they can be only used for a network with a small number of transmitters and receivers. This is due to the fact that, as the number of transmitters and receivers increases, the number of neurons in the input, output, and hidden layers increases. Since FNNs need to calculate the gradients of all of the neurons (in contrast to ESNs that only need to update the output weight matrix), the training complexity will significantly increase. 

\item The presented IoT application is restricted to a very simple mapping of IoT devices via an FNN. However, the IoT domain is much richer than this application and one can envision a plethora of resource management, physical layer enhancement, and network optimization problems that can be addressed using more elaborate ANNs such as those presented in Section III (and in the previous applications).


\end{itemize} {\color{black}Note that, the first, second, and third bullets observations above can be generalized to other works that rely on FNNs for solving wireless communication problems. } }
\subsubsection{\textbf{Future Works}}
ANNs are undoubtedly an important tool for solving a variety of problems in the IoT, particularly in terms of intelligent data analytics and smart operation. In fact, beyond using FNNs to map the IoT devices hence optimizing the connections between the IoT devices as discussed above, FNNs can also be used to map other systems. For example, one can map the input layer of an FNN to the IoT devices and the output layer to the computing centers. Then, one can find an optimal allocation of computational tasks via FNN mapping.
Moreover, ANNs can be used for data compression and recovery so as to reduce both the size of the transmitted data and end-to-end devices latency. To compress the data, an ANN needs to extract the most important features from the data and, then, these features can be used to present the compressed data.
In particular, CNNs can be used for data compression and recovery in the spatial domain while RNNs can be used for data compression and recovery in the time domain. This is because CNNs are effective at extracting patterns and features from large amounts of data while RNNs are suitable for extracting the relationships from time-dependent series data. In addition, DNNs can be used for human identification. An IoT ecosystem that can identify different individuals can  pre-allocate spectral or computational resources to the IoT devices that a certain individual often uses. DNNs are suitable here because they have multiple hidden layers to store more information related to a user compared to other ANNs and, hence, DNNs can use one user's information such as hairstyle, clothes, and oral patterns to identify that individual so as to provide services tailored to this user.
A summary of key problems that can be solved by using ANNs in IoT system is shown in Table \ref{ta:UAV} along with the challenges and future works.

{\subsection{Summary}
\begin{table*}
{\color{black} 
\centering
  \newcommand{\tabincell}[2]{\begin{tabular}{@{}#1@{}}#2\end{tabular}}
\renewcommand\arraystretch{1}
 \caption{
    \vspace*{-0.1em}Summary of the use of ANN-based Learning Algorithms for Existing Works in Specific Application }\label{ta:EX}\vspace*{-0.6em}
\centering
\begin{tabular}{|c |l|l |l |c |c |c |}
\hline
 \multicolumn{1}{|c|}{\multirow{2}{*}{\textbf{Applications}}} &  \multicolumn{2}{|c|}{\multirow{1}{*}{\textbf{ Existing Works }}} &   \multicolumn{1}{|c|}{\multirow{2}{*}{\textbf{ ANN Tool}}}   &   \multicolumn{2}{|c|}{\multirow{1}{*}{\textbf{Data Analytics }}}& \multicolumn{1}{|c|}{\multirow{2}{*}{\textbf{ RL }}}  \\
 \cline{5-6}
  \cline{2-3}
 &\textbf{Problems}&\textbf{Reference}&&\textbf{Supervised}&\textbf{Unsupervised}&\\
\hline
\multirow{8}{*}{\textbf{UAV}} & \multirow{2}{*}{$\bullet$ UAV control.} &$\bullet$ \cite{8432464}&\multirow{1}{*}{$\bullet$ FNNs.}&  &&$\surd$\\
&&$\bullet$ \cite{nodland2013neural} & $\bullet$ FNNs & $\surd$&&\\
\cline{2-7}
& {$\bullet$ Position estimation.}& $\bullet$  \cite{braga2016image} & $\bullet$ FNNs & $\surd$ &&\\
& {$\bullet$ UAV detection.}& $\bullet$ \cite{8353152}& \multirow{1}{*}{$\bullet$ RNNs.} & $\surd$ &&\\
\cline{2-7}
&\multirow{2}{*}{$\bullet$ Resource allocation.}&$\bullet$ \cite{cui2018multi} & $\bullet$ RNNs. &  &&$\surd$ \\
&&$\bullet$ \cite{chen2018liquid} & $\bullet$ SNNs. & && $\surd$ \\
\cline{2-7}
& {\multirow{2}{*}{$\bullet$ UAV deployment.} }&$\bullet$  \cite{7451189} &$\bullet$ FNNs. &  $\surd$ && \\
& &$\bullet$ \cite{chen2017caching} & $\bullet$ RNNs. &$\surd$ &&$\surd$ \\
\hline

\multirow{3}{*}{\textbf{VR}} &$\bullet$  Head movement prediction. &$\bullet$ \cite{chen2018federated}  &$\bullet$ RNNs.  & $\surd$ &&$\surd$ \\
&$\bullet$ Resource allocation. &$\bullet$ \cite{VRTL,VROWNchen}  &$\bullet$ RNNs. & &&$\surd$. \\
&$\bullet$ VR content caching and transmission. &$\bullet$ \cite{chen2018echo}  &$\bullet$ DNNs. & &&$\surd$. \\
\hline
\multirow{5}{1.6cm}{\textbf{Caching and\\~Computing}} & \multirow{3}{*}{$\bullet$ Cache replacement.}&$\bullet$ \cite{cobb2008web}  &\multirow{1}{*}{$\bullet$ FNNs.}&   &$\surd$& \\
&&$\bullet$ \cite{8478380} &$\bullet$ DNNs.  & $\surd$&& \\
&&$\bullet$ \cite{8513863} &$\bullet$ DNNs.  & &&$\surd$ \\
\cline{2-7}
& {\multirow{1}{*}{$\bullet$ Content popularity prediction.}}&$\bullet$ \cite{7873292}& \multirow{1}{*}{$\bullet$ FNNs.} & $\surd$ && \\
& {\multirow{1}{*}{$\bullet$ Content request distribution prediction.}} &$\bullet$ \cite{chen2017caching}, \cite{chen2017echo}   &\multirow{1}{*}{$\bullet$  RNNs.} &  $\surd$&&\\

\hline
\multirow{7}{1.5cm}{\textbf{ Multi-RAT}} &\multirow{2}{*} {$\bullet$ Resource management. }&$\bullet$ \cite{fuzzy}  &\multirow{1}{*}{$\bullet$ DNNs.}&   &&$\surd$ \\
&&$\bullet$ \cite{Chen2016Echo} &  $\bullet$  RNNs.  & &&$\surd$ \\
\cline{2-7}
& {$\bullet$ RAT selection. }& $\bullet$  \cite{8353153}&  $\bullet$  CNNs.    & $\surd$ &&\\

&\multirow{1}{*}{$\bullet$ Transmission technology classification.}&$\bullet$ \cite{MLP_RAT}& \multirow{1}{*}{$\bullet$ FNNs.   } & $\surd$&&\\
&$\bullet$ Multi-radio packet scheduling. &$\bullet$  \cite{hopfield} & $\bullet$  RNNs. &  &&$\surd$\\
& {$\bullet$ Mode selection.}&$\bullet$ \cite{8468000} & $\bullet$  FNNs.  &  &&$\surd$ \\
& {$\bullet$ Automatic root cause analysis.}&$\bullet$ \cite{you2019ai} & $\bullet$  RNNs.  &  &$\surd$& \\
\hline
\multirow{8}{0.7cm}{\textbf{IoT}} &$\bullet$ Model IoT as ANNs. &$\bullet$ \cite{kaminski2017neural}, \cite{ning2011future} &\multirow{1}{1.5cm}{$\bullet$ FNNs.}&   $\surd$ &$\surd$& \\

&$\bullet$ Failure detection. &$\bullet$  \cite{naidu1990use} & $\bullet$ FNNs.    & $\surd$&&\\

&\multirow{1}{*}{$\bullet$ User activities classification.}&$\bullet$ \cite{alam2016analysis} & \multirow{1}{1.5cm}{$\bullet$  DNNs. } & &$\surd$&. \\

&  \multirow{1}{*}{$\bullet$ Tracking accuracy improvement.} & $\bullet$ \cite{luo2016laguerre}   &  $\bullet$  DNNs.   & &$\surd$& \\
 &$\bullet$ Image detection.& $\bullet$  \cite{du2017reconfigurable} &$\bullet$ CNNs.&$\surd$&&\\
 &{$\bullet$ Data sampling.}&$\bullet$ \cite{8496746} &$\bullet$ PNNs. &&$\surd$&\\
 & {$\bullet$ Entity state prediction.} &$\bullet$ \cite{8502822} & $\bullet$ DNNs.&$\surd$ &&\\
  & {$\bullet$ Target surveillance.} &$\bullet$ \cite{8463616} & $\bullet$ FNNs.  & $\surd$ && \\
\hline

\end{tabular}
 \vspace{-0.2cm}
}
\end{table*}

\begin{table*}
  {\color{black}
\centering
  \newcommand{\tabincell}[2]{\begin{tabular}{@{}#1@{}}#1\end{tabular}}
\renewcommand\arraystretch{1}

 \caption{
    \vspace*{-0.1em}Summary of the use of ANNs for Specific Wireless Problems}\label{ta:UAV}\vspace*{-0.6em}
\centering
\begin{tabular}{|c| l | c | c | c | c | c | c | c |c|c|}
\hline
 \multicolumn{1}{|c|}{\multirow{3}{2.8cm}{\textbf{Wireless networking related problems}}} &  \multicolumn{1}{|c|}{\multirow{3}{*}{\textbf{ Challenges }}} &   \multicolumn{1}{|c|}{\multirow{3}{0.7cm}{\textbf{ ANN\\ $\;$Tools}}} &\multicolumn{2}{|c|}{\multirow{1}{*}{\textbf{  Use Case }}}  &   \multicolumn{6}{|c|}{\multirow{1}{*}{\textbf{ Relevant applications }}}  \\
 \cline{4-11}
 &&& {\multirow{2}{*}{{DA}}}& {\multirow{2}{*}{{RL}}}& {\multirow{2}{*}{{UAV}}} & {\multirow{2}{*}{{VR}}} & {\multirow{2}{*}{{MECC}}} & {\multirow{2}{0.7 cm}{{Multi\\~RAT}}}& {\multirow{2}{*}{{IoT}}} & {\multirow{2}{1 cm}{{Physical\\~Layer}}} \\
 &&&&&&&&&&\\
 \hline
 
     {\multirow{5}{3.2cm}{{  Resource allocation}}} & {\multirow{1}{4.3cm}{{$\bullet$ Large networks and action spaces.}}}     & {\multirow{5}{0.7cm}{{RNNs DNNs}}}&& {\multirow{5}{*}{{ $\surd$}}}  &  {\multirow{5}{*}{{ $\surd$}}}& {\multirow{5}{*}{{ $\surd$}}}& {\multirow{5}{*}{{ $\surd$}}}& {\multirow{5}{*}{{ $\surd$}}} & {\multirow{5}{*}{{ $\surd$}}}& {\multirow{5}{*}{{ $\surd$}}} \\         
  &$\bullet$ Need for self-organizing solutions. &  &&& &&&&&\\ 
   &{\multirow{1}{5.1cm}{{$\bullet$  Resource allocation variables are coupled. }}}   &  &&& &&&&&\\ 
      & $\bullet$  Need for self-organizing solution.  &  &&& &&&&&\\ 
         & $\bullet$ Non-convex optimization problems.  &  &&& &&&&&\\ 
   \hline
        {\multirow{4}{3.2cm}{{Wireless-aware path planning for autonomous systems (e.g., UAVs)}}} & {\multirow{1}{4.3cm}{{$\bullet$ Involves time-dependent locations.}}}     & {\multirow{4}{0.7cm}{{RNNs DNNs}}}&& {\multirow{4}{*}{{ $\surd$}}}  &{\multirow{4}{*}{{ $\surd$}}} & && && \\         
  &$\bullet$ Driven by environmental data. &  &&& &&&&&\\ 
   & $\bullet$ Need for adaptation to dynamic settings. &  &&& &&&&&\\ 
      & $\bullet$ Require distributed solutions. &  &&& &&&&&\\ 
   \hline
           {\multirow{4}{3.2cm}{{Channel modeling and estimation}}} & {\multirow{1}{5cm}{{$\bullet$ Unknown relationship between the received and transmitted signals.}}}     & {\multirow{4}{*}{{SNNs }}}&{\multirow{4}{*}{{ $\surd$}}} & &{\multirow{4}{*}{{ $\surd$}}} &{\multirow{4}{*}{{ $\surd$}}} && {\multirow{4}{*}{{ $\surd$}}}&{\multirow{4}{*}{{ $\surd$}}}&{\multirow{4}{*}{{ $\surd$}}} \\         
  & &  &&& &&&&&\\ 
   & $\bullet$ Need for estimation of wireless channels.  &  &&& &&&&&\\ 
      &{\multirow{1}{4.3cm}{{$\bullet$ Need for modeling solutions that can adapt to time-varying channels.}}}   &  &&& &&&&&\\ 
         &  &  &&& &&&&&\\ 
   \hline
   
              {\multirow{4}{*}{{Handover}}} & {\multirow{1}{5cm}{{$\bullet$ Handover often involves dynamic mobility thus requiring adaptive solutions.}}}     & {\multirow{4}{0.7cm}{{RNNs}}}&& {\multirow{4}{*}{{ $\surd$}}}  &{\multirow{4}{*}{{ $\surd$}}} &{\multirow{4}{*}{{ $\surd$}}} && {\multirow{4}{*}{{ $\surd$}}}&{\multirow{4}{*}{{ $\surd$}}}& \\         
  &  &  &&& &&&&&\\ 
   & $\bullet$ Need for on-the-fly decisions.  &  &&& &&&&&\\ 
    & $\bullet$ Optimized variables are binary.  &  &&& &&&&&\\ 
   \hline
   
                 {\multirow{4}{3.2cm}{{Wireless user behavior estimation}}} & {\multirow{1}{4.3cm}{{$\bullet$ User behavior is correlated in time.}}}     & {\multirow{4}{0.7cm}{{RNNs DNNs}}}&{\multirow{4}{*}{{ $\surd$}}}   &  &{\multirow{4}{*}{{ $\surd$}}} &{\multirow{4}{*}{{ $\surd$}}} &{\multirow{4}{*}{{ $\surd$}}}&{\multirow{4}{*}{{ $\surd$}}} &{\multirow{4}{*}{{ $\surd$}}}&{\multirow{4}{*}{{ $\surd$}}} \\         
  & {\multirow{1}{5 cm}{{$\bullet$ User behavior involves underlying factors that must be characterized.}}}   &  &&& &&&&&\\ 
   & &  &&& &&&&&\\ 
     &  {\multirow{1}{5 cm}{{$\bullet$ User behaviors vary across time scales.}}}   &  &&& &&&&&\\ 
   \hline

     {\multirow{3}{3.2cm}{{Wireless content prediction }}} & {\multirow{1}{4.4cm}{{$\bullet$ Content is time and user dependent.}}}     & {\multirow{3}{0.7cm}{{SNNs DNNs}}}&{\multirow{3}{*}{{ $\surd$}}} &  &{\multirow{3}{*}{{ $\surd$}}} &{\multirow{3}{*}{{ $\surd$}}} &{\multirow{3}{*}{{ $\surd$}}}& &{\multirow{3}{*}{{ $\surd$}}}& \\         
  &   {\multirow{1}{4.3cm}{{$\bullet$ Content requests is often arbitrary. }}}                 &  &&& &&&&&\\ 
   & {\multirow{1}{5cm}{{$\bullet$ Predictions is focused on data analytics.}}}    &  &&& &&&&&\\ 
   \hline
   
        {\multirow{3}{3.2cm}{{Content delivery format and method (e.g., $360^\circ$ or $120^\circ$ contents) }}} & {\multirow{1}{5cm}{{$\bullet$ Need to consider users' requirements.}}}     & {\multirow{3}{0.7cm}{{RNNs DNNs}}}&& {\multirow{3}{*}{{ $\surd$}}}  &{\multirow{3}{*}{{ $\surd$}}} &{\multirow{3}{*}{{ $\surd$}}} &{\multirow{3}{*}{{ $\surd$}}}& && \\         
  &   {\multirow{1}{4.3cm}{{$\bullet$ Optimized variables are discrete. }}}                 &  &&& &&&&&\\ 
   & {\multirow{1}{4.5cm}{{$\bullet$ Content requests are time varying.}}}    &  &&& &&&&&\\ 
   \hline
          {\multirow{2}{3.2cm}{{Users and computational tasks clustering }}} & {\multirow{1}{5cm}{{$\bullet$ High complexity to scan all of users
and computational tasks.}}}     & {\multirow{2}{0.7cm}{{CNNs}}}&{\multirow{2}{*}{{ $\surd$}}} &  &{\multirow{2}{*}{{ $\surd$}}} &{\multirow{2}{*}{{ $\surd$}}} &{\multirow{2}{*}{{ $\surd$}}}& &{\multirow{2}{*}{{ $\surd$}}}& \\         
  &              &  &&& &&&&&\\ 
  
     \hline
           {\multirow{4}{3.2cm}{{Computational time and demand predictions of each task requested by each user}}} & {\multirow{1}{5cm}{{$\bullet$ Computational time and demands are time-dependent and continuous.}}}     & {\multirow{4}{*}{{SNNs }}}&{\multirow{4}{*}{{ $\surd$}}} &  & & &{\multirow{4}{*}{{ $\surd$}}}& && \\         
  &                &  &&& &&&&&\\ 
   & {\multirow{1}{4.5cm}{{$\bullet$ Predictions driven by users' other behaviors and information.}}}    &  &&& &&&&&\\ 
     &                &  &&& &&&&&\\ 
   \hline
              {\multirow{3}{*}{{ Detection of LoS links}}} & {\multirow{1}{5cm}{{$\bullet$ LoS links are dynamic and time-varying.}}}     & {\multirow{3}{0.7cm}{{SNNs DNNs }}}&{\multirow{3}{*}{{ $\surd$}}} &  &{\multirow{3}{*}{{ $\surd$}}} &{\multirow{3}{*}{{ $\surd$}}} &&{\multirow{3}{*}{{ $\surd$}}} &{\multirow{3}{*}{{ $\surd$}}}&{\multirow{3}{*}{{ $\surd$}}} \\         
  &  {\multirow{1}{5cm}{{$\bullet$ Need to observe the physical channel.}}}                  &  &&& &&&&&\\ 
   & {\multirow{1}{4.5cm}{{$\bullet$ Need to track users' mobility.}}}    &  &&& &&&&&\\ 
   \hline
              {\multirow{5}{*}{{ Antenna tilting}}} & {\multirow{1}{5cm}{{$\bullet$ Must estimate the angle of the receiver's antennas.}}}     & {\multirow{5}{*}{{SNNs }}}&{\multirow{5}{*}{{ $\surd$}}} & {\multirow{5}{*}{{ $\surd$}}}  & & && &&{\multirow{5}{*}{{ $\surd$}}} \\         
  &                &  &&& &&&&&\\ 
   & {\multirow{1}{5cm}{{$\bullet$ Requires intelligent tracking of transmitter-receiver coupling.}}}    &  &&& &&&&&\\ 
     &                &  &&& &&&&&\\ 
          &  {\multirow{1}{5cm}{{$\bullet$ Must be executed in a short time.}}}               &  &&& &&&&&\\ 
   \hline
{\multirow{3}{3.2cm}{{Data compression and recovery for data transmission and caching}}} & {\multirow{1}{5 cm}{{$\bullet$ High complexity of data scanning.}}}     & {\multirow{3}{*}{{CNNs}}}&{\multirow{3}{*}{{ $\surd$}}}  &  &  {\multirow{3}{*}{{ $\surd$}}}& {\multirow{3}{*}{{ $\surd$}}}& {\multirow{3}{*}{{ $\surd$}}}& & {\multirow{3}{*}{{ $\surd$}}}&\\         
  &$\bullet$ Correlation among user data. &  &&& &&&&&\\ 
   & {\multirow{1}{5 cm}{{$\bullet$ Lack of prior models on user identities.}}}    &  &&& &&&&&\\ 

   \hline
           {\multirow{4}{3.2cm}{{User and device identifications}}} & {\multirow{1}{*}{{$\bullet$ A large amount of input data..}}}     & {\multirow{4}{*}{{DNNs }}}&{\multirow{4}{*}{{ $\surd$}}} &  & & &{\multirow{4}{*}{{ $\surd$}}}& && \\         
  &  $\bullet$ Large-scale nature of the network.              &  &&& &&&&&\\ 
   & {\multirow{1}{4.5cm}{{$\bullet$ Presence of large volumes of data.}}}    &  &&& &&&&&\\ 
     &  $\bullet$   High churn and dynamics           &  &&& &&&&&\\ 
   \hline
              {\multirow{3}{*}{{ IoT device management}}} & {\multirow{1}{5cm}{{$\bullet$ Diversity of IoT devices.}}}     & {\multirow{3}{*}{{ DNNs }}}&&{\multirow{3}{*}{{ $\surd$}}}   &{\multirow{3}{*}{{ $\surd$}}} &{\multirow{3}{*}{{ $\surd$}}} &&&{\multirow{3}{*}{{ $\surd$}}}& \\         
  &  {\multirow{1}{5cm}{{$\bullet$ Large-scale nature of the IoT system.}}}                  &  &&& &&&&&\\ 
   & {\multirow{1}{4.5cm}{{$\bullet$ High churn and dynamics in IoT.}}}    &  &&& &&&&&\\ 
   \hline
              {\multirow{3}{3.2cm}{{Wireless network modeling}}} & {\multirow{1}{5cm}{{$\bullet$ Diversity of mobile devices.}}}     & {\multirow{3}{*}{{DNNs }}}&{\multirow{3}{*}{{ $\surd$}}} &  & {\multirow{3}{*}{{ $\surd$}}} && & {\multirow{3}{*}{{ $\surd$}}} &{\multirow{3}{*}{{ $\surd$}}} &\\         
  &    {\multirow{1}{5cm}{{$\bullet$ Need to identify various mobile devices.}}}                   &  &&& &&&&&\\ 
   &  {\multirow{1}{5cm}{{$\bullet$ Need to adapt to dynamic environment.}}}         &  &&& &&&&&\\ 
\hline

             {\multirow{4}{3.2cm}{{ Autonomous vehicle (e.g., UAV) trajectory prediction}}} & {\multirow{1}{5cm}{{$\bullet$ A UAV's trajectory is continuous.}}}     & {\multirow{4}{*}{{SNNs }}}&{\multirow{4}{*}{{ $\surd$}}} &   & {\multirow{4}{*}{{ $\surd$}}} && & & &\\         
  &    {\multirow{1}{5cm}{{$\bullet$ Trajectory is time-dependent.}}}                   &  &&& &&&&&\\ 
   &    {\multirow{1}{5cm}{$\bullet$ Trajectory depends on wireless\\~$\;\;$parameters (e.g., interference).}}        &   &&& &&&&&\\ 
    &       &   &&& &&&&&\\  
\hline
  {\multirow{3}{*}{{Wireless data correlation }}} & {\multirow{1}{5cm}{{$\bullet$ Data is correlated in time and space\\$\;\;$ domain.}}}     & {\multirow{3}{0.7cm}{{RNNs CNNs}}}&{\multirow{3}{*}{{ $\surd$}}} &  &{\multirow{3}{*}{{ $\surd$}}} &{\multirow{3}{*}{{ $\surd$}}} &{\multirow{3}{*}{{ $\surd$}}}& &{\multirow{3}{*}{{ $\surd$}}}& \\  
     &    &   &&& &&&&&\\       
   & {\multirow{1}{4.3cm}{{$\bullet$ Need to process large sized data. }}}    &  &&& &&&&&\\ 
   \hline
\end{tabular}
 \vspace{-0.2cm}
}
\end{table*} }

 In summary, for wireless communications, ANNs have two important use cases: 1) ANN-based RL algorithms for network control, resource management, user association, and interference alignment, and 2) intelligent data analytics for signal detection, spectrum sensing, channel state detection, energy prediction, as well as user behavior predictions and classifications. In this subsection, we first summarize the advantages, challenges, and limitations of ANN based RL algorithms for wireless communication applications. Then, we introduce the advantages, challenges, and limitations of using ANNs for data analytics in wireless networks.

\subsubsection{\color{black} Advantages of ANN-based RL Algorithms}
In general, RL algorithms based on ANNs can be used for wireless network control and resource management as the wireless network states and conditions are unknown, as shown in the example of co-existence of multiple radio access technologies. 
Moreover, RL algorithms can be used to solve non-convex optimization problems or problems in which the optimization variables are coupled, as shown in the example of wireless virtual reality.
\subsubsection{\color{black}Challenges and Limitations of ANN-based RL Algorithms}
 Implementing ANN-based RL algorithms in wireless networks also faces many challenges. First, for RL algorithms, the training complexity increases quickly as the number of BSs or users that implement RL algorithms increases. In consequence, one needs to find a smart training method to decrease the training complexity. Moreover, the complexity and convergence of RL algorithms that rely on ANNs can be challenging to characterize analytically.
 Recently, most of the existing works use models based on Markov decision processes (MDPs) and game theory to analyze the convergence of RL algorithms. In fact, RL algorithms can also be used for the problems that cannot be modeled by MDP or game theory models. However, the convergence of these problems is often challenging to ascertain analytically and, thus, one has to rely on simulations. 
  In addition, one must reduce the computational resources and power needed for the ANN-based RL algorithms that must be implemented at wireless devices. In fact, for ANN-based RL algorithms, the number of actions and states must be finite. In this case, ANN-based RL algorithms need to be carefully designed if they are to be used to solve the problems that have continuous states and actions. 

{\color{black}
\subsubsection{\color{black}Advantages of ANN-based Data Analytics Algorithms} The second important use case of ANNs in wireless networks is data analytics. In wireless networks, most of the collected data will be time-dependent. For example, mobile user behaviors, wireless signals, and novel energy are all time-dependent. In consequence, wireless operators can use RNNs for user behavior prediction, signal detection, channel modeling, and energy prediction. In particular, due to the unique neuron connection method (each neuron in one layer can connect to the neurons in previous layers) of RNNs, they are effective in dealing with time-dependent data. Moreover, one can use CNNs, a type of DNNs, for modulation classification, as done in~\cite{o2017introduction}. CNNs can also be used to analyze the images captured by the mobile devices such as VR devices and UAVs so as to extract the features of captured images. The  features extracted by CNNs can be used for the users movement identification, environment identification, and data compression and recovery which can be used for wireless network control and data traffic offloading.  
For example, one can use CNNs for data compression at the transmitters and data recovery at the receivers so as to reduce the traffic load over the transmission links between transmitters and receivers. 
Meanwhile, since SNNs consist of spiking neurons, they are effective in dealing with continuous data. In consequence, one can use SNNs for signal detection, channel modeling, channel state detection, and wireless device (aerial or ground) identification. For example, one can use both continuous flying trajectory and radio frequency signals as the input of SNNs to identify UAVs and then tweak their transmission parameters.
 \subsubsection{\color{black} Challenges and Limitations of ANN-based Data Analytics Algorithms} 
 Implementing ANNs for data analytics in wireless networks also faces many challenges. First, the data related to the behavior of mobile users is not easy to collect due to privacy concerns. For instance, a network operator such as Verizon can collect only partial datasets related to the mobile users. 
 Due to this partial availability of datasets, the prediction accuracy of ANNs can be compromised. 
 Second, for data analytics, existing ANN-based learning algorithms cannot be readily implemented at the mobile devices such as smartphones due to high training complexity and energy consumption. In fact, small IoT or wearable devices such as watches and IoT sensors, or even smartphones, can record more data related to the users' environment compared to BSs that are located far away from the users. In consequence, if an ANN learning algorithm can be implemented at wearable and carriable devices, it can use more data related to the users' behaviors for training purpose and, hence, the prediction accuracy can be improved, while also alleviating privacy concerns. One possibility to overcome this challenge is to train at a BS or cloud then implement the trained ANNs at the users' device. 
 Third, distributed ANN learning algorithms are needed for wireless networks. In particular, mobile users will connect to the different BSs as they move from one cell to another. In this case, the data related to such mobile user may be located at different BSs and the BSs may not be able to exchange the collected data due to limited capacity of backhaul links. In consequence, a distributed ANN learning algorithm is needed for data analytics as the users' data is located at different BSs. One possibility to overcome this challenge is to leverage the emerging idea of federated learning \cite{SmithFederated2017} that enables distributed learning. Moreover, the training complexity of ANN-based data analytics algorithms can be higher than other ML tools such as ridge regression. In consequence, one must balance the tradeoff between prediction accuracy and training complexity. Finally, training ANNs may require a large amount of training data (depending on the application) and such data may not be always readily available in a wireless network.
 
  {\color{black}Table \ref{ta:EX} summarizes the type of ANNs and learning algorithms used for each existing work in each application. Based on this table, one can identify the advantages, disadvantages, and limitations of each learning algorithm for all types of problems encountered in the literature.} Table \ref{ta:UAV} provides a summary of the key wireless networking problems that can be solved by using ANNs along with the challenges and relevant applications. }

\section{Conclusion}
In this paper, we have provided one of the first comprehensive tutorials on the use of artificial neural networks-based machine learning for enabling a variety of applications in tomorrow's wireless networks. In particular, we have presented an overview of a number of key types of neural networks such as recurrent, spiking, and deep neural networks. For each type, we have overviewed the basic architecture as well as the associated challenges and opportunities. Then, we have provided a panoramic overview of the variety of wireless communication problems that can be addressed using ANNs. In particular, we have investigated many emerging applications including unmanned aerial vehicles, wireless virtual reality, mobile edge caching and computing, Internet of Things, and multi-RAT wireless networks. For each application, we have provided the main motivation for using ANNs along with their associated challenges while also providing a detailed example for a use case scenario. Last, but not least, for each application, we have provided a broad overview on future works that can be addressed using ANNs. Clearly, the future of wireless networks will inevitably rely on artificial intelligence and, thus, this paper provides a stepping stone towards understanding the analytical machinery needed to develop such a new breed of wireless networks. 





\def\baselinestretch{0.94}
\bibliographystyle{IEEEbib}
\bibliography{references}
\end{document}